\documentclass[notitlepage,showpacs,aps,floatfix,13pt]{revtex4-1}
\input{revstyle.tex}

\makeatletter \fussy 

\input epsf 
\usepackage{graphicx}
\usepackage[outdir=./]{epstopdf}
\usepackage{caption} 
\usepackage{subcaption}

\begin{document}
\title{Meson modes  of probe $D7$-branes in Klebanov-Strassler background}   
\author{M. Chemtob} \email{marc.chemtob@ipht.fr}   
\affiliation{Universit\'e Paris-Saclay, CNRS, CEA, Institut de physique
th\'eorique, 91191, Gif-sur-Yvette, France} 
\thanks {\it Supported by 
Direction  de la Recherche Fondamentale et aux  Energies  Alternatives
Saclay} \date{\today} 
\begin{abstract}
We examine the properties of 4-d bosonic and fermionic  (mesons
and mesinos)  modes on $D7$-branes wrapped over  Kuperstein 4-cycle of
the  Klebanov-Strassler background.  The wave equations are derived in a
convenient parameterization of the warped deformed conifold
metric. The mass spectra, wave functions and couplings of bosonic
modes are evaluated by means of semi-classical tools.  We explore
the possible existence of fermionic zero modes.
\end{abstract}     
\maketitle \renewcommand{\thefootnote}{alph{footnote}}
\tableofcontents\addtocontents{toc}{\protect\setcounter{tocdepth}{1}}


\section{Introduction} 
Branes and fluxes are necessary ingredients~\cite{giveon98} in
building solutions of superstring theories dual to confining gauge
theories that include flavour degrees of
freedom~\cite{PolchGrana,karchkatz02,gimon02,kruczenski03,marche08,marche10}
and/or chiral
fermions~\cite{acharya99,conlon08,hartnoll02,acharya06,abe15}.  The
pioneering applications to hadronic physics considered configurations
of Dirichlet
$D3/D7,\ D4/D6$-branes~\cite{kruczenski03,hongler04,kirsch06,mateos07}
or $ D4/D8$-branes~\cite{witten98,Sakaito04,aharony06} embedded in
their own backgrounds.  (See the
reviews~\cite{peetklar07,erdmenger07,mateosR07}.) For
superbranes embedded in curved spacetimes~\cite{minasian99},
useful  help came  from  the  constraints imposed by $\kappa
$-symmetry~\cite{bergshoe96,bergshoe97}  for
internal space manifolds of calibration type
supporting magnetized or instantonic
backgrounds~\cite{marino99,gauntlett03,martucci05,lustmartsim08}.
The Klebanov-Strassler
solution~\cite{klebstrass00} for type $ II\ b$ supergravity on the
deformed conifold  offered an ideal setting for this task.
The addressed  issues   included the classification
of holomorphic 4-cycles~\cite{arean04,arean06}, stability of
supersymmetric embeddings~\cite{chenyangshiu08}, mesons
spectroscopy~\cite{gimon02,ouyang03,kuperstein04,leviyang05,cotroneDK10},
chiral symmetry breaking~\cite{sakaisonn03,kupsonn08,dymarkupson09},
duality correspondence to Klebanov-Witten gauge theory with quark
flavours~\cite{dymarsky09,benini11}, supersymmetry breaking and its
mediation~\cite{benini09}, back-reaction on background
geometry~\cite{benillo06,beninI07}, ultraviolet
instabilities~\cite{paredes06,casero06,bigazzi0809}, and connection to
the Randall-Sundrum mechanism~\cite{gherghetta06}.

In this work we pursue this research thread by extending our previous
work on the Kaluza-Klein theory for Klebanov-Strassler
solution~\cite{chemtob22} to the flavoured background including probe
$D7$-branes.  Our analysis follows the Giddings-Kachru-Polchinski
(GKP) approach~\cite{gkp01} to flux compactification on $ M_4 \times
X_6 $ spacetimes,  involving a conic Calabi-Yau orientifold $X_6 $ with
an attached warped deformed conifold throat $\calc _6
$~\cite{candelas90}.  We introduce in this background $ D7$-branes
that extend over Minkowski spacetime $ M_4$ and wrap over
Kuperstein~\cite{kuperstein04} 4-cycle $\S _4\subset \calc _6 $.  The
supergravity solution with 5-flux  $ N= MK$ 
and 3-fluxes  $(M,\ K )$ and  a $ N_f \  D7$-brane stack
is dual to the flavoured version of
Klebanov-Witten~\cite{klewit98} supersymmetric field  theory.  This is
the  4-d gauge     theory on $(N \ D3 + M\  D5) $-branes near the conifold apex
with $N_f $ conjugate pairs of quark flavours $Q,\ \tilde Q $,  which
decouples from gravity in the 't Hooft-Veneziano limit, $N_f << N >>
1$.  Along with the glueball modes from fluctuations of the
supergravity multiplet (closed strings) fields in $\calc _6$, the
theory comprises meson modes from fluctuations of ($ (D3,D7) +
(D7,D3)$ open strings) fields in $\S _4$.

The bosonic (scalar and vector) and fermionic (spinor) modes are
examined within the streamlined
approach~\cite{kuperstein04,casero06,bigazzi0809} using the deformed
conifold parameterization as a foliation by Kuperstein 4-cycle
leaves~\cite{benini09}. The wave equations are solved by means of the
semi-classical JWKB (Jeffreys-Wentzel-Kramers-Brillouin) method.  The
information on the modes masses and wave functions is used to compute
the self-couplings of meson modes along with their couplings to
graviton modes. An attempt is also made to test the existence of
fermionic zero modes.

The contents of this paper are organized into four sections.  The
formalism for the flavoured Klebanov-Strassler background is presented
in Section~\ref{sec1}.  Subsection~\ref{sec1sub1} reviews the 
supergravity-gauge theory  dual descriptions and Subsection~\ref{sec1sub2}
discusses a simplified version of the $D7$-brane world volume action
which will be used in later sections to compute the mesons properties.

Section~\ref{sec2} discusses the Kaluza-Klein reduction of the bosonic
action for $D7$-branes wrapped on Kuperstein 4-cycle of the warped
deformed conifold.  We derive the wave equations for scalar and vector
meson modes and present numerical results for the modes wave
functions, mass spectra, self-couplings and couplings to bulk graviton
modes, all evaluated by means of the JWKB method.
Subsection~\ref{sec2sub1} discusses algebraic properties of the
4-cycle embedding in the conifold.  Subsections~\ref{sec2sub2}
and~\ref{sec2sub2p} discuss the reduced action for scalar moduli and
gauge vector boson fields on $D7$-branes.  Subsection~\ref{sec2sub3}
starts with a presentation of the free parameters and next presents
results of numerical applications for the mass spectra and
interactions of scalar and vector meson modes.

Section~\ref{sec3} focuses on the fermionic action of $D7$-branes in
the approach of~\cite{marche08}.  Subsection~\ref{sec3sub1} discusses
the reduced action and Kaluza-Klein decomposition of spinor fields and
Subsection~\ref{sec3sub2} the zero modes wave functions.
Section~\ref{sect4} summarizes the main conclusions.  Useful tools and
intermediate results are presented in
Appendix~\ref{appwdcsubD7}. Subsection~\ref{appD7sub0} discusses
geometrical properties of the 4-cycle embedding,
Subsection~\ref{appD7subB1} the $D7$-brane bosonic action,
Subsection~\ref{appD7subFe1} the Dirac operator restriction to $\S _4$
and Subsection~\ref{appD7subsing} collects intermediate results
appropriate to the singular conifold limit.

\section{Type $II \ b  $ string theory on
warped deformed   conifold with $Dp$-branes}  
\label{sec1} 

We  consider the  GKP~\cite{gkp01}  flux  compactification of
type $II\ b$ superstring theory  on the 10-d target
spacetime orientifold $ M_{10} = M_4 \times X_6 $  with
the warped deformed conifold $ \calc _6$ glued at
a conic singularity in the moduli space  of $ X_6$.  
The   classical background   in $ \calc _6$ is  described by  Klebanov-Strassler
solution~\cite{klebstrass00}  for type $II\ b$ supergravity   theory.
We   introduce in this background
probe  $D7$-branes  wrapped over an  holomorphic 4-cycle of  $ \calc _6$ 
preserving  $\caln = 1$ supersymmetry in $M_4$.
At low energies  and small  curvatures relative  to the string scale and 
up to back-reaction effects on the curved background that we neglect,
the  solution provides   a  strong coupling description (in the sense of the
anti-de Sitter conformal field theory (AdS/CFT) or the gravity-gauge duality)
of the flavoured Klebanov-Witten gauge theory~\cite{klewit98,benini11}. 
We   shall begin  the  discussion
with a brief review of the Klebanov-Strassler background~\cite{klebstrass00}.   
A simplified version of the bosonic and fermionic action of $
Dp$-branes  is presented  next in view of later applications to
the Kaluza-Klein reduction  of the $D7$-brane action.

\subsection{Flavoured Klebanov-Strassler background} 
\label{sec1sub1}

The deformed conifold $ \calc _6$ is commonly defined as the locus of
the quadratic algebraic equation in $ C^4$, \bea && Det (W) = - \sum
_{a=1}^4 w _a^2 = -z_1 z_2 + z_3 z_4 = - \e ^2 , \ [W= \pmatrix{ w^3 +
    i w^4 & w^1-i w^2 \cr w^1+i w^2 & -w^3 + i w^4 } \equiv \pmatrix{
    z_3 & z_1 \cr z_2 & z_4 } ] \label{sec1.dc} \eea where $\e \in C $
is the complex structure modulus and $w_a $ and $z_a ,\ [a=1,\cdots ,
  4]$ are  linearly related   sets of complex coordinates, $
({w_1 , - iw_2 }) = \ud (z_1 \pm z_2),\ ({w_3 , iw_4}) = \ud (z_3 \mp
z_4)$.  The conifold admits the isometry group $ SO(4) \simeq SU(2) _1 \times SU(2) _2 $
 acting   on the coordinates matrix $ W$ by left and right
 multiplication,  $ W \to g_1 W g_2 ^\dagger $.  Its intersections
with the spherical manifolds, \bea && \ud Tr (W^\dagger W )
= \sum _a \vert w _a \vert ^2 = \sum _a \vert z _a \vert ^2 = \vert \e
\vert ^2 \cosh \tau ,\label{sec1.dcI} \eea labelled by the parameter
$\tau \in R$, define radial sections of the conifold isomorphic to the compact
coset space $ T^{1,1} \simeq SU(2)_{1} \times  SU(2)_{2} / U(1) $.
The conifold parameterization combining the radial
coordinate $\tau \in [0,\infty ] $ with the five independent angle
coordinates $\T ^\a = (\t _1, \phi _1, \t _2, \phi _2, \psi )$
of $ T^{1,1}$~\cite{minasian99}, \bea && w_1= \e [\cosh {S } \cos
  \t _+ \cos \phi _+ + i\sinh {S } \cos \t _- \sin \phi _+ ] ,\ w_2=
\e [- \cosh {S } \cos \t _+ \sin \phi _+ + i\sinh {S } \cos \t _- \cos
\phi _+ ] ,\ \cr && w_3=\e [ - \cosh {S } \sin \t _+ \cos \phi _- +
i\sinh {S } \sin \t _- \sin \phi _- ] ,\ w_4= \e [- \cosh {S } \sin
\t _+ \sin \phi _- - i\sinh {S} \sin \t _- \cos \phi _- ] ,\cr &&
[S={\tau + i\psi \over 2} ,\ \t _\pm = \ud (\t _1 \pm \t _2),\ \phi
 _\pm = \ud (\phi _1 \pm \phi _2),\ (\t \in [0, \pi ] , \ \phi
 _i\in [0, 2\pi ] , \ \psi \in [0, 4\pi ] ) ]\label{app1.eq6} \eea
is encoded in the representation  of  the matrix $W$,
\bea && W= g_1 W _0 (\tau )
g_2 ^\dagger ,\ [W _0 (\tau ) = \overline{diag} (e ^{\tau /2} , e
  ^{-\tau /2} ) ,\ \varphi _1 - \varphi _2 =0 ,\ \psi   = \varphi _1 +
  \varphi _2] \eea   with  the left and right action of
$g_{i} (\t _i, \phi _i, \varphi _i) \in SU(2)_{i} ,\ [i=1, 2]$  
on the    anti-diagonal  matrix  $W _0 (\tau ) $. 
The Klebanov-Strassler solution of type $II\ b $
supergravity~\cite{klebstrass00} on $ M_4\times \calc _6$ involves
3-fluxes, $\hat l_s ^2 \int _{B, A} ( H_3,\ F_3 ) = (- K , \ M ),\ [N=
  MK,\ \hat l _s = 2\pi \sqrt {\a '} ]$ across the pair of Poincar\'e
dual 3-cycles $ B,\ A $ of $\calc _6$, with the resulting 5-flux $\hat
l_s ^4 \int _{T^{1,1} } F_5= N $ sourcing a warped spacetime of
classical metric, \bea && ds ^2 _{10} = h^{-\ud } (\tau ) d\tilde s ^2
_4 + h^{\ud } (\tau ) d \tilde s ^2 (\calc _6) , \ [h (\tau ) \equiv e
  ^{-4 A (\tau ) } =2 ^{2/3} (g_s M \a ' )^2 \e ^{-8 /3} I (\tau ) ,
  \cr && d\tilde s^2_6 (\calc _6) = {\e ^{4/3} K(\tau ) \over 2} (
      {1\over 3 K ^3(\tau ) } (d \tau ^2 + (g^{(5)} )^2 ) + \cosh ^2
      ({\tau \over 2}) ((g^{(3)})^2 + (g^{(4)} )^2 ) + \sinh ^2 ({\tau
        \over 2}) ((g^{(1)})^2 + (g^{(2)})^2 ) ) ,\cr && K(\tau )= {
        (\sinh (2\tau ) - 2 \tau) ^{1/3} \over 2^{1\over 3}\sinh \tau
      } ,\ I (\tau ) = \int _\tau ^\infty dx {x \coth x -1\over \sinh
        ^2 x } (\sinh (2 x) - 2 x) ^{1\over 3} ] \label{sec1.ks} \eea
asymptotic to the direct product metric of the spacetime $ AdS _5
\times T^{1,1}$,     describing  the background for
$ N D3 + M D5$-brane stacks embedded  near the  conifold apex.
We follow same notations
as~\cite{chemtob16,chemtob22} where $h (\tau ) $ is the warp profile,
$ K(\tau ),\ I(\tau )$ are auxiliary radial functions, $ g^{(a)} (\T )
$ is the diagonal basis of left-invariant 1-forms on $T^{1,1}$ and $
g_s ,\ \a ' = 1/m_s^2 $ denote the string theory coupling constant and
scale parameters. The classical solutions for the 2- and 3-form fields
are quoted in Eq.~(\ref{eq.pullJBC}) below.

In the limit $ N>> 1$, fixed $ g_s N$, the dual gauge theory on the
regular $ N D3 $-  and  fractional  $M D5$-branes
is the Klebanov-Witten 4-d $\caln =1 $ supersymmetric theory with local 
symmetry group $ \calg = \calg_1 \times \calg _2= SU(N+M) \times
SU(N)$~\cite{klewit98} of coupling constants $ g_{1,2} $ and global
(flavour) symmetry group $ G= SU(2)\times SU(2)\times U(1)_b\times Z_2
^r ,\ [ Z_2 ^r \subset U(1)_r] $.  The small curvature regime $ g_s N
>> 1 $, where classical supergravity theory is valid, corresponds to
't Hooft limit $\l = g_{1,2} ^2 N >> 1$ of the gauge theory.  The
gauge sector consists of vector supermultiplets $ V_{1,2} $ in the
adjoint representations of $\calg _{1,2} $ and the (matter) flavour
sector consists of two pairs of chiral supermultiplets in the
bi-fundamental representations, $ A_i \sim ({N+M} ,
\overline{N}),\ B_i \sim (\overline{N+M} , {N}), \ [i=1,2] $ of $
\calg $ and the representations $ (2, 1)_{1, \ud } , \ (1, 2)_{-1, \ud
} $ of $ G $.  The renormalization group evolution features a
cascading type flow towards the infrared through self-similar Seiberg
type dualities $ SU(N+M) \times SU(N) \to SU(N-M) \times SU(N) $.  For
$ N= kM ,\ [k\in Z]$ the successive duality steps $\d k = -1 $ end
into the confining pure gauge theory $ SU(M)$~\cite{dymklebseib05}.
In the presence of the interaction superpotential $ W_{KW} = {h\over
  2} \e ^{ik} \e ^{jl} Tr (A_i B_j A_k B_l ) ,$ the renormalization
group flow (for $ N>> M $) drives the theory to a superconformal
(infrared) fixed submanifold $h= h (g_1, g_2 )$ in the space of
coupling parameters $g_1,\ g_2 ,\ h $~\cite{leighstrassler95}.  The
fields and superpotential are assigned in the limit $M\to 0 $ the
scaling dimensions $\D (\phi )= d (\phi )+ \g (\phi ) / 2$ and
(unbroken) R-symmetry charges $R$ (including quantum contributions),
\bea && R (A_i) = 2 \D (A_i ) / 3 = 1/2 ,\ R (B_i) = 2 \D (B_i ) / 3 =
1/2 ,\ \D (\l ) =3/2 ,\ R(\l ) = 1 ,\ R (W) = 2 \D (W) /3 = 2 .\eea

The probe $N_f \ D7$-branes  wrapped over a 4-cycle $\S _4
\subset \calc _6$, add   modes of open $F1$-strings
stretched between the $ N_f D7$- and $ (N D 3 + M D5)$-branes.
Massless quark modes are present if $\S _4 $ touches the conifold apex
where the $D3$-branes are located.  The gauge dual theory is then the
flavoured Klebanov-Witten gauge theory with $ N_f $ conjugate pairs of
quarks and anti-quarks $Q,\ \tilde Q$~\cite{karchkatz02} in
bi-fundamental representations of the colour and flavour groups.  One
must consider holomorphic and topologically trivial $\S _4 $ in order
to preserve $\caln =1 $ 4-d supersymmetry along with $D3$-charge
conservation and $C_4$-tadpole cancellation.  We shall specialize to
the Kuperstein 4-cycle~\cite{kuperstein04}, $ w_4 =\mu $, sub-locus of
the conifold $\sum_{a=1} ^4 w_a ^2 = \e ^2$, where the complex
parameter $\mu $ (of same  energy dimension $ E^{-3/2}$ as $\e $) sets the
minimal classical radial distance between the apexes of $\calc _6$ and
$\S _4 $.  Note that our $\e ^2 ,\ \mu $ correspond to the $ \e ,\ \mu
/\sqrt 2 $ of~\cite{benini11}.

In full generality, the $ D7$-branes stack can be split into left
and right sub-stacks,  $ N_f = N_{f_L} + N_{f_R} $. The gauge theory is
then described by an horizontal type quiver diagram with $ Q_{L}
,\ \tilde Q_L$ in bi-fundamentals of $SU(N_{f_L}) \times SU(N+M) $ and
$ Q_{R} ,\ \tilde Q_R$ in bi-fundamentals of $ SU(N) \times
SU(N_{f_R})$.  (For comparison, the Ouyang embedding~\cite{ouyang03},
$ z_4 = \mu $, involves instead the vertical type quiver diagram with
$ Q_{L} ,\ \tilde Q_R$ charged under $ SU(N+M)$ and $ Q_{R} ,\ \tilde
Q_L$ charged under $ SU(N)$.)  The dual theory of gauge symmetry group
$\calg = SU (N_{f_L} )\times SU(N+M) \times SU(N) \times SU (N_{f_R}
)$ includes,   in addition to $A_i,\ B_i $, a quark flavour sector
with left and right (conjugate pairs) of quarks supermultiplets in the
$\calg $ group representations \bea && Q _L \sim (N_{f_L} , \overline{
  N +M } , 1, 1 ) ,\ \tilde Q _L \sim ( \bar N_{f_L} , N +M , 1, 1 )
,\ Q _R \sim (1, 1, \bar N , N_{f_R} ) ,\ \tilde Q _R \sim (1, 1, N,
\bar N_{f_R} ) .  \eea The total quartic order superpotential with
(implicit) traces over the gauge and quark flavour quantum numbers, \bea && W= h
(A_1 B_1 A_2 B_2 - A_1 B_2 A_2 B_1) - \eta _L \sqrt {h} \tilde Q _L
(A^i B_i - {\mu \over \sqrt h } ) Q_L + {\eta _L ^2 \over 2} (\tilde Q
_L Q_L)^2 \cr && - \eta _R \sqrt {h} \tilde Q _R (B^iA_i - {\mu \over
  \sqrt h } ) Q_R - {\eta _R^2 \over 2} (\tilde Q _R Q_R)^2
, \label{eq.supfl} \eea drives the theory (in the quarks massless
limit $\mu \to 0 $) to a superconformal fixed submanifold on which $
\D (Q_{L,R} ) = 3/4,\ R (Q_{L,R} ) =1/2 $.  One can always rescale the
superfields $ Q_{L, R}$ so as to set $\eta _{L, R} = \sqrt h$ in the
superpotential $W$ which then depends on the overall coupling constant
$ h$ and the quarks bare mass parameter $ m_Q \simeq \mu / \sqrt h .$
The correspondence between the dimensional parameters on the string
theory side $\e,\ \mu $ and the dynamical and bare quark mass scales
$\L _{ir},\ m_Q $ on the gauge theory side, $ \e ^{2/3} \simeq \a ' \L
_{ir} $ and $ \mu ^{2/3} \simeq \a ' m _Q $, is set via dimensional
analysis.

 The structure of the superpotential in Eq.~(\ref{eq.supfl}) is
 formally similar to that of the $SU(N_c),\ \caln = 2 $ supersymmetric
 gauge theory~\cite{CarlKonishi00} with a massive adjoint superfield
 $\phi $ coupled to massive $ N_f (Q, \tilde Q )$ fundamental quarks
 hypermultiplets. The decoupling of $ \phi $ along the Higgs branch,
 labelled by the scalar field component VEV of $ \phi $, deforms the
 $\caln = 2 $ theory to the $\caln = 1 $ theory and changes the
 superpotential as \bea && W _{\caln = 2} = {s\over 2} Tr (\phi ^{2} )
 + Tr (\phi \tilde Q Q) + Tr (m Q \tilde Q) \ \Longrightarrow \ W
 _{\caln = 1} = -{1\over 2 s} Tr ( (\tilde Q Q)^2 ) +Tr (m Q \tilde Q)
 .\eea

An even closer analog~\cite{benini11} to the flavoured gauge theory is
provided by the type $ II \ b$ supergravity on the (flat orbifold)
spacetime $ M_4\times (C^2 _{1,2}/ Z_2) \times C _3 $ (coordinates $
(x^\mu , w_{1,2} , w_3 ) $) with $N\ D3 $-branes at the fixed point $
w_{1,2,3} = 0 $ and $ N_f \ D7$-branes over $ M_4\times (C^2
_{1,2}/ Z_2) $ at $w_3 = m$. The 4-d supersymmetric gauge theory $
SU(N) \times SU(N_f) ,\ \caln = 2 $ consists (in $\caln = 1$ language)
of one vector and three complex chiral supermultiplets $\Phi
_{i=1,2,3} $, adjoints of $SU(N)$ (descending from the supermultiplet
of the $\caln = 4 $ theory on $ N\ D3$-branes) and $ N_f$ quark
flavour hypermultiplets $ Q = diag (Q_L,\ Q_R) ,\ \tilde Q = diag (
\tilde Q_L,\ \tilde Q_R) $. The total superpotential, including mass
terms for $\Phi _3 $ and quarks,
\bea && W= - Trace [[\Phi _1, \Phi _2 ] \Phi _3 + {s\over 2}  \Phi
  _3 ^2  + \tilde Q \Phi _3 Q - \mu ( \tilde Q_L Q_L + \tilde Q_R Q_R
  ) ], \eea yields, after the Higgs field $ \Phi _{3}= diag ( \phi _3
, - \tilde \phi _3)$ is integrated out, the $\caln = 1 $ flavoured
Klebanov-Witten gauge theory superpotential in Eq.~(\ref{eq.supfl}),
with the anti-diagonal matrix $\Phi _{1,2}= \overline {diag} (A_{1,2}
, \ \pm B^{2,1} ) $ and $ m \simeq (B^i A_i - \mu )$.

In order that the moduli space of vacua of the supergravity theory
matches that of the flavoured gauge theory, one may need to activate
classical fields in the $D7$-branes gauge
sector~\cite{douglasmoore96,dymarsky09,benini11}.  The case where no
classical gauge fields are present corresponds to having only
$Q_R$-type quarks coupled to the lower rank gauge theory factor
$SU(N)$.  We shall specialize here to this case which then involves
the single species of quarks $ Q \sim ( 1, \overline{N}, N_f) ,
\ \tilde Q (1, N, \overline{N_f}) $ of $SU(N+M)\times SU(N) \otimes
SU(N_f) $ with the total interaction superpotential $ W = W_{KW} - h[
  \tilde Q ( B^i A_i - {\mu \over \sqrt h} ) Q + {1 \over 2} (\tilde Q
  Q)^2 ].$

\subsection{$Dp$-brane action}
\label{sec1sub2}

The dimensional reduction of the $D7$-brane action on Kuperstein
4-cycle gives rise to an effective field theory in $ M_4$ with meson
modes added to the glueball modes from the supergravity multiplet.
The geometrical tools needed in applying the Kaluza-Klein approach for
submanifolds embedded in curved manifolds, are well
documented~\cite{sorokin99,willmore12}.  To prepare the ground for
this discussion, we  introduce  here general notations that will
be used throughout the text. The target spacetime $ M_{10} = M_4
\times X_6 $ is parameterized by world coordinates $X^M$ with a metric
tensor $ g_{MN}$ describing the (diffeomorphisms invariant) proper
distances squared, \bea && d s^2 ( M_{10})= g_{MN} d X^M d X^N ,\ [X^M
  = (X^\mu ,\ X^m ),\ M= 0, \cdots , 9,\ \mu =0, \cdots , 3 ,\ m=1,
    \cdots , 6 ] .
  \label{eq.indMA} \eea
  The  coordinates $ X^A $ for the  (flat) tangent spaces of (constant)
  metric tensor $\eta _{AB} = diag (-1, +1, \cdots , +1)$ are related
  to the $ X^M$ by the transformations $X^A= e^A_M X^M ,\ X^M= e_A^M
  X^A$ involving the components $e^A _M,\ e_A^M $ of the dual bases of
  (local inertial vielbein) frame vectors $(e^A,\ e_A )$ which satisfy
  the properties \bea && d s^2 (M_{10} ) = \sum _{A=0}^9 (e^A)^2
  ,\ [e^A = e ^{A} _M d X^M ,\ e_A= e _A ^M \dh _M,\ \eta _{AB} e ^{A}
    _M e ^{B} _M = g _{MN} ].  \eea Similar definitions are adopted
  for the world and flat bases of Dirac gamma matrices obeying the
  Clifford-Dirac algebra, \bea && \{\G _A ,\G _B\} = 2 \eta _{AB}
  ,\ \{\G _M ,\G _N\} = 2 g _{MN} ,\ [\G ^A= e^A_M \G ^M ,\ \G ^M=
    e_A^M \G ^A] .\eea
  
The world volume $ M_{p+1}$ of parallel $Dp$-branes in $ M_{10}$ is
parameterized by the intrinsic coordinates $\xi ^\a = (\xi ^\mu ,\ \xi
^r ),\ [\a =0, \cdots ,p ,\ \mu = 0, \cdots , 3 ,\ r =1, \cdots , p-3]
$.  Its embedding along the directions of $ M_4 $ and the
$(p-3)$-cycle $\S _{p-3} \subset \calc _6 $ splits up the coordinates
into tangential and normal subsets, $ X^M  = (X^\a ,\ X^u) = (X^{\mu }
,\ X^r , \ X^u ),\ [\a =0, \cdots ,p ,\ \mu = 0, \cdots , 3 ,\ r =1,
  \cdots , p-3,\ u =1, \cdots , 9-p ]$  regarded  as fields
$ X^M  (\xi ^\a )$  on the  world volume $ M_{p+1}$. 
On  the   $N_f \ Dp$-branes stack, the  coordinates fields
are promoted to $N_f\times N_f$ matrices in the
adjoint representation of $ U(N_f)$ with the $X^\a / \a ' $ identified
to gauge fields $ A ^\a $ and the $X^u $ to  moduli
fields. We shall restrict consideration to diagonal matrices in the
Abelian subgroup $ U(1)^{N_f}$  corresponding   to $ N_f$ copies
of single $Dp$-branes.

 
The $ D7$-brane action includes the familiar Dirac-Born-Infeld and
Chern-Simons bosonic terms, built from the embedding coordinates $X^M
$, the brane pull-backs of bulk fields $ B_2,\ C_r $ and the brane
1-form gauge field $ A_1 = A_ \a d \xi ^\a $.  The fermionic action is
constructed as an expansion in powers of the pull-back of the spinor
doublet field $\T $ of 10-d type $II \ b$
supergravity~\cite{martuccipro05}.  We restrict hereafter to the
quadratic order action in $\T $ coupled to the NSNS 2-form and RR
4-form background fields and the brane gauge field strength $F_2 = d
A_1 $.  The combined bosonic and fermionic action for $ Dp /\bar
Dp$-branes in Einstein frame \bea && S (Dp /\bar Dp) = - \tau _p \int
_{M_{p+1}} d ^{p+1} \xi e ^{(p-3 ) \phi /4 } (- Det (\g + e^{- \phi /2
} \calf ) )^{1/2} \cr && \times [1 - i \bar \T P_\mp ^{Dp} ( (\calm
  ^{-1} )^{\a \b } \G_\b D _\a + {i\over 8}\g ^{\a \b } \G_\b \Fslash
  _5 \G_\a \otimes \s _2 ) \T ] \pm \mu _p \int _{M_{p+1}} [\sum _q C
  _{q} \wedge e ^{\calf _2} ]_{p+1} ,\cr && [\tau _p ={\mu _p} = {2\pi
    \over \hat l_s ^{p+1} } ,\ \calf _2 = B_2 + F_2 ,\ \ F_2 = d A
  (\xi ) ,\ F_{q+1} = d C_q ,\
       \T (\xi ) = {\t _1\choose \t _2 },\ \bar \T = \T ^\dagger \G^0
       , \cr && \calm _{\a \b } = \g _{\a \b } + e ^{-\phi /2} \calf
       _{\a \b } \G _{(10)}\otimes \s _3, \ P_\pm ^{Dp} (\calf )= \ud
       (1 \pm \G _{Dp} (\calf ) ) ] \label{eq.branesdp} \eea is
expressed here in same notations as~\cite{marche08,chemtob22} with
$\tau _p = 2\pi /\hat l_s ^{p+1} $ and $g_{Dp} ^2 = 4 \pi g_s \hat l_s
^{p-3}$ denoting the brane tension
and gauge theory coupling constant.  The action is built from the
covariant derivatives along $M_8:\ D_\a = \nabla _\a + A_\a ,\
[ \nabla _\a =\dh _\a + \o _\a ] $  and the pull-back
transforms of the bulk spacetime metric and Dirac
matrix fields $g _{MN} ,\ \G _M$,  the  axio-dilaton scalar
fields $(\phi ,\ C_0) $ and  the $q$-form fields $B_2 ,\ C_{2,4} ,\ [F_5 =
  d C_4] $ \bea && \pmatrix{\g _{\a \b } \cr B _{\a \b } } =\dh _{\a }
X^M (\xi ) \dh _{\b } X^N (\xi ) \pmatrix{g_{MN} (X (\xi ) ) \cr B
  _{MN} (X (\xi ) )} ,\ \G _\a = \dh _\a X^M \G _M = E^M_A \dh _\a X^A
\G _A.\eea The self-dual field strength 5-form has the classical value
$ F_5^{cl} = \dh _\tau h ^{-1} (\tau ) d \tau \wedge vol (M_4).$
The (anticommuting, Grassmannian) bulk spinor field, $\T (X (\xi ) ) = (\t
_1 (X (\xi ) ) ,\ \t _2 (X (\xi ) ) ) $, of Dirac conjugate, $ \bar \T
= \T ^\dagger \G ^{\underline 0} = \T ^T \G ^{\underline 0} $,
consists of a pair of Majorana-Weyl spinor fields $\t _i (X) \in 16 $
of $ SO(9,1)$ doublet in the Pauli matrices $\s _{1,2,3} $ space. The
Weyl spinor fields of fixed 10-d chirality, $ \G _{(10)} \t _i = \t _i
,\ [\G _{(10)} = \G^{01\cdots 9} ]$ satisfy the Majorana reality
condition, \bea && B \t _i = \t _i^\star ,\ [B \G _M B^\star = \G _M
  ^\star ,\ B^\star B =1] \ \Longrightarrow \ \bar \t _i = \t ^T _i C
, \ [\G _A ^T =- C \G _A C^{-1} ,\ C= C^\dagger = C^{-1}] \eea where
the matrices $B$ and $C$, which transform the Dirac matrices to their
complex conjugate and transpose versions, respectively, are related
(up to convention dependent choices of relative signs) as, $ C=B^T \G
_0 = -B^T \G ^0 ,\ B^\star = \G _0 C = \G _0 C^{-1} $.  A
comprehensive presentation of tools for supersymmetry and Dirac
matrices in various dimensions can be found in~\cite{vanProeyen99}.

Finally, we hope  that no confusion should  arise between
spinor indices of gamma
matrices $(\G _M )_{\a \b }$ and the induced brane metric components
$\g _{\a \b } $ in the effective metric $\calm _{\a \b } $ and
likewise in our use of same symbols for 1-forms and differentials in
the cotangent and tangent spaces. Note also that
the transition from our Einstein frame to the alternative Einstein
frame in~\cite{benini09} is realized by replacing $ g_{MN} \to g_{MN}
/ g_s ^{1/2} ,\ \sqrt {\g } \to \sqrt {\g } / g_s ^2 , \ \calf \to g
_s ^{1/2} \calf ,\ \tau _p \to \tau _p / g_s ^{(p+1)/4 }. $


\section{Bosonic sector of $D7$-branes}    
\label{sec2}

We examine in this section the bosonic modes on $D7$-branes wrapped
over Kuperstein 4-cycle~\cite{kuperstein04} of the deformed conifold.
After discussing properties of the classical embedding, we derive the
wave equations for wave functions of mode fields in $M_4$ arising from
fluctuations of the scalar moduli and gauge vector brane fields
decomposed on harmonics of radial sections of the 4-cycle compact base
of geometry $ S^3/Z_2$.  We apply the semi-classical JWKB method to
evaluate the mass spectra and wave functions of meson modes. This
information is used to compute their self couplings and boundary
couplings to graviton modes.  (The fermionic action will be examined
in Section~\ref{sec3}.)

\subsection{Classical embedding in  the warped  deformed conifold}  
\label{sec2sub1}

The Kuperstein 4-cycle~\cite{kuperstein04} $\S _4: \ w_4 =\mu $, is
invariant under the diagonal group $ SO(3)_D \subset (SU(2) _L \times
SO(4)/ Z_2 $ of the conifold isometry group which comprises rotations
of the complex coordinates $ w _{1,2,3}$.  We see from
Eq.~(\ref{app1.eq6}) that this cycle intersects the fixed-$\tau $
subloci of the conifold, $\t _- =0,\ \phi _- =0 \ \Longrightarrow \ \t
_1 = \t _2 = \t ,\ \phi _1=\phi _2= \phi $, generated by the
restricted form of the coordinates matrix $ W= g_1 W_0 (\tau )g_2
^\dagger,\ [g_1= g_2] $~\cite{chemtob22}.  The initial
study~\cite{kuperstein04} invoked this observation to parameterize the
deformed conifold by combining the coordinates $\tau ,\ \t ,\ \phi $
with the angles $\g ,\ \d $ of the $ SU(2)$ matrix, $ S (\g , \d ) =
g_1 W_0 (\tau ) g_1 ^\dagger $.  Instead, we shall consider the
alternative parameterization of the deformed conifold as a foliation
by $\S _4$ leaves~\cite{benini09}, \bea && w_1 =i \eta (\chi ) (\hat
C_\rho c_\phi c _\t + i \hat S_\rho s _\phi ) ,\ w_2 = i \eta (\chi )
( -\hat C_\rho s_\phi c _\t + i \hat S_\rho c _\phi ) ,\ w_3 =- i \eta
(\chi ) \hat C_\rho s _\t ,\ w_4 = (\mu +\chi ) ,\cr && [c_\phi = \cos
  \phi , \ c _\t = \cos \t ,\ s_\phi = \sin \phi , \ s _\t = \sin \t
  ,\ \hat C_\rho = \cosh ({\rho + i \g \over 2} ) ,\ \hat S_\rho =
  \sinh ({\rho + i \g \over 2} ), \cr && \chi \in C ,\ \rho \in [0,
    \infty ],\ \a ^k =(\t \in [0, \pi ], \ \phi \in [0, 2\pi ] , \ \g
  \in [0, 4 \pi ]) ] .  \label{eq.conufol} \eea

The $D7$-brane splits up the spacetime $M_4 \times \calc _6$
coordinates $X^M =(x^\mu , y ^m ) $ into 8 longitudinal real
coordinates $ X^\a = (X^\mu ,\ X^a = (\rho ,\ \a ^k ) ) ,\ [\a =
  0,\cdots , 7, \ a = 1, \cdots , 4,\ \ k=1,2,3 ]$, and the single
transversal complex coordinate, $\chi $.  The bulk bosonic and
fermionic coordinates, $X^M$ and $\T ^\a $, are regarded as brane
fields on the world volume $ M_{8}$ of intrinsic coordinates $ \xi ^\a
$. The invariance of the theory under coordinates diffeomorphisms can
be used to choose the so-called static gauge, $X^\a (\xi ) = \xi ^\a
$, with the complex modulus field $ \chi (\xi ^\a )= \chi (\rho , \a
^m )$ describing the brane location along the transversal directions.
Upon substituting Eq.~(\ref {eq.conufol}) into Eqs.~(\ref{sec1.dc})
and~(\ref{sec1.dcI}) for the conifold and its radial fixed-$\tau $
sections, one obtains the pair of matching equations linking the
conifold and 4-cycle coordinates, \bea && \sum _a w_a ^2 = \e ^2 =
(\mu +\chi )^2 -\eta ^2 (\chi ) ,\ \sum _a \vert w_a \vert ^2 = \vert
\e \vert ^2 \cosh \tau
= \hat r ^3 = \vert \eta (\chi )\vert ^2 \cosh \rho + \vert \mu +\chi
\vert ^2 , \ [\hat r ^3 = ({2\over 3})^{3/2} r ^3 ] \label{eq.cof1}
\eea where the first equation defines the complex function $\eta (\chi
)=\pm (\mu ^2 -\e ^2 + 2\mu \chi +\chi ^2 )^{1/2}$ and the second
defines the 1-to-1 correspondence between the radial variables $ \rho
$ and $\tau $ at fixed $\chi $, with $ \vert \mu \vert $  measuring 
the distance between the apexes  of $\calc _6$  and $\S _4$. 
To lighten the formalism, we introduce in the sequel the
 dimensionless complex parameter $L= \mu / \e $ along with the
 auxiliary complex functions $Y  (\chi ) = 1/ X  (\chi )
 =\eta _\chi / \mu _\chi  ,\  [\eta _\chi \equiv  \eta (\chi ) ,\
\mu _\chi \equiv \mu + \chi ] $  which satisfy  the relations 
 \bea && \eta _\chi = {\mu _\chi \over X (\chi )} = \mu _\chi Y (\chi )
 \ \Longrightarrow  \ d \eta _\chi = X d \chi ={d  \chi \over Y} ,
 \  [X (\chi    )\equiv {1 \over Y (\chi )} =
{1\over (1-\e ^2 / \mu _\chi ^2)^{1/2}   } =
(1- {1 \over (L+ \chi / \e )^{2} })^{-1/2} ] .\label{eq.DefXY}\eea

For real values of the parameters $\mu , \ \e $ and the variable
$\chi$, one finds that $ \eta _\chi \simeq \chi / X (\chi ) \to \chi
(1-\e ^2 /\chi ^2 )^{1/2} $ at $\chi >> \mu $ and $ \eta _\chi \to
\mu (1-\e ^2 /\mu ^2 )^{1/2} $ at $\chi << \mu $.  In the parameters
interval, $\mu < \e $, these functions develop imaginary parts with $
\eta _\chi \to \pm i \vert \e /\mu \vert $ for values of the
coordinate variable in the ranges $\e > \chi >> \mu $ and $ \chi <<
\mu < \e $.  The matching equations in Eq.~(\ref{eq.cof1}) can be
combined into the $\chi $-dependent functional relation, linking the
radial variables $ \tau $ and $\rho $, \bea && \cosh \tau = \vert
{\mu _\chi \over \e } \vert ^2 (\vert Y \vert^2 \cosh \rho + 1) =
\vert { \mu ^2 _\chi \over \e ^2} - 1 \vert \cosh \rho + \vert { \mu
^2 _\chi \over \e ^2 } \vert .\label{eq.Deftaurho} \eea


Since the classical solution for the brane modulus field is expected
to coincide with the 4-cycle locus, $\chi (\rho , \a ^m) =0$, as we
show below, then at the on-shell value $\chi =0$ the change of radial
variable from $ \rho \to \tau $ and vice versa can be written as,
\bea && \cosh \tau = \vert L\vert ^2 (\vert 1- L^{-2} \vert \cosh
\rho +1) ,\ \cosh \rho = \vert 1- L^{-2} \vert ^{-1} \vert ^2 ({\cosh
\tau \over \vert L \vert ^2 } -1) ,\ [{ d\tau \over d \rho }
 = \vert L^2 -1\vert {\sinh \rho \over \sinh \tau
}] \label{eqmatch2} \eea where $L^2 = {\mu^2 / \e ^2 }$ and $Y\equiv
{1 / X} = (1 - L^{-2} ) ^{1/2} $ are now regarded as complex
parameters.  Note that $ \lim _{\rho \to 0} d^2 \tau / d \rho ^2
\simeq \cosh \rho /\sinh \rho $ diverges as $ 1/ \rho .$ Along
the real $ L^2$-axis, $X$ is pure imaginary inside the segment $
0< L^2 < 1 $ (with $ X \simeq \pm i L$ for small $L^2$), and
real elsewhere.  At large $\vert L^2 \vert $, the limits $X \to
1 ,\ \cosh \rho \to (\cosh \tau / \vert L \vert ^2 -1 ) $ show
that the radial region $\tau < \tau _0 ,\ [\cosh \tau _0 =\vert
L \vert ^2]$ is forbidden.  The minimal radial distance $\tau
_{min} $ from the conifold apex (at $\tau =0$) to the 4-cycle
apex (at $\rho \to \rho _{min} =0$), \bea && \cosh \tau _{min} =
\vert {\mu \over \e } \vert ^2 ({1 \over \vert X \vert ^2 } + 1)
= \vert L \vert ^2 ( 1 + \vert 1 - {1\over L^2 } \vert ) ,\ [{d
  \tau \over d \rho } \vert _{\tau _{min}} \simeq {\vert L^2
  -1 \vert \rho \over (4 \vert L^2 -1 \vert -1 )^{1/2} } ]
\eea measures the extent to which $\S _4$ protrudes inside
$\calc _6$.  Inside the three contiguous intervals along the
real axis of the $ L ^2 $-plane this can be expressed by the
formulas \bea && \tau _{min} = [\cosh ^{-1} (- 2 L ^2 +1 ) ,\ 0,
\ \cosh ^{-1} (2 L^2 -1) ] ,\ [L^2<0,\ 0< L^2 < 1,\ L^2> 1]
\eea which can be rewritten in a more familiar way using the
elementary relationship, $\cosh ^{-1} A = \ln (A + \sqrt {A^2 -1
} )$.  For $ L^2 >1:\ \cosh \tau _{min} = 2 L^2 -1 ,\ \sinh \tau
_{min} = 2 L ( L^2 -1)^{1/2} .$ Inside the real axis segment $
0< L^2 \leq 1 $, we see that $ \tau _{min } =0$ saturates the
lower bound,  $ \tau \geq 0 $,  and that its value elsewhere, $
\tau _{min } = \ln ( 2 \vert L \vert ^2 -1 + 2 \vert L \vert (
\vert L \vert ^2 \mp 1 )^{1/2} ) $, grows as $ \tau _{min }
\simeq 2 \ln (\vert L \vert ) $ at large $\vert L^2 \vert $.
Note that $ \tau _{min} (L) $ is a regular function of the
variable real $L$ except for a discontinuous derivative, ${d
\tau _{min} / d L } \simeq 2\sqrt 2 L / (L^2 -1)^{1/2} $, near
$ L =1$, forecasting a possible cusp behaviour there.

The undeformed conifold limit $\e \to 0 ,\ \tau \to \infty $ must be
taken at fixed $r$ and fixed $\mu $.  In the change of variable $
\tau \to r $, the residual dependence on $\e $ gets absorbed inside
the product $\e L \to \mu $.  The minimal distance to the conifold
apex sets at the conic radial variable value, \bea && \hat r^3 _{min}
\equiv ({2 \over 3} )^{3/2} r^3 _{min} = \vert\e \vert ^2 \cosh \tau
_{min} = (\vert \mu ^2 -\e ^2 \vert + \vert \mu \vert ^2 ) \simeq 2
\vert \mu \vert ^2. \eea
With the definition of the conic radial
variable $r$ in Eq.~(\ref{eq.cof1}), the radial warping profile
function reads $h(\tau ) \to h (r) \simeq (\calr / r)^4 $.  Our
rescaled conic variable $\hat r $ coincides with the $r$
of~\cite{benini09} while our $ h ( r) =(4/9 ) h (\hat r) $ is $(4/9 )
$ times their $ h (r) $.


Having dealt with the algebraic properties of the 4-cycle embedding,
we now focus on its geometrical properties. The unwarped metric is
evaluated by substituting the parameterization in
Eq.~(\ref{eq.conufol}) into the defining formula~\cite{candelas90},
\bea && d \tilde s^2 ({\calc _6 }) = {1 \over 2} {\dh \calf \over \dh
  \hat r ^3 } Tr (dW^\dagger d W ) + {1 \over 4} {\dh ^2 \calf \over
  \dh (\hat r ^3 )^2 } \vert Tr (W^\dagger d W ) \vert ^2 = ( 2 F_1 \d
_{ab} + 4 F_2 \bar w_{a} w_{b} ) d w^{a} \bar dw^{b}
 \label{eq.cnfD7} ,\eea 
 where the coefficient functions $ F_1 (\rho ) ,\ F_2(\rho ) $
 correspond to the single and double trace terms for the differential
 $ dW $ of the coordinates matrix $W$ in Eq.~(\ref{sec1.dc}).  To take
 advantage of the isometry group $SO(3) = SU(2)/Z_2\sim S^3/ Z_2$ of
 the 4-cycle fixed-$\rho $ sections, it is useful to trade the angular
 1-forms $ d \a ^m = (d \t , d \phi , d \g ) $ for the left invariant
 1-forms $ \hat h_k ,\ [k=1,2,3]$ of $ SO(3)$, satisfying
 Maurer-Cartan relations, $ d \hat h_i = {1\over 4} \e _{ijk} \hat h_j
 \wedge \hat h_k $.  The resulting basis of left invariant
 1-forms~\cite{benini09} is found to be given by the (angle dependent)
 linear combinations of angular differential forms $ d \a _m $, \bea
 && \hat h _1 = -2(\cos {\g \over 2} d \t + \sin {\g \over 2} \ \sin
 \t \ d\phi ) ,\ \hat h _2 = 2( -\sin {\g \over 2} \ d \t + \cos {\g
   \over 2} \ \sin \t \ d\phi ),\ \hat h _3 = 2({d \g \over 2} + \cos
 \t \ d\phi ) , \cr && [d \t = - \ud (\cos {\g \over 2} \ \hat h _1 +
   \sin {\g \over 2} \ \hat h _2 ),\ d\phi = {1\over 2 \sin \t }
   (-\sin {\g \over 2} \ \hat h _1 + \cos {\g \over 2} \ \hat h
   _2),\ d\g = \hat h _3 -\cot \t \ (-\sin {\g \over 2} \ \hat h _1 +
   \cos {\g \over 2} \ \hat h _2) ] .  \label{eq.conufol2} \eea
 Note that our sign conventions for the $ w_a $ in Eq.~(\ref{eq.conufol})
 and for the above $\hat h _i $ differ from those of~\cite{benini09}
 by the replacements $ \t \to - \t$ and $ \phi \to - \phi $ and $ w _3
 \to - w_3 .$ One can now use the above relations to express the
 conifold metric as a diagonal quadratic form in $ \hat h_i $ (up to
 cross terms $d\chi ( d \rho - i \hat h _3 ) $) weighted by $\rho
 $-dependent functions, \bea && d \tilde s^2 (\calc _6)
= F_1 \bigg [2 (d \bar \chi d \chi + d \bar \eta d \eta \cosh \rho ) +
  {1\over 4} \bar \eta _\chi \eta _\chi (\hat h_1 ^2 - \hat h_2 ^2 +
  (2 d \rho ^2 + \hat h_1^2 + \hat h_2 ^2 + 2 \hat h_3 ^2 ) \cosh \rho
  ) \cr && + d \eta (d \rho - i \hat h_3) \bar \eta \sinh \rho + d
  \bar \eta (d \rho + i \hat h_3) \eta \sinh \rho + R \ \vert 2 \mu
  _\chi d \bar \chi + (2 \cosh \rho d \bar \eta + \bar \eta _\chi
  \sinh \rho ) (d \rho - i \hat h_3) \vert ^2 \bigg ] , \cr && [F_1
  ={1 \over 2} {\dh \calf \over \dh \hat \rho ^2 } = {\e ^{-2/3}K(\tau
    ) \over 2} ,\ R \equiv  {F_2 \over F_1} =- {1 \over 2 \e ^2 \sinh ^2
    \tau } ( \cosh \tau - {2 \over 3 K ^3(\tau ) } ) ].
\label{ap.conmet} \eea  
The factor of $R= {F_2 / F_1} $ in the last term inside
brackets has the expected perfect square form.  We rewrite  for
convenience the  expressions  for the unwarped metric tensor components,
\bea && \tilde g _{\rho \rho } =\tilde g _{\hat h_3 \hat h_3 } =
 {F_1\over 2} \vert \eta _\chi \vert ^2 A_1,\ [\tilde g _{\hat h
     _1 \hat h_1},\ \tilde g _{\hat h _2 \hat h_2}] = {F_1\over 4}
 \vert \eta _\chi \vert ^2 (\pm 1 + C _\rho ) ,\ \tilde g _{\chi
   \bar \chi } = F_1 A_3 ,\ \tilde g _{\rho \chi } = i \tilde g
 _{\hat h_3 \chi } = \ud F_1 \bar \eta _\chi S_\rho A_2 ,\cr &&
   [A_1=C_\rho + 2 \eta _\chi \bar \eta _\chi R S_\rho
 ^2,\ A_2= X + 2 \bar \mu _\chi \eta _\chi R (1 + C_\rho
 {X \over \bar X} ) ,\ A_3 = 1+ C_\rho X\bar X + 2 \mu
 _\chi \bar \mu _\chi R\ \vert 1+ C_\rho { X \over \bar X
 } \vert ^2 ] \label{eqINDm} \eea
 where we introduced the useful auxiliary functions $ A_1,\ A_2,\ A_3$ and the
abbreviations, $ C_\rho = \cosh \rho ,\ S_\rho = \sinh \rho
$.  For $ 0 < L^2 < 1$, $X$ becomes complex and $A_2 $ pure
imaginary.  The restriction
to $\S _4 $ at constant $\chi $  removes  the  derivatives $ d
\chi ,\ d \eta $ and  reduces the metric to a diagonal
quadratic form in the $ S^3 $ tangent vectors $\hat h_{1,
 2,3} $ with $\rho $-dependent coefficients, \bea && d
\tilde s^2 (\calc _6 ) \vert _{\S _4} =\bar \eta _\chi \eta
_\chi [{1\over 4} F_1 (\tau ) (\hat h_1 ^2 - \hat h_2 ^2 +
 (2 d \rho ^2 + \hat h_1^2 + \hat h_2 ^2 + 2 \hat h_3 ^2 )
 \cosh \rho ) + R (\tau ) (d \rho ^2 +\hat h_3 ^2 ) \sinh
 ^2 \rho ] \cr && = {\vert \eta _\chi \vert ^2 K (\tau )
 \over 2 \e ^{2/3} } [ K_2 (\rho ) (d\rho ^2 + \hat h_3 ^2
 ) + \cosh ^2 {\rho \over 2} \hat h_1^2 + \sinh ^2 {\rho
   \over 2} \hat h_2^2 ] ,\cr && [K_2 (\rho ) = \cosh
 (\rho ) + 2 \vert \eta _\chi \vert ^2 \sinh ^2 (\rho ) R
 (\tau ) =\cosh \rho - {\vert \eta _\chi \vert ^2 \sinh ^2
   \rho \over \vert \e \vert ^2 \sinh ^2 \tau } (\cosh
 \tau - {2 \over 3 K ^3(\tau ) } ) ] . \label{eq.cond1}
\eea

One must not confuse the c-number frame vectors $ \hat h_k = \hat h_k
^m d\a ^m $ above with the corresponding differential 1-forms $ h
^{\hat k} = h ^{\hat k } _m d\a ^m $.  These quantities admit similar
decompositions on the angles differentials, $ h ^{\hat k } _m d\a ^m
$, but satisfy different c-number and wedge product composition laws,
$ \hat h_k \hat h_l $ and $ h ^{\hat k } \wedge h ^{\hat l} $.  The
basis of derivative operators $ h _{\hat k } \equiv \dh _{ h _k } = h
_{\hat k } ^m \dh /\dh \a ^m $ is dual to the basis of 1-forms $ h
^{\hat k }$, as made explicit by the expressions of components $h
^{\hat k } _m ,\ h _{\hat k } ^m $ in Eqs.~(\ref{eq.cof2}).  In most
parts of the text we choose $\rho $ as the independent radial
variable.  The transformation from $ \tau \to \rho $ via the
change of variables in Eqs.~(\ref{eqmatch2}) must  then be accompanied by
the substitutions, $g _{\rho \rho } \to \dot \tau ^2g _{\tau \tau },
\ \dh _\rho \to \dot \tau \dh _{\tau }, \ Det (\g ) \to \dot \tau ^2
Det (\g ),\ [ \dot \tau \equiv \dh \tau / \dh \rho ].$ More details on
the formalism are provided in Appendix~\ref{appD7sub0}.

\subsection{Meson   scalar modes from   brane  moduli fields} 
\label{sec2sub2}

The bosonic brane action in Eq.~(\ref{eq.branesdp}) depends on the
moduli field $\chi $ and its spacetime derivatives through the bulk
fields pull-back transforms.  The induced metric on the brane \bea &&
d s^2 (D7) = \g _{\a \b } d \xi ^\a d \xi ^\b = \g _{\mu \nu } d x^\mu
d x ^\nu + \g _{\rho \rho } d\rho ^2 + \g _{h_i h_j } \hat h_i \hat
h_j + 2 ( \g _{\mu \rho } d x^\mu d \rho + \g _{\mu h_i } d x^\mu \hat
h_i + \g _{\rho h_i } d\rho \hat h_i ) , \eea can be calculated from
the bulk metric in Eq.~(\ref{eq.metS0}) either using the definition, $
\g _{\a \b } = \dh _{\xi ^\a } X^M \dh _{\xi ^\b } X^N g_{MN} $, or
substituting in $ d s ^2 (M_{10}) = g_{MN} d X^M d X^N$ the
decomposition of the spacetime coordinates differentials, $ d X^M =
\sum _\a (\dh X^M /\dh \xi ^\a ) d \xi ^\a $ in $ d s ^2 (M_{10}) =
g_{MN} d X^M d X^N$. In the static gauge, $ \xi ^\a = X^\a $, the
induced metric acquires in addition to the bulk metric restriction $
\g ^{(0)}_{\a \b } = g _{\a \b } $ the extra part $\a _{\a \b } = \g
_{\a \b } - \g ^{(0)}_{\a \b } $ comprising linear and quadratic order
terms in $d \chi (x, \rho , \hat h_a) $. The unwarped induced metric
components $\tilde \g _{\a \b } $ (distinguished by tilde symbols) are
related to the warped ones as, $ \g _{\mu \nu } = h^{-1/2} \tilde \g
_{\mu \nu } ,\ \g _{\mu a } = h^{1/2} \tilde \g _{\mu a } ,\ \g _{a b}
= h^{1/2} \tilde \g _{a b} $.  The general formulas for the induced
metric including the dependence on $\eta _\chi =\mu _\chi Y=\mu _\chi
/ X $ and $ \dh _{ \mu } \chi ,\ \dh _{\rho } \chi ,\ \dh _{h_a} \chi
$ are given in Eqs.~(\ref{eq.metS3}).  We select from these results
the following formulas for the induced metric components along $M_4 $
and radial direction of $\S _4$, that will be of direct relevance to
us in the applications, \bea && \tilde \g _{\mu \nu } = \tilde g _{\mu
  \nu } + 2 F_1 h (\tau ) A_3 \dh _\mu \chi \dh _\nu \bar \chi ,
\ \tilde \g _{\rho \rho }= \tilde g _{\rho \rho }
+ F_1 [(\bar \eta _\chi \dh _\rho \chi S_\rho A_2 + H.\ c. ) + 2 \dh
  _\rho \chi \dh _\rho \bar \chi A_3 ] ,\cr && \tilde \g _{\mu \rho }=
\ud F_1 ( \bar \eta _\chi \dh _\mu \chi S _\rho A_2 + 2 \dh _\rho \bar
\chi \dh _\mu \chi A_3 + H.\ c. ) ,\ [C_\rho = \cosh \rho ,\ S_\rho =
  \sinh \rho ].  \label{eq.metsinglet3} \eea The auxiliary functions $
A_1,\ A_2,\ A_3 $ were defined in Eq.~(\ref{eqINDm}) and the terms
linear in $ \dh _{\mu } \chi ,\ \dh _{\rho } \chi $ originate from the
off-diagonal $ d \rho d \chi $ components of the conifold metric.  The
pull-back transforms to $\S _4$ of the classical solutions for the
K\"ahler and potential 2-forms $J_2,\ B_2 $ and the complex field
strength 3-form $ G_3 $ are quoted in Eq.~(\ref{eq.pullJBC}) of
Appendix~\ref{appD7subB1} where one can find complementary details on
the formalism.


The determinant factor in the $D7$-brane action can be evaluated as a
power expansion in field derivatives $ \dh _{\mu , \rho , h_i } \chi $
using Eq.~(\ref{eq.expdet}).  Since the leading order part of the
induced metric is diagonal, $ g _{\a \b } \propto g _{\a \a } \d _{\a
  \b } $, one can make use of the simplified formula \bea &&
{ Det ^{1/2} (g +\a ) \over Det ^{1/2} (g) }= 1 + {1\over 2} \sum_{\a
} {\a_{\a \a } \over g _{\a \a } } - {1\over 2} \sum_{\a ,\b } {\a_{\a
    \b } \a _{\b \a } \over g _{\a \a } g _{\b \b }} +{1\over 8} (
\sum_{\a } {\a_{\a \a } \over g _{\a \a } } )^2 + O( \a ^3)
. \label{eqdet1}\eea

We ignore momentarily the contributions from $ B^{cl} _2$ and, instead
of the general intricate formula up to $ O(\dh _{\mu , \rho , h_i
}\chi )^2 $ in Eq.~(\ref{app.indactP}), quote below the approximate
expression retaining the dependence on $ \eta _\chi $ but neglecting
contributions of $ O( \e / \mu ) $ (relative to unity),
\bea && S_B ^{(2)} (D7)= - {\mu _7} \int d^4 x \int d\rho \hat h_1
\wedge \hat h_2 \wedge \hat h_3 e ^{ \phi } \sqrt {-\tilde \g ^{(0)} }
\bigg (1 + {T_2 \over \vert \eta _\chi \vert ^2 T_3} (\eta _\chi \dh
_\rho \bar \chi + \bar \eta _\chi \dh _\rho \chi ) \cr && +F_1 h (\tau
) ( T_1 - {T_2 ^2 \over T_3} ) \tilde g^{\mu \nu } \dh _\mu \chi \dh
_\nu \bar \chi + {2 \over \vert \eta _\chi \vert ^2 T_3} ( T_1 - {T_2
  ^2 \over T_3} ) \dh _\rho \chi \dh _\rho \bar \chi
 \bigg ) , \cr && [T_1= (1 + C_\rho ) T ,\ T_2= S_\rho T ,\ T_3=
   C_\rho + 2 \vert \eta _\chi \vert ^2 R S_\rho ^2 , \ T =( 1 + 2
   \vert \eta _\chi \vert ^2 R (1 + C_\rho ) ) , \cr && F_1 = \ud \e
   ^{-2/3} K (\tau ) ,\ R= {F_2 \over F_1} = {1 \over 2 \e ^2\sinh ^2
     \tau } ({2 \over 3 K^3 (\tau )} - \cosh \tau ) = {K '(\tau )\over
     2 \e ^2 K (\tau ) \sinh ^2 \tau } ,\cr && V (\rho ) \equiv
   \sqrt {- \tilde \g ^{(0)} } = {F_1^2 \over 16 } \sinh (2\rho )
   \vert \eta _\chi \vert ^4 (1+ 2 R \vert \eta _\chi \vert ^2 {\sinh
     ^2 \rho \over \cosh \rho }) = {\vert \eta _\chi \vert ^4 \over 8
   } F_1 ^2 (\tau ) T_3 (\rho ) \sinh (\rho ) ] .  \label{eqbracts}
 \eea The radial integral measure, $ V (\rho ) = (- \tilde \g ^{(0)}
 )^{1/2}$, includes the contributions from the sub-determinants of the
 induced metric along $ M_4 $ and $\S _4 $.  Only the    real
 quadratic order  terms $\vert \dh _{\mu , \rho } \chi \vert ^2 $ are present
 while the  complex    terms  $ ( \dh _{\mu , \rho } \chi
 )^2 $ cancel out.  The field components in the decompositions, $\chi
 = \phi _1 + i \phi _2 = \vert \chi \vert e ^{i \Phi } $, are
thus  associated to decoupled degenerate modes in leading order of the
 derivatives expansion.  (Note that the first and second terms in the
 combination $(T_1 - T^2_2 / T_3) \vert \dh _{\mu , \rho } \chi \vert
 ^2$ originate from the determinant correction terms, $\a _{\a \a }
 g^{\a \a } $ and $\a _{\a \b } \a _{\b \a } g^{\a \a } g^{\b \b } $
 in Eq.~(\ref{eqdet1}), respectively, and that the second term inside
 parentheses is missing in the results of~\cite{benini09}.)


 In the limit $ \vert \mu / \e \vert >> 1:\ Y \equiv 1/X = \eta _\chi
 /\mu _\chi = ( 1 - (L+\chi /\e )^{-2} ) ^{1/2} \to ( 1 - (L)^{-2} )
 ^{1/2} \simeq 1 .$ The results prior to taking the limit of large
 $L^2$ are obtained from Eq.~(\ref{app.indactP}) by replacing $ (T_1 -
 T^2 _2 / T_3 ) \to (A_3 - \vert A_2 \vert ^2 S_\rho ^2 / T_3) $ in
 Eq.~(\ref{eqbracts}).  The correspondence between the auxiliary
 functions $ A_{1,2,3}$ in Eq.~(\ref{eq.metsinglet3}) and the real
 functions $ T, \ T_{1,2,3} $ is given by $A_1 \to T_3, \ A_2 \to T,
 \ A_3 \to T_1,\ A_2 \sinh \rho \to T_2 $.  In order to describe
 accurately the off shell dependence on $\chi (\rho ) \leq \mu $ of
 the fluctuation field derivative in the wave equation, $ \dh _\rho
 \chi / \eta _\chi \simeq \dh _\rho \chi / (\mu + \chi (\rho )) $, it
 is useful to retain the dependence on $\chi $ in $ \eta _\chi $.

We are now in a position to discuss the $D7$-brane stability.  The
brane effective scalar potential is expected to vanish at its minimum
due to the preserved supersymmetry.  This is verified upon combining
the $O ((d \chi )^0)$ contribution from the scalar potential $V (\rho
)$ with the linear order (tadpole) term $ O(\chi ) $ from integration
by parts of the $\rho $-integral in Eq.~(\ref{eqbracts}), and
expanding in powers of $\chi$, \bea && V (\rho ) = V_0 (\rho ) -
({\chi \over \eta _\chi } + {\bar \chi \over \bar \eta _\chi }) \dh
_\rho ({V_0 (\rho ) T_2 \over T_3} )\cr && \simeq V_0(\rho ) + ({\chi
  \over \eta _\chi } + {\bar \chi \over \bar \eta _\chi }) {\vert \mu
  \vert ^ 4 \over 8} [2 F_1 ^2S_\rho (C_\rho + 3 R \vert \mu \vert ^2
  S_\rho ^2) -\dh _\rho (F_1 ^2 S_\rho ^2 (1 + 2 R \vert \mu \vert ^2
  (1 + C_\rho ) ) ) ] + O(\chi ^2) , \eea where $ V_0 (\rho ) = V(\rho
) \vert _{\chi =0} $.  The verification is easier in the singular
conifold limit where the net effective potential admits the expansion
in powers of $\chi $, \bea && V (\rho ) = \mu _7 e ^{\phi } [ \sqrt {
    -\g ^ {(0)}} + \sqrt { -\g ^{(0)} } \g ^{ab} \g _{\chi a } \dh
  _{b} \chi ] = \mu _7 e ^{\phi } [ V_0 (\rho ) - ({\chi \over \eta }
  + {\bar \chi \over \bar \eta } ) {\dh \over \dh \rho } ({V_0 (\rho )
    T_2 \over T_3 } ) _{\chi =0} ] \cr && \simeq \mu _7 e ^{\phi } [
  {\vert \mu + \chi \vert ^{8/3} \over 96 } - {4 \over 3 } { \vert \mu
    \vert ^{8/3} \over 96 } ({\chi \over \mu } + {\bar \chi \over \bar
    \mu } ) ] ( 2 r _\chi^3 -1) ({ r _\chi^3 -2 \over r _\chi })^{1/2}
+ O(\chi ^2), \eea showing explicitly how the linear order terms in $
V (\rho ) \propto \vert \mu \vert ^{8/3} \chi $ cancel out after
integrating by parts the $O(\chi ) $ source term.  Since the finite
term $ V_0 $ can always be removed through a constant shift of the
action, the net brane potential of $ O(\chi ^2) $ is minimized at the
stable vacuum (on shell) value for constant $\chi (\rho ) =0$.
Whether the cancellation still holds upon switching on the background
2-form field $ B_2 ^{cl} $ is verified by combining the Born-Infeld
and Chern-Simons actions and replacing the metric determinant $\tilde
\g ^{(0)} \to \tilde q $, as detailed in Appendix~\ref{appD7subB1},
\bea && V_0(\rho ) + V_0 ^{CS} (\rho ) = \mu _7 [e ^{\phi } \sqrt
  {-\tilde q} - {1 \over 2} C_4 \wedge B^2 ] = \mu _7 [ e ^{\phi }
  (\sqrt {-\tilde \g ^{(0)} } +e ^{-\phi } {Pf (B)\over h (\tau ) } )
  - {Pf (B)\over h (\tau ) } ] = \mu _7 e ^{\phi } V_0 (\rho
).  \label{eq.eq.textB22} \eea
       
We  now return  to the diagonal terms in the reduced $D7$-brane
action  in Eq.~(\ref{eqbracts})  and express the
quadratic order in  $\vert d \chi \vert $   in terms of 
the effective brane inverse metric  components $ G^{\mu \nu } , \ G^{\rho \rho } $,
\bea && \d S ^{(2)} _B (D7) = - \mu _7 \int d^4 x
\int d\rho \hat h_1 \wedge \hat h_1 \wedge \hat h_3 e ^{\phi } V _0
(\rho ) ( G^{\mu \nu } \dh _\mu \chi \dh _\nu \bar \chi + G ^{\rho
  \rho } \dh _\rho \chi \dh _\rho \bar \chi ) ,\cr && [G^{\mu \nu } =
  \tilde g ^{\mu \nu } Q(\rho ) = \tilde g ^{\mu \nu } F_1 (\rho ) h
  (\tau ) A(\rho ) ,\ G^{\rho \rho } = {2 P(\rho ) \over \vert \eta
    _\chi \vert ^2} = {2 A(\rho ) \over \vert \eta _\chi \vert ^2 T_3}
  , \ A(\rho ) = (A_3 - {\vert A_2 \vert ^2 S_\rho ^2 \over A_1 })
].  \label{eq.fol4}\eea
The Kaluza-Klein decomposition of the complex moduli field \bea &&
\chi (\xi ) = \sum _{m} \chi ^{(m)} (x) e ^{ik_m\cdot x }\Psi
_{m}(\rho ,\a )= \sum _{m, j} \chi ^{(m)} (x) e ^{ik_m\cdot x } f_{m,
  j} (\rho ) \Phi ^{j } _{m_l, m_r} (\a ) , \label{eq.fol5} \eea
introduces mode fields $\chi ^{(m)} (x) $ in $ M_4$, labelled by the
radial and angular excitation indices $k_m $, whose wave functions $
\Psi _{m}(\rho ,\a ), \ [\a = (\t , \ \phi ,\ \g ) ] $ are linear
combinations with radial coefficient functions $ f_{m, j } (\rho) $ of
the $S^3 = SO(4)/SO(3) $ scalar harmonic functions $ \Phi ^{j} _{m_l,
  m_r} (\a )\sim \caly ^{{l/ 2}, {l/ 2} } _{m_l, m_r} (\a ),\ [j= {l
    \over 2} = 0, 1/2 ,\dots ,\ (m_l,\ m_r) \in [-j,\cdots , j] ] $.


Substituting the decomposition in Eq.~(\ref{eq.fol5}) into the perturbed
action in Eq.~(\ref{eq.fol4}) yields a coupled system of second order
differential equations of Sturm-Liouville type for the radial wave
functions $ f_{m, j} (\rho) $.  We shall restrict our study to the
(angle independent) singlet modes described by radial wave functions $
f_m(\rho ) =f_{m, j=0} (\rho ) $.  The volume integral over the
compact base $ \int \hat h_1 \wedge \hat h_1 \wedge \hat h_3 = 64 \pi
^2 $ can then be absorbed into the overall constant factor including
the brane tension, $ C = {64 \pi ^2 \mu _7} $.  One convenient way to
derive the wave equations is by adding and subtracting the diagonal
terms $ \o ^2_m \d _{mn} $ so as to separate the quadratic order
action for the $\chi ^ {(m)} (x) $ into kinetic and mass terms and
on-shell constraint terms, \bea && \d S _B^{(2)} (D7) (\chi )= C \int
d^4 x d \rho e ^{\phi } \sqrt {- \tilde \g ^{(0)} } \sum _{m, n} \bar
\chi ^{(m)} (x) f^\dagger _m (\rho ) [ Q (\rho )(\tilde \nabla ^2_4 -
  \o _m^2 \d _{mn} ) \cr && + ( Q (\rho ) \o _m^2\d _{mn} + {1\over
    \sqrt {-\tilde \g ^{(0)}}} \dh _\rho \calg \dh _\rho ) ] \chi
^{(n)} (x) f_n (\rho ) ,\cr && [Q(\rho ) = F_1(\rho ) h (\tau ) A(\rho
  ) ,\ \calg = G^{\rho \rho } \sqrt {- \tilde \g ^{(0)} } = {2 \sqrt
    {- \tilde \g ^{(0)} } \over \vert \eta _\chi \vert ^2} P(\rho )
  ,\ P(\rho ) = {A(\rho ) \over T_3},\ \o _m^2 = - k_m^2]
.  \label{eq.wechi1} \eea The classical field $ B_2 ^{cl}$ can be
included at this stage by simply replacing $A(\rho ) \to [A_3 -
  ({\vert A_2 S_\rho \vert ^2 / (A_1 R_S) }) ] $, as explained near
Eqs.~(\ref{eq.termsB2}) or~(\ref{eq.recipeNSNS}) in
Appendix~\ref{appD7subB1}.  The rescaled fields $\chi ^{(m)} (x) \to
\chi ^{(m)} (x) / \sqrt {C} $ with canonical kinetic actions, $ \d S
^{(2)} = \int d^4 x \bar \chi ^{(m)} (x) (\tilde \nabla _4 ^2 - \o _m
^2 ) \chi ^{(m)} (x)$, are associated to orthonormal modes of squared
masses $ \o ^2_m $ and wave functions $ f_m (\rho ) $ satisfying the
wave equations and normalization conditions, \bea && 0= \sqrt {-
  \tilde \g ^{(0)}} ( Q (\rho ) \o _m^2 + {1\over \sqrt {- \tilde \g
    ^{(0)}} } \dh _\rho \calg \dh _\rho ) f_m (\rho ) = \calg ( {Q
  (\rho ) \over G^{\rho \rho }} \o _m^2 - {1\over \calg } \dh _\rho
\calg \dh _\rho ) f_m (\rho ) ,\cr && [\int d \rho \sqrt {- \tilde \g
    ^{(0)}} Q(\rho ) f ^\dagger _m(\rho ) f_{n'} (\rho ) = \d _{mm'} ]
.\label{eq.wechi2} \eea One could have reached the above results also
by requiring the stationary condition $\dh S_B (D7) /\dh \bar \chi
^{(m)} =0 $, subject to the mass shell constraint, $\nabla ^2_4 \to \o
_m^2 $.  The wave equation can be transformed via the wave function
rescaling $f_m = {\tilde f_m / \calg ^{1/2} (\rho ) }$ to the
Schrodinger type form \bea && (\dh _\rho ^2 - V_{eff} ) \tilde f _m
(\rho ) =0 ,\ \ [V_{eff} = \calg_1 - {\o _m ^2 Q (\rho ) \over G^{\rho
      \rho } }
= \calg_1 - \hat \o _m ^2 L^2 {\hat Q (\rho ) \over 2 P(\rho )}
,\ \calg_1 = {(\calg ^{1/2} )'' \over \calg ^{1/2} } ,\ P (\rho ) =\ud
\vert \eta _\chi \vert ^2 G^{\rho \rho } ,\cr && \o _m ^2 = {\e ^{4/3}
  \hat \o _m ^2\over (g_s M \a ' )^{2}}, \ Q (\rho ) = { (g_s M \a '
  )^2 \over \e ^{10/3} } \hat Q (\rho ) ,\ \hat Q (\rho ) ={1\over 2
  ^{1/3}} K (\tau ) I (\tau ) A(\rho ) ] \label{eq.fol3}\eea where the
prime denotes the $\dh /\dh \rho $ derivative and the normalization
integral for wave functions is changed to, $\int d \rho \sqrt {-
  \tilde \g ^{(0)}} ( Q (\rho ) / \calg ) \tilde f ^\dagger _m(\rho
)\tilde f _{m'} (\rho ) = \d _{mm'} $.  The dimensionless effective
potential $ V_{eff}$ depends on the ratio parameter $ L=\mu /\e $ and
the rescaled masses $ \hat \o _m ^2 = \e ^{-4/3} (g_s M \a ' )^2 \o _m
^2 $.  Note the naive energy dimensions, $ [f_m (\rho ) ] = E^{-3/2}
,\ [\tilde f_m (\rho ) ] = E^{0 } ,\ [\mu ] = [\e]= E^{-2/3} $ and the
fact that $ A(\rho ) $ cancels out in the mass dependent term, $ \o
_m^2 Q(\rho ) V(\rho ) / \calg =\o _m^2 F_1 h \vert \eta \vert ^2 T_3
/ 2 $, but not in the term $\calg _1 $.

The discrete mass spectrum for the ground state and radially excited
modes is evaluated by applying the semi-classical Bohr-Sommerfeld
quantization rule, \bea && (n- \ud +{\d _W \over 4} ) \pi = \int _{
  \rho _{min} } ^ {\rho _0 } d \rho (-V_{eff} (\rho ))^{1/2} ,\ [n =
  1, 2 , \cdots ,\ V_{eff} (\rho _0) =0,\ \rho _{min}
  =0] \label{eq.wkbBS} \eea with corresponding wave functions in the
inner well region before the turning point $\rho _0 $ of form \bea &&
\tilde f_m (\rho ) = {C_m \over (- V_{eff}(\rho ) )^{1/4} } \sin ({\pi
  \over 4} + \int _{\rho } ^{\rho _0} d\rho ' (- V_{eff}(\rho ')
)^{1/2} ) .\eea The barrier parameter $\d _W $ in the phase integral
is set to $ 0$ if the potential slopes at the turning point near
threshold $\rho _{min} =0 $ and to $ 1$ if the potential is finite
there which amounts to impose a hard wall forcing the wave function to
vanish at the origin.  At large $\rho $, the two terms in the
effective potential $ V_{eff}$ vanish exponentially.  Near the origin
$\rho = \rho_{min} = 0$, the second (mass dependent) term is regular
but the first (mass independent) term diverges as $ \calg _1 \to - {1
  \over 4 \rho ^2 } $.  The quantization rule in Eq.~(\ref{eq.wkbBS})
is thus invalidated due to the divergent contribution to the phase
integral near the origin diverges, $\int _{\rho _{min} } d\rho (-
V_{eff} (\rho ) )^{1/2} = 1/ (4 \rho _{min} ).$ The divergence is
still present if one used $\tau $ as the independent radial
variable~\cite{cotroneDK10} in place of $\rho $.  Analyzing the limit
in the case $ L^2 >1$, in terms of the limited expansions in powers of
$\tau -\tau _{min} $, one finds that the contribution to the phase
integral near the origin diverges logarithmically, \bea && \rho = 2 \a
(\tau -\tau _{min})^{1/2} , \ {d \tau \over d \rho } = \a ^{-1} (\tau
-\tau _{min})^{1/2} ,\ \calg _1= - {1 \over (4 \a )^2 (\tau -\tau
  _{min}) },\ [\a = {L^{1/2} \over (L^2 -1 )^{1/4} } ] \cr &&
\Longrightarrow \ \int _{\tau _{min} } {d\tau \over d \tau / d \rho}
(- \calg _1 )^{1/2} = \int _{\tau _{min} } d \tau {1 \over 4 ( \tau -
  \tau _{min}) } = -{1\over 4} \ln ( \tau - \tau _{min}) .  \eea Since
the divergence is mode independent, one could consider the artificial
cure of imposing an infrared cutoff $ \rho \geq \rho '_0 $.
However, the natural way to remove the singularity is by
choosing an appropriate  boundary condition on the radial wave
functions. Proceeding along same lines as
the modified semi-classical approach of~\cite{minahan99,russo99}.
we are led to   consider  the     wave function rescaling  and  change
of radial variable, $ f _m =( \rho / \calg )^{1/2} \hat f_m , \ \rho =
e ^y, \ [y \in (-\infty , \infty ) ] $ transforming  Eq.~(\ref{eq.fol3})
to the equivalent wave equation, \bea && 0= e ^{-3 y/2} ( \dh _y ^2 -
W_{eff} ) \hat f_m ,\ [W_{eff} = {1\over 4 } + e ^{2y } V_{eff} = \rho
  ^2 ( \hat\calg _1 - \o _m^2 {Q (\rho ) V_0 (\rho ) \over \calg } )
  ,\ \hat\calg _1 = \calg _1 + {1 \over 4 \rho ^2 }] \label{eq.wkbMRS}
\eea
where the modified effective potential $ W_{eff} $ is everywhere
regular, thanks to the shift transformation $\calg _1 \to \hat \calg
_1 $ with $\hat \calg _1 (0)=0 $.  (For the sake of completeness, we
note the alternative choice for the radial variable made
in~\cite{bigazzi0809} $e ^y = e^\rho - 1$.  We also mention the
similar kind of singular wave equation encountered in the analysis of
scalar glueballs in~\cite{benna07}, where taking advantage of the
supersymmetric type structure of the wave operator as a product of
first order operators, $ Q_1 Q_2 w = m^2 w $, the authors replaced the
initial wave equation by the regular one, $ Q_2 Q_1 \hat w = m^2 \hat
w $, retaining the same mass spectrum.)
Substituting the power expansion near $\rho \to 0 $ of the wave
function $ \tilde f _m  $ into the wave equation, 
\bea && 0= (\dh _\rho ^2 - V_{eff} ) \tilde f_m \simeq
(\dh _\rho ^2 +{1\over 4 \rho ^2 } ) (\rho ^{1/2} \hat f_m ),\
[\tilde f_m \equiv f_m \sqrt {\calg } \equiv  \rho ^{1/2}
\hat f_m  =  \rho ^{1/2} (x_0 + x_1 \rho + x_2 \rho ^2 +\cdots ) ] \eea
requires imposing $ x_1 =0 $  while leaving $ x_0,\ x_2 $  arbitrary.
It follows that $ f_m (0) \ne 0 ,\ \dh _\rho f_m (0) = 0 $ are allowed boundary
conditions at the origin.  The quantization rule for the modified wave
equation in Eq.~(\ref{eq.wkbMRS}), \bea && (n- \ud +{\d _W \over 4} )
\pi = \int _{-\infty } ^ {y_0} d y (- W_{eff} (y) )^{1/2} = \int _{0 }
^ {\rho _0 } d \rho (- \hat V_{eff} (\rho ))^{1/2} ,\ [\hat V_{eff}
  (\rho _0) =0 ,\ n=1, 2,\cdots , \cr && \hat V_{eff} (\rho ) = (\calg
  _1 + {1\over 4 \rho ^2 }) - \o _m^2 {Q (\rho ) \over G^{\rho \rho }
  } = \hat\calg _1 - \o _m^2 {Q (\rho )V _0 (\rho ) \over \calg }
  ,\ G^{\rho \rho } = {\calg \over V _0 (\rho )} ] \label{eq.wkbBSp}
\eea is seen to take the standard form when expressed in terms of the
auxiliary potential, $ \hat V_{eff} (\rho ) = W_{eff} (y) / \rho ^2 $.
We can then compute the mass spectra by applying the standard
numerical procedure~\cite{krasnitz00}, used in our earlier
work~\cite{chemtob22}, to the modified potential $ \hat V_{eff} (\rho
) $ and evaluate the rescaled wave functions using \bea && \hat f_m
(\rho ) = {C_m \over \rho ^{1/2} (- \hat V_{eff}(\rho ) )^{1/4} } \sin
({\pi \over 4} + \int _{\rho } ^{\rho _0} d\rho ' (- \hat V_{eff}(\rho
') )^{1/2} ) ,\ [\hat V_{eff} = {W_{eff} \over \rho ^2} ].  \eea


We now examine  the  couplings for scalar modes $\chi ^{(m)} (x)$
inferred fron expansion of the action in powers of $\chi $ of the
overall factor in the effective potential \bea && V _{eff} (\rho )
\simeq V_0 ( \rho ) \vert \mu +\chi \vert ^4 = V_0 ( \rho ) ( 1 + ( 2
\chi /\mu + (\chi /\mu )^2 + H.\ c. )  + 4 \vert \chi /\mu \vert ^4
+\cdots ,\ [V_0 ( \rho ) = {\vert \mu \vert ^2 \over 8} F_1^2 S_\rho
  A_1 ( \rho )] . \label{eqpowV} \eea This is seen to contribute mass
terms to the meson modes, replacing in the wave equation, $ \hat \o
_m ^2 \to \hat \o _m ^{'2}= \hat \o _m ^2 -{4 / ( \vert L \vert ^2
  \hat Q(\rho ) )} ,\ [\hat Q(\rho ) = 2 ^{-1/3} K(\tau ) I(\tau )
  (A_3 - \vert A_2 S_\rho \vert ^2 / A_1 ) ] $.  The correction to the
squared mass parameter has a negative sign and is sizeable at small
$L$ where $\hat Q(\rho ) \to 0 $ at large $\rho $.  The kinetic energy
term in Eq.~(\ref{eq.wechi1}) produces couplings of meson modes to
massive gravitons. These are derived by substituting $ \tilde g _{\mu
  \nu } \to \tilde g _{\mu \nu } + h_{\mu \nu } $ and decomposing the
graviton field into modes $h_{\mu \nu } ^{(m)} (x) $, \bea && h_{\mu
  \nu } (X) = {1 \over \sqrt {\l _2 } } \sum _{m} N _m h_{\mu \nu }
^{(m)} (x) R_m (\tau ) \Phi _m (\T ) ,\
[N_m = {2^{13/6} \sqrt 3 \e ^{-2/3} \over g_s M \a ' J _{(m)}}
  ,\ {1\over \sqrt {\l _2 }}= {2 \sqrt {V_W} \over M_\star} = {2 (2\pi
    )^3 \eta _W ^3 \calr ^3 \over M_\star } ,\cr && \eta _W = {L_W
    \over \calr } ,\ J _{(m)}= (\int d \tau {I(\tau )\over K^2(\tau )
  } B_m ^\dagger B_m (\tau ) )^{1/2} ,\ R_m = {B _m \over \tilde G
    ^{1/2}}] \label{eq.grvKKdc} \eea in same notations
as~\cite{chemtob16}.  Starting from the kinetic action, \bea && \d S
^{(2)} _{B,1 } (D7) = - \int _{M_4} d^4 x \int d \rho W (\rho )
(\tilde g^{\mu \nu } - h^{\mu \nu } ) \dh _\mu \chi ^{\dagger } \dh
_\nu \chi ,\cr && [W(\rho ) = C ' V_0 (\rho ) Q (\rho ) = {C' \vert
    \mu _\chi \vert ^4 \over 8} F_1 ^3 h (\tau ) S_\rho T_3 A(\rho ),
  \ C ' = C e ^\phi = 64 \pi ^2 \mu _7 g_s = {g_s m_s ^8 \over 2 \pi
    ^5} ] \label{eq.grmes1} \eea and expanding the factor $ \vert \mu
_\chi \vert ^4 $ in powers of $ \chi $, as detailed in
Eq.~(\ref{eqpowV}), yields interactions for graviton modes coupled to
canonically normalized mesons mode fields, $ \chi ^{(m)} \to \chi
^{(m)}/ (\int d \rho W_0 f_m ^2 )^{1/2} $, represented by operators of
dimensions $4, 5, 6 $, \bea && \d S ^{(2)} _{B,1} (D7) = + \int _{M_4}
d^4 x \bigg (- \tilde g ^{\mu \nu } \dh _\mu \chi ^{(m) \dagger } \dh
_\nu \chi ^{(m)} (x) + {N _p \over \sqrt {\l _2 } } h ^{ (p) \mu \nu }
(x) \dh _\mu \chi ^{(m) \dagger } \dh _\nu \chi ^{(n)} (x) \int d \rho
W_0 (\rho ) {f_m f_n ( \rho) R_p ( \tau ) \over \prod _{x=m,n} (\int d
  \rho W_0 f_x ^2 )^{1/2} } \cr && + h ^{ (p) \mu \nu } (x) \dh _\mu
\chi ^{(m) \dagger } \dh _\nu \chi ^{(n)} (x) \chi ^{(l) } (x) \int d
\rho W_1 (\rho ) {f_m f_n f_l ( \rho) R_p ( \tau ) \over \prod
_{x=m,n,l} (\int d \rho W_0 f_x ^2 )^{1/2} } \bigg ) ,\cr && [W_0 =
{C' \vert \mu \vert ^4 \over 8} F_1 ^3 h (\tau ) S_\rho (T_1 T_3 -
T^2 _2 ),\ W_1 = {4 W_0 \over \mu },\ W_2 = {6 W_0 \over \mu ^2 }]
. \label{eq.grmes2} \eea

The kinetic energy term in Eq.~(\ref{eq.grmes1}) for the $D7$-brane
action also contributes self couplings for the $\chi ^{(m)} (x)$
represented by operators of dimension $5 $ \bea && \d S ^{(2)} _{B,2}
(D7) = - \int _{M_4} d^4 x \tilde g ^{\mu \nu } \dh _\mu \chi ^{(m)
\dagger } \dh _\nu \chi ^{(n)} (x) \chi ^{(p) } (x) {\int d \rho
W_1 f_m f_n f_p \over \prod _{x=m,n,p} (\int d \rho W_0 f_x ^2
)^{1/2} } ,\eea where $ W_1 $ is specified in
Eq.~(\ref{eq.grmes2}).  Substituting $ \chi \to \chi ^{(m)} $ in the
radial derivative part of the $D7$-brane action, \bea && \d S ^{(2)}
_{B,3} (D7) = - \int _{M_4} d^4 x \int d\rho S(\rho ) \dh _\rho \chi
^{\dagger } \dh _\rho \chi , \cr && [S(\rho ) = {C' \vert \mu _\chi
 \vert ^2 \over 4} F_1^2 S_\rho X(\rho ) = S_0 \vert 1 + {\chi
 \over \mu } \vert ^2 = S_0 + ({\chi \over \mu } + H.\ c. ) S_1 +
\vert {\chi \over \mu } \vert ^2 S_2 ] \eea yields, via the factor
$\vert \mu _\chi \vert ^2 $, local self couplings represented by
operators of dimensions $ 2, 3, 4 $, \bea && \d S ^{(2)} _{B,2} (D7)
= - \int d^4 x \chi ^{(m)\dagger } \chi ^{(n)} (x) \int d \rho [ S_0
(\rho ) \dh _\rho f_m ^\dagger \dh _\rho f_n + ({ \chi ^{(p)} (x)
 \over \mu } S_1 (\rho ) \dh _\rho f_m ^\dagger \dh _\rho f_n f_p
+ H.\ c. ) \cr && + {\chi ^{(p)} (x) \chi ^{(q)\dagger }(x) } {S_2
 (\rho ) \over \vert \mu \vert ^2} \dh _\rho f_m \dh _\rho f_n
f_pf_q^\dagger ] , \label{eq.selfmes2} \eea where we absorbed the
normalization factors inside the wave functions  by substituting,
$f_x / (\int d \rho W_0 f_x ^2 )^{1/2} \to f_x$.

\subsection{Meson    modes from $D7$-brane gauge sector} 
\label{sec2sub2p}

We here examine the mesons modes descending from the $D7$-brane gauge
connection 1-form field, $A = A_\a d \xi ^\a $. We shall  include  
from the outset the 2-form classical field $B_2 $   inside 
the brane effective metric tensor, $ q _{\a \b } = \g _{\a \b } + e
^{-\phi /2 } \calf ^{cl} _{\a \b } $.   The contributions from
the Born-Infeld and Chern-Simons parts in the Einstein frame action
\bea && S_B (D7) = \mu _7 \int d ^8 \xi [ - e ^{\phi } Det ^{1/2} (-(
  \g + e ^{-\phi /2 } \calf )) + (C_8 + C_6 \wedge \calf + \ud C_4
  \wedge \calf {^2} ) ] , \eea depend on the 2-form combination of the
gauge field strength $F = d A $ and NSNS sector $B_2 $ potential,
$\calf = B + F,\ F = d A $, and the RR sector $ C_{8, 6, 4}$-form
potentials~\cite{herzog01}.  In the present classical background
solution, $ C_8$ is absent, $ C_6$ vanishes identically, $ F_7 \equiv
d C_6 = \star _{10} F_3 - C_4 \wedge H_3 =0 $~\cite{kuperstein04}, and
$ C_4$ takes the simple form \bea && C_4 (\tau ) = \int _0 ^\tau d
\tau ' \tilde F_5 (\tau ') = [4\e ^{-8/3} (g_s M \a ' )^2 \int ^\tau
  _0 d \tau ' (l(\tau ') / (K (\tau ') \sinh (\tau ' ))^{2} ) ]^{-1}
vol (M_4) = vol (M_4) / h (\tau ) . \eea
The determinant of the matrix $
q$ separates into two factors associated to the two diagonal blocks in
the $M_4 $ and $\S _4 $ vector spaces, \bea && Det ^{1/2}( - ( \g + e
^{-\phi /2 } \calf )) = Det ^{1/2}( -q)_{\a \b } Det^{1/2}(1 + e
^{-\phi /2 } q ^{\a \b } F_{\b \a } ) \cr && = Det ^{1/2}(-\tilde \g
)_{\mu \nu } \ Det ^{1/2} ( \tilde \g + e ^{-\phi /2 } h ^{-1/2} B
)_{a b } \ Det ^{1/2} ( 1 + e ^{-\phi } q ^{-1} F)_{\a \b} ,\cr && [q
  _{\mu \nu } = h^{-1/2} \tilde \g _{\mu \nu } + e ^{-\phi /2 } B
  _{\mu \nu } ,\ q_{a b } = h^{1/2} \tilde \g _{a b } + \calf_{ab}
  ,\cr && Det^{1/2} (-\tilde q )_{\a \b} = Det^{1/2} (-\tilde \g
  )_{\mu \nu } Det^{1/2} (\tilde q ) _{ab} = \sqrt {-\tilde \g ^{(0)}
  } (Det ^{1/2} (\tilde \g ) + e ^{-\phi } h^{-1} (\tau ) Pf (B)
  )_{ab} ] \eea where we have ignored in the final line entry above
the components $B _{\mu \nu } $ along $M_4 $.  The expansions of the
determinant $Det ^{1/2} ( 1 + e ^{-\phi /2 } q ^{-1} F) $ and
exponentional factor $e ^{F}$ in powers of $F $
yields the  4-d gauge action of quadratic order in
$F$ \bea && \d S _B^{(2)} (D7) = \mu _7 \int d^4 x \wedge d \rho \wedge
\hat h_1 \wedge \hat h_2 \wedge \hat h_3 [-\sqrt {-\tilde q } ( e
  ^\phi - {1\over 4} Tr ( q^{-1} F q^{-1} F ) + {1\over 8} (Tr(\tilde
  q^{-1} F ))^2 ) + \ud C_4 \wedge \calf \wedge \calf ]
    \label{eqbrjauge1}  \eea

The Kaluza-Klein ansatz for the gauge connection components, $A _\a
(X) = (A_\mu ,\ A _a ) = (A_\mu ,\ A _\rho ,\ A_k ) $, \bea && A_\mu
(X) = \sum _{m , j} a^ {(m)}_\mu (x) a _{m,j} (\rho ) \Phi ^{j} (\a ),
\ A _{[\rho , k]} (X) = \sum _{m , j} a^ {(m)}_{[\rho , k]} (x)
g_{m,j} (\rho ) \Phi ^{j} _{[\rho , k] } (\a ), \label{eq.SVhar} \eea
introduces fields in $M_4$ belonging to a single tower of vector modes
and three towers of scalar (gauge invariant) modes $ [a^ {(m)}_\mu (x)
  ,\ a^ {(m)}_{ \rho } (x) ,\ a^ {(m)}_{k } (x) ] $. The wave
functions are given by linear combinations of radial functions $ [ a
  _{m , j } (\rho ) ,\ g_{m , j } (\rho ) ] $ times scalar and vector
harmonics $ \Phi ^{j} (\a ), \ \Phi ^{j} _k (\a ) $ of the compact
base manifold $ S^3 / Z_2$.  Upon removing the classical part $ - \mu
_7 \int d^4 x \sqrt {- \tilde q} e ^{\phi }$ and using the relations,
$q ^{\mu \nu } = h ^{1/2} \tilde g ^{\mu \nu } ,\ q ^{\rho \rho } = h
^{-1/2} \tilde q^{\rho \rho } ,\ \sqrt { - q } = \sqrt { - \tilde q
}$, the reduced action for vector modes $a _\mu ^{(m)}$ in Lorentz
gauge $ \dh _\mu a^{(m) \mu} = 0$ becomes \bea && \d S _B ^{(2)} (
a_\mu ) = + {\mu _7 \over 4} \int d^4 x \int d \rho \hat h_1 \wedge
\hat h_2 \wedge \hat h_3 e ^\phi \sqrt { - q } ( q ^{\a \b } F_{\b \g
} q ^{\g \d } F_{\d \a} -\ud (q ^{\a \b } F_{\b \a } )^2 ) \cr && = -
     {\mu _7 \over 2} \int d^4 x \int d \rho \hat h_1 \wedge \hat h_2
     \wedge \hat h_3 e ^\phi \sqrt { - \tilde q } [\tilde g ^{\mu \nu
       } \tilde g ^{\l \rho } h(\tau ) \dh _\mu A_\l \dh _\nu A_\rho +
       \tilde g ^{\mu \nu } \tilde q ^{\rho \rho } \dh _\rho A_\mu \dh
       _\rho A_\nu ] . \label{eq.redBI} \eea Selecting in
     Eq.~(\ref{eq.SVhar}) only the singlet modes $ A_\mu \to a _\mu
     ^{(m) } (x) a _m (\rho ) $ of squared masses $ \tilde \nabla ^2_4
     \to \o _m ^2$, and separating the terms in the sum over modes
     into kinetic and on-shell constraint parts (by subtracting and
     adding the mass terms $\o _m^2 \vert a_\mu ^{(m)} \vert ^2 $)
     yields \bea && \d S _B ^{(2)} ( a_\mu ) = + {C \over 2} \int d^4
     x \int d \rho \sqrt {-\tilde q} \tilde g ^{\mu \nu } a _\mu ^{(m)
       \dagger } (x) a _m ^\dagger (\rho ) [h (\tau ) (\tilde \nabla
       ^2_4 - \o _m^2 ) \cr && + (h(\tau ) \o _m^2 + {1 \over \sqrt {-
           \tilde q} } \dh _\rho \tilde q^{\rho \rho } \sqrt {- \tilde
         q} \dh _\rho ) ] a _\nu ^{(m)} (x) a _m (\rho ) ,\ [C= 64 \pi
       ^2 \mu _7 = {1\over 2 \pi ^5 l_s ^8} ] . \label{eq.weamu1} \eea
     One can now deduce by inspection the equations for the modes wave
     functions and their normalization conditions, \bea && 0= [ {\o _m
         ^2 \vert \eta _\chi \vert ^2 Q (\rho ) \over 2 P(\rho )} + {1
         \over \calg _a } \dh _\rho \calg _a \dh _\rho ] a _m(\rho ) =
     [ {1\over 2 ^{4/3} } \hat \o _m ^2 L^2 K(\tau ) I(\tau ) R_S + {1
         \over \calg _a } \dh _\rho \calg _a \dh _\rho ] a _m(\rho ) ,
     \cr && [\o _m ^2 = \e ^{4/3} (g_s M \a ' )^{-2} \hat \o _m ^2
       ,\ \ Q = h (\tau ) ,\ P (\rho ) = {\vert \eta _\chi \vert ^2
         \tilde q ^{\rho \rho } \over 2 } = {\vert \eta _\chi \vert ^2
         \tilde \g ^{\rho \rho } \over 2 R_S },\cr && \calg _a (\rho
       )= \sqrt {-\tilde q } \tilde q ^{\rho \rho } ,\ R_S = {\sqrt
         {-\tilde q } \over \sqrt {-\tilde \g } }, \ \tilde \g _{\rho
         \rho } = {1\over \tilde \g ^{\rho \rho }} \ud F_1 \vert \eta
       _\chi \vert ^2 A_1 ,\ \int d \rho \sqrt {- \tilde q} Q(\rho ) a
       ^\dagger _m(\rho ) a _{m'} (\rho ) = \d _{mm'}] \eea where only
     the first term involving the mass parameter $\o _m $ depends on
     the classical field $ B_2 ^{cl} $, unlike the second term, $\calg
     _a (\rho )= \sqrt { - \tilde q } \tilde q ^{\rho \rho } = \sqrt {
       - \tilde \g } \tilde \g ^{\rho \rho } $, as inferred from
     Eq.~(\ref{eq.termsTXB3}).  The wave functions rescaling $ a
     _m(\rho ) \to \tilde a _m(\rho ) / \calg _a ^{1/2} $ transforms
     the wave equations to the Schr\"odinger form, $(\dh _\rho ^2 -
     V_{eff} ) \tilde a _m (\rho ) =0,$ \bea && V_{eff} ^{(m )} (\rho
     )=\calg _{a,1} - {Q \o _m ^2 \over \tilde q ^{\rho \rho }} =
  \calg _{a,1} -\vert \eta _\chi \vert ^2 { Q \over 2 P} \o _m ^2 =
     \calg _{a,1} - {\hat \o _m ^2 R_S \over 2 ^{4/3} } \vert {L }
     \vert ^2 I(\tau ) K(\tau ) A _1 (\rho ),\ [\calg _{a,1} = {
         (\calg _a ^{1/2 } (\rho ) )'' \over \calg _a ^{1/2 } (\rho )
     } ]. \label{eq.fol7} \eea
The dimensionless effective potential $V_{eff}^{(m )} $ is formally
similar to that of $\chi $ modes in Eq.~(\ref{eq.fol3}) but with
different auxiliary functions $ Q (\rho ),\ P (\rho ) $.  The mass
independent contribution again diverges near the origin, $ \calg _a
\sim 1/\rho ,\ \calg _{a,1} \sim -1/ (4 \rho ^2) $, so the
semi-classical method must be modified by changing the independent
variable $ \rho \to y = \ln \rho $ and the wave functions $ a _m (\rho
) \to (\rho /\calg _a )^{1/2} \hat a _m (\rho ) $, as in
Eq.~(\ref{eq.wkbMRS}).  One can then evaluate the masses by applying
the familiar quantization rule to the modified effective potential,
similarly to Eq.~(\ref{eq.wkbBSp}), \bea && \hat V_{eff}^{(m )} (\rho
) = (\calg _{a,1} + {1\over 4 \rho ^2 }) - \o _m^2 {Q (\rho ) \over
  \tilde q ^{\rho \rho } } = \hat\calg _{a,1} - \o _m^2 {Q (\rho )
  \sqrt{ -\tilde q} \over \calg _a } ,\ [\hat\calg _{a,1} = \calg
  _{a,1} + {1\over 4 \rho ^2}] . \eea

The coupling of gravitons to pairs of brane modes can be obtained by
substituting in Eq.~(\ref{eq.redBI}) for the reduced action the
decomposition on singlet modes, $a^{ (m) } _{\mu } ,\ [a^{ (m) } _\mu
  k ^\mu =0] $.  The resulting 4-d effective gravitational action is
given in same notations as Eq.~(\ref{eq.grvKKdc}) by \bea && \d S
_{B,1} ^{(2)} (D7) =+\int d^4 x \l ^h_{mnp} h^{(p) \mu \nu } \dh _\l a
_{\mu } ^{(m)} \dh ^\l a _{\nu } ^{(n)} ,\ [\l _{mnp} ^h = { C' \over
    2 \sqrt {\l _2 } } { \int d \rho V (\rho ) R_S h(\tau ) \tilde a_m
    ^\dagger \tilde a_n (\rho ) R_p (\tau ) \over (\prod _{x=m,n} \int
    d \rho ' W _A (\rho ') \vert \tilde a_x \vert ^2 )^{1/2}
}]. \label{eq.grAmu} \eea

We consider next the components of the gauge field 1-form $A = A_{\a }
d \xi ^\a =A_{\mu } d x^\mu + A_{\rho } d \rho +A_{\hat k } h^{\hat k}
$ along the radial and angular directions of the tangent space of $\S
_4 $ in $ M_4 \times \S _4 $ and settle again to the static
gauge. Note that these are linearly related to the tangent frame
vielbeins components, $ A = A_{\hat \rho } e^{\hat \rho } +A_{\hat e^k
} e^{\hat k } = A_{\hat \rho } \eta ^{\hat \rho } d \rho +A_{\hat k }
\eta ^{\hat k } h ^{\hat k } _m d \a ^m $, as $ A_{\rho } = A_{\hat
  \rho } \eta ^{\hat \rho } ,\ A_{\hat k } = A_{\hat e^k } \eta ^{\hat
  k},\ A_m = A_{\hat k } h^{\hat k} _m .$ Substituting the
decomposition for the field strength 2-form $ F = d A = \ud F_{\a \b }
d \xi ^\a \wedge d \xi ^\b $, \bea && F ={1\over 2! } F_{\nu \mu } d
x^\nu \wedge d x^\mu +F_{\rho \mu } d \rho \wedge d x^\mu +F_{\rho
  \hat k } d \rho \wedge h ^{\hat k} +{1\over 2! } F_{\hat k \hat l }
h ^{\hat k} \wedge h ^{\hat l} ,\cr && [F_{\nu \mu } = 2 \dh _{ [\nu }
    A_{\mu ]} ,\ F_{\rho \nu } = 2 \dh _{ [\rho } A_{\nu ]} ,\ F_{\rho
    \hat k } = 2 \dh _{ [\rho } A_{\hat k ]} ,\ F_{\hat k \hat l} = 2
  \dh _{ [\hat k } A_{\hat l ]} +\ud \e _{mkl} A_{\hat m}
] \label{eq.decAF} \eea yields the diagonal form for the reduced
action, \bea && \d S ^{(2)}_{B,2} (D7) = - { \mu _7 \over 2} \int d^8
\xi e ^{\phi } \sqrt {-\tilde q } [q ^{\mu \mu } q ^{ \rho \rho } F^2
  _{\mu \rho } +q ^{\mu \mu } q ^{ \hat k \hat k } F^2 _{\mu \hat k}
  +q ^{\hat k \hat k } q ^{ \rho \rho } F^2 _{\rho \hat k} + q ^{\hat
    l \hat l } q ^{\hat m \hat m} F^2 _{\hat l \hat m} + q ^{\hat l
    \hat l } q ^{\hat m \hat m} ] . \label{eq.totacAk} \eea We
consider the $U(1)$ theory in the gauge $ A _{\hat \rho } =0 $,
substitute the Kaluza-Klein decompositions of the fields $ A_{\hat k
}$ in Eq.~(\ref{eq.decAF}) analogous to those in Eqs.~(\ref{eq.SVhar})
and restrict again to the (purely radial) singlet modes $ A_{\hat k }
(\rho )$, hence ignoring angular derivatives.  The contributions to
the reduced $D7$-brane action from the symmetric part of the metric
tensor $ q ^{(\a \b )} $ in Eq.~(\ref{eqbrjauge1}) takes the diagonal
quadratic form, \bea && \d S ^{(2)}_{B,2} (D7) = - {C\over 2} \int d^4
x d \rho [ \sqrt {-\tilde q } (q ^{\mu \nu } q ^{\hat k \hat k} \dh
  _\mu A _{\hat k } \dh _\nu A _{\hat k } + q ^{\hat \rho \hat \rho }
  q ^{\hat k \hat k} \dh _\rho A _{\hat k } \dh _\rho A _{\hat k } +
  {1\over 2} q ^{\hat l \hat l} q ^{\hat m \hat m} A ^2_{\hat k } ) +
  {1\over h (\tau ) } A _{\hat k } \dh _\rho A _{\hat k } ]
, \label{eq.actionAk} \eea where the third term inside parentheses is
summed over cyclic permutations of the indices $ (\hat k, \hat l,\hat
m) $ running over the basis $ [\hat 3, \hat 1,\hat 2] $.  The
Chern-Simons contribution in the last term above is expressed through
integration by parts (and neglect of boundary contributions) by the
effective mass term, $ - (C/4) \dh _\rho ( h^{-1} (\tau )) A ^2 _{\hat
  k} $. After integration by parts and substitution of the expression
for the inverse effective metric tensor, one finds the action for
singlet fields, \bea && \d S ^{(2)}_{B,2}(D7) = - {C \over 2} \int d^4
x d \rho \calg _{k} [ -{\sqrt {-\tilde q }\over \calg _{k} (\eta
    ^{\hat k} )^2 R_S } A_{\hat k} \tilde \nabla _4 ^2 A_{\hat k} -
  {1\over \calg _{k} } A_{\hat k} \dh _\rho \calg _{k} \dh _\rho
  A_{\hat k} \cr && + {\sqrt {-\tilde q } \over 2 h (\tau ) \calg _{k}
    R_S ^2 } {1 \over (\eta ^{\hat l} \eta ^{\hat m} )^2 } A_{\hat k}
  ^2 + {1\over 2 \calg _{k} h (\tau ) } \dh _\rho ( A_{\hat k} ^2) ]
,\ [\calg _k = {\sqrt {-\tilde q } \over h (\tau ) ( \eta ^{\hat \rho
    } \eta ^{\hat k}) ^2 R_S ^2 } ,\ R_S = {\sqrt {-\tilde q } \over
    \sqrt {-\tilde \g } } ].  \eea The wave functions of singlet modes
$a _{\hat k, m} (\rho ) $ satisfy the wave equations \bea && 0= [\o _m
  ^2 {\sqrt {- q } \over \calg _k \vert \eta ^{\hat k} \vert ^2 R_S} +
  {1 \over \calg _k } \dh _\rho \calg _k \dh _\rho - {\sqrt {- q }
    \over 2 h (\tau ) \calg _k R_S^2 (\eta ^{\hat l} \eta ^{\hat m}
    )^2 } + { \dh _\rho h ^{-1}(\tau ) \over 2 \calg _k } \dh _\rho ]
a_{\hat k, m} \cr && = [ \o _m ^2 Q R_S + {1 \over \calg _k } \dh
  _\rho \calg _k \dh _\rho - {1 \over 2} ({\eta ^{\hat \rho } \eta
    ^{\hat k} \over \eta ^{\hat l} \eta ^{\hat m} } )^2 + { \dh _\rho
    h ^{-1}(\tau ) \over 2 \calg _k} \dh _\rho ] a_{\hat k, m} , \ [Q=
  h(\tau ) (\eta ^{\hat \rho })^2 ] . \eea where the summation over
cyclic indices $\hat k, \ \hat l, \ \hat m $ in the third term inside
the brackets yield $ - (\eta ^{\hat \rho } \eta ^{\hat k} / (\eta
^{\hat l} \eta ^{\hat m} ) )^2 ,$ for the modes $a_{\hat 3, m} $.  The
rescaled wave functions $\tilde a_{\hat k, m} = a_{\hat k, m } \calg
^{1/2} _{k} $ satisfy the Schr\"odinger type wave equations, \bea &&
(\dh _\rho ^2 - V^{\hat k}_{eff} ) \tilde a _{\hat k, m} =0, \ [
  V_{\hat k}^{eff} = -Q \o _m ^2 R_S + \calg _{k,1} + \ud ({\eta
    _{\hat l} \eta _{\hat m} \over \eta _{\hat \rho } \eta _{\hat k} }
  )^2 - {\dh _\rho (h^{-1} ) \over 2 \calg _k } ,\ \calg _{k,1} =
  {(\calg _{k} ^{1/2} )'' \over \calg _{k} ^{1/2} } ]
.  \label{eq.weA} \eea The singularity near $\rho =0$ in $V_{eff}
(\rho )$ from $ \calg _{k,1} $ is again removed by the changes of
variable and wave functions, $ \rho = e^y ,\ a _{\hat k ,m} = (\rho
/\calg _k ) ^{1/2} \hat a _{\hat k ,m} $. The resulting Schr\"odinger
wave equation with the regular effective potential $ W_{eff} (y) =
{1\over 4 } + e ^{2y } V_{eff} $ and the quantization rule for the
modes potential $ \hat V_{eff} $ and masses $\o _m ^2 $, \bea && ( \dh
_y ^2 - W ^{\hat k}_{eff} ) \hat a_{\hat k ,m} =0 ,\ \ [\hat V ^{\hat
    k}_{eff} \equiv {W ^{\hat k}_{eff} \over \rho ^2} = -Q \o _m ^2
  R_S + \hat \calg _{k,1} + \ud ({\eta _{\hat l} \eta _{\hat m} \over
    \eta _{\hat \rho } \eta _{\hat k} } )^2 - {\dh _\rho (h^{-1} (\tau
    ) ) \over \calg _{k} } ,\ \hat \calg _{k,1} = \calg _{k,1} +
  {1\over 4 \rho ^2} , \cr && (n - 1/2 + \d _W / 4 )\pi = \int
  _{-\infty } ^{y_0} d y (- W^{\hat k}_{eff} (y) )^{1/2} = \int _{0}
  ^{\rho _0} d \rho (-\hat V^{\hat k}_{eff} ( \rho ) )^{1/2} ,\ \hat
  V^{\hat k}_{eff} \equiv {W^{\hat k}_{eff} \over \rho ^2} = {1 \over
    4\rho ^2 } +V^{\hat k}_{eff} ] \label{eq.weASII} \eea are
derived in a similar fashion as Eq.~(\ref{eq.wkbMRS}).

We turn finally to the contributions from the antisymmetric part of
the effective metric of quadratic order in $ q ^{[\a \b ]}= h ^{-1/2}
(\tau ) \tilde q ^{[\a \b ]} $ and linear order $ q ^{[\a \b ]} q
^{(\d \d )} $, given in the notations of Eq.~(\ref{eq.termsTXB3}) by $
\tilde q ^{[\a \b ]} = (B^{-1} ) _{\a \b } / R_A ,\ \tilde q ^{(\a \b
  )}=(\tilde \g )^{\a \b } / R_S$.
Retaining for simplicity the leading contributions at large $\mu $
which are included in the metric components, $ q ^{ [\rho \hat h_2 ]}
,\ q ^{ [\hat h_1 \hat h_3 ]} $, adds the following part to the
reduced action, \bea && \d S ^{(2)}_{B,3} (D7) = - {\mu _7 } \int d ^8
\xi \sqrt {-\tilde q } e^\phi q ^{ [\rho \hat h_2 ]} q ^{ [\hat h_3
    \hat h_1 ]} ( F_{\hat h_2 \rho } F_{\hat h_1 \hat h_3 } + F_{\hat
  h_3 \rho } F_{\hat h_2 \hat h_1 } + F_{\hat h_1 \rho } F_{\hat h_3
  \hat h_2} ),\cr && \simeq - {C \over 2} \int d^4 x \int d \rho \sqrt
    {-\tilde q } q ^{ [\rho \hat h_2 ]} q ^{ [\hat h_3 \hat h_1 ]} A
    _{\hat k} \dh _\rho A _{\hat k} , \label{eq.antis1} \eea where the
    second entry is obtained by restricting to the singlet modes.

\subsection{Properties  of  meson modes}  
\label{sec2sub3}

\subsubsection{Parameter space}
\label{sec2sub3I}

The data needed to evaluate the properties of meson modes consists of
three sets of inputs: (1) the string theory and internal manifold
volume parameters, $ g_s ,\ \a ' = 1/m_s^2$ and $ V_W = (2\pi
L_W)^{1/6}$, (2) the Klebanov-Strassler throat parameters $ \e ,\ \mu
$, associated to the confinement scale and the quarks bare masses of
the dual gauge theory, $\L _{ir} \sim \e ^{2/3} /\a ' $ and $ m _Q
\sim \mu ^{2/3} /\a ' $ and (3) the throat ultraviolet cutoff $\tau
_{uv}$.  To ease the contact with phenomenology, we trade $ m_s $ for
the Planck mass $M_\star $ and $\e $ for the mass hierarchy warping
factor $w= m_{eff} / M_\star = w_s m_s / M_\star $, using the
relations \bea && m_s = ({ \pi M_\star ^2 \over L_W^6 } )^{1/8} =
M_\star \sqrt {\pi \over \calv } ,\ \ \ m_s ^{3/2} \e = ({2 ^{1/3} a_0
  ^{1/2} g_s M \over \pi } )^{3/4} ({w \calv ^{1/3} })^{3/2} =({2
  ^{1/3} a_0 ^{1/2} g_s M } )^{3/4} ({ w_s \over \calv ^{1/6} })^{3/2} , \eea
where the  dependence of $\e $   on the compactification
manifold $ X_6$ in string units, $\calv \equiv e^{6u } = V_W /\hat
l_s^6 =(L_W /l_s)^6 $, results from our  definition  for   
the warping factor~\cite{chemtob22}, $w_s = e^{u} h^{-1/4} (0) .$
(Note that the power  index change from $\calv ^{-1/4} \to \calv ^{1/2}
$ upon going from $ w_s  \to w$.) The dimensionless mesons masses
$\hat \o _m $,    depending   on the parameters  ratio  $ L $, 
\bea && \hat \o _m = \e ^{-2/3}
(g_s M \a ') \o _m= \mu ^{-2/3} (g_s M \a ') L^{2/3} \o _m = (g_s
M)^{1/2} \calv ^{1/6} \o _m / (2 ^{1/6} a_0 ^{1/4} M_\star w ) , \ [L
  \equiv \mu / \e \simeq ({m _Q \over  \L _{ir}}) ^{3/2} ] \eea
are  defined with the same  rescaling as for  glueball masses $ \hat  E_m $.

The complex fields $ \chi $ and $ A_{\hat 1} +i A_{\hat 2} $ belong to
chiral supermultiplets  of $\caln =1 $ supersymmetry
and the real fields $ A_\mu $ and $A _{\hat 3}$ to vector supermultiplets.  The
components in each supermultiplet have equal masses but not  necessarily  
same radial wave functions~\cite{benini11}.  The gravity-gauge
holographic correspondence between bulk and boundary theories in
$AdS_{d} $ and $ M_{d-1} $ spacetimes relates the mass $ \o $ of
normalizable modes to the scaling dimensions $\D $ of operators of
same spin $s$ and quantum numbers) by $ \o ^2= ( \D -s ) (\D -d
+s) $.  For bulk spacetimes asymptotic to $AdS_{d} \times X_{10-d} $,
one can use the asymptotic expansions near the boundary for the
non-normalizable (NN) and normalizable (N) classical fields of fixed
mass, \bea && \d \Phi _{\o , s} (x,r) \sim r ^{s} (c_{NN} r^{\D -d } +
c_N r^{-\D } ) ,\ [({2 \over \e ^2 })^{1/3} \hat r \simeq e ^{\tau /3}
  \simeq L^{3/2} e ^{\rho /3} ] \eea to determine the scaling
dimensions $\D $ from the radial profiles and hence identify the
structure of the dual gauge theory composite operators in the quark
superfields.  The following table shows the correspondence between the
superspace $\t ,\ \bar \t $ components of vector and chiral bilinear
quark operators and the asymptotic radial wave functions $ \Phi _{\o ,
  s} (x,r) $ of the modes $ A^{(m)} _{\hat 3} $ and $A_\mu ^{(m)} $
part of the same vector supermultiplet and the modes $A ^{(m)} _{\hat
  1 , \hat 2} $ and $\chi ^{(m)} $ part of the same chiral
supermultiplet.

\begin{center} \begin{tabular}{|c|c|c||c|c|} \hline
Operator $ O $&$ O^V _{\t \bar \t } = q ^{\dagger} \dh _\mu q - \tilde
q ^{\dagger} \dh _\mu \tilde q $&$ O^V _{\t ^0\bar \t ^0} = q^{\dagger
} q - \tilde q^{\dagger } \tilde q $&$ (\tilde Q Q) _{\t ^0} $&$
(\tilde Q Q) _{\t ^2}$ \\ $\D , s $ & $ 3, 1 $&$ 2, 0 $&$ 3/2 , 0 $&$
5/2 , 0$ \\ \hline AdS Field & $ A_\mu $ &$ A _{\hat 3} $&$ (A _{\hat
  1} , A _{\hat 2}) $&$ \chi $ \\ $\lim _{r\to \infty } \d \Phi _{\D , s}
(x,r) $ & $ c_{NN} r ^{0} + c_N r^{-2} $&$ c_{NN} r ^{-2} + c_N r^{-2}
\ln r $&$ c_{NN} r ^{-5/2} + c_N r^{-3/2} $& $ c_{NN} r ^{-3/2} + c_N
r^{-5/2}$ \\ \hline \end{tabular}\end{center}

The  independent (non-normalizable  and normalizable) solutions
for the classical bulk fields perturbations $ \d \Phi
_{\o , s} (x,r) = (r^{-3/2} \chi ,\ A_a ,\ A_\mu ) (x,r)$
have asymptotic    series expansions   
at  large   radial distances  $ r ^{3/2} / \mu >> 1 $
whose   leading terms are   assigned the
constant coefficients $c_{NN} ,\ c_N $. The correspondence to
the  gauge   field theory bare action associates  $c_{NN} $
to the source perturbation, $\d L = c_{NN} O _{\D , s} (x) $, and $ c_N $
to the  perturbed operator VEV,  $ c_{N} \sim < O _{\D , s} > $.
(Note that the radial  variable   rescaling  factor for $\chi $ is necessary to
transform it to a canonically normalized field  and  that the
components $ c_{NN}  $ and $c_N $ are leading and subleading for $\chi
,\ A_3 ,\ A_\mu $ with the reverse holding for $A_{1,2} $.)  The
fields $ A^{(m)} _{\hat 3} $ and $A_\mu ^{(m)} $ are part of a vector
supermultiplet whose dual counterpart is the flavor $ U(N_f) $ current
superfield operator, $ O^V= Q^\dagger e^{V} Q - \tilde Q^\dagger
e^{-V} \tilde Q ,\ [Q = q +\t \psi _q + \t ^2 F_q] $.  Since 
current conservation entails non-renormalization (or vanishing
anomalous dimension), one is led  to assign the unperturbed scaling
dimension, $ \D ( O^V _{\t ^0 \bar t ^0} ) = 2.$ The chiral operators
are renormalized with their dimensions  determined from the
superconformal symmetry by the additive rule, $\D (Q^n)_{\t ^0} = 3n
/4 $.  The fields $A ^{(m)} _{\hat 1 , \hat 2} $ and $\chi ^{(m)} $
thus correspond to the superspace components of the chiral superfield
operator $ \tilde Q Q ,\ [\D (\tilde Q Q) _{\t ^0} =3/2 ] $.


In order to gather   existing  information on the supergravity parameter
$\mu $, we consider first the possibility~\cite{cotroneDK10} that the
mesons in the flavoured Klebanov-Strassler background correspond to
the $q-\tilde q $ quarkonia of QCD.  To pursue with this comparison
one must specify the matrix $ \mu $ in the quarks flavour space of
dimension $N_f$.  We  specialize to  the simplified case of an $ U(1)^{N_f}$
group for which $ L = \mu /\e $ is a diagonal matrix with entries
related to the bare quark masses and the dynamical scale as, $ L
_i\equiv \mu _i /\e = (m_{Q _i} / \L _{ ir, N_f } )^{3/2 } \simeq
(m_{Q } / \L _{ ir } )^{3/2 } $, assuming equality of the dynamical
scales for the pure and flavoured gauge theories.  The extension to a
non-Abelian flavour symmetry group involves tools whose implementation
exceeds the scope of the present study~\cite{babing03,erdmenger23}.
The lattice simulations of 3-flavour QCD using chiral perturbation
theory  assign the following values to  the up, down and strange quark masses, $ \bar
m _q= (m_u + m_d)/2 = (3.39 \pm 0.04) \ MeV, \ m_s = (92.09\pm
0.7)\ MeV $ for  the (renormalization scheme dependent) scale $\L =
200 \ MeV $. The experimental   data  for heavy flavour quarkonia
set the charm and bottom quarks masses at $ m_c = (1.280\pm 0.025)
\ GeV,\ \ m_b = (4.18\pm 0.03)\ GeV $.
The  resulting values for $ L_i $ for the two
choices $\L =200 \ - \ 770 $ MeV,  utilized   in
lattice simulations and holographic quark models,  are displayed in the
following table. Note that the  estimates for $ L_i $ in the two line entries
differ by a constant rescaling factor.  
 
\begin{center}  \begin{tabular}{c|cc||cc}
$\L $ \ (MeV) & $ L_{u,d} $ & $ L_s$ &  \ \ \ $ L_c$ & \ \ \ $ L_b$ \\ \hline
200 & $2.21 \times 10^{-3} $ & 0.312 & \ \ \  16.2 &
\ \ \  95.5 \\ 770 & $ 3.4\times  10^{-4}$ & 0.0413 &
\ \ \  2.14 & \ \ \  12.6 \\ \hline
\end{tabular}\end {center}

We postpone  a further  discussion  of this application  to
Subsection~(\ref{sec2sub3II}) below  and  proceed now to  the
cosmological type  constraints on the parameter $\mu $.
The   analyses~\cite{baumann06,baumann08}  of the inflation model of KKLT~\cite{kklt03}   are   concerned  with
the slow-roll of  a $D3$-brane in Klebanov-Strassler throat
attracted to a $\bar D3$-brane located near the
warped deformed conifold tip.  The
potential energy  arises from three sources: (1) The
superpotentials $W_{flux} (S) $ and $ W_{np} (\rho ) $ from the $ \int
F_3\sim M, \ \int H_3 \sim -K $
fluxes~\cite{gkp01,kapLouis94,dougshelba07} and the non-perturbative
Euclidean $ \cale 3 $-branes ($ n_f =1$) or gaugino condensation on $
n_f \ D7$-branes, necessary to stabilize the axio-dilaton and deformed
conifold complex structure and Kahler volume moduli $\tau ,\ S ,\ \rho
$.  (2) The $D3-\bar D3$-branes attractive Coulomb potential $V
_{Coul} $.  (3) The $N_{\bar D3} \bar D3$-branes potential energy $
V(\bar D3 ) = 2 e ^{4 A(\tau ) } \tau _{D3} $ necessary to produce a
de Sitter metastable vacuum with spontaneously broken supersymmetry.
We quote for definiteness schematic expressions for these three
contributions \bea && W_{flux} (S) = \hat l_s ^2 M_\star ^ 8 S
({M\over 2\pi i } (\ln {\L _{uv}^3 \over S } + 1) + i {K \over g_s } )
,\ W_{np} (\rho ) = A_0 f^{1/n_F} (w_a) e ^{-a \rho } ,\cr && V
_{Coul} = 2 N \mu _3 w^4 (1 -{4 \over 9 N} ({w \calr \over r })^4 )
,\ V(\bar D3 )\simeq {S^{4/3}\over (g_s M )^2 }, \eea
but refrain from quoting   the Kahler potentials  for the  moduli
superfields~\cite{wolfegiddings02}  which are   significant  inputs
in  this problem~\cite{baumann06,baumann08}.  The  $\bar D3 $-branes
play a critical r\^ole  regarding the supersymmetry breaking
mechanism and the adverse possibility that their  back-reaction on the
metric and 3-form fluxes produces non-physical
singularities.  Favourable answers to both issues were reported in the
extensive literature on this subject.  Firstly, it appears that the de
Sitter uplifting can be described in terms a linear realizations of a
(global or local) supersymmetry coupled to a nilpotent
superfield~\cite{kallosh22}.  Secondly, it seems that a proper account of
the constraints from warping and off-shell effects on the scalar
potential of the modulus $S$ can   remove singularities~\cite{LustRandall22}.

There are  two  possible (angular and radial)
inflation scenarios depending on whether the $D3$-brane moves
along an angle direction of the $S^3$ at the conifold tip or
along the radial $\tau $ direction. Both were analyzed for $
n_f\ D7$-branes embedded in Kuperstein 4-cycle $ f (w) = w_4 /\mu -1$
and in  Ouyang 4-cycle~\cite{ouyang03}, $f(w)= z_4 /\mu -1 = (-w_3 + i
w _4)/\mu -1 $.   For angular inflation, the condition that the $\bar D3
$-brane mass term from compactification effects~\cite{aharony05}, $
m^2_{comp} \simeq w ^{3.28} / (g_s M \a ' ) $, is dominated by the
mass term from moduli stabilization $ m^2 _{infl} $, yields the bound
on $\mu $~\cite{dewol07} \bea && m^2 _{infl} \simeq w ^2 \e /(\mu n_f
g_s M \a ' ) >> m^2_{comp} \ \Longrightarrow \ L n_f \equiv \mu n_f
/\e << w ^{-1.28}.\eea The alternative analysis using the large volume
limit $ \Re (\rho ) >> 1 $~\cite{krausepa07} finds
a comparable  bound for both Ouyang and Kuperstein embeddings, $ n_f L
\equiv n_f \mu /\e << (g_s M \a ' / \e ^{4/3} ) ^{2/3 } \approx w_s
^{-4/3} $ (setting for convenience $ w ^{-0.32 } \to w ^{-1/3} $).


The status of the radial inflation scenario  is  inconclusive
due to insufficient cancellations in the
radial scalar potential $ V(D3-\bar D3 ) $ to guarantee  slow-roll. 
An independent  information on $\mu $ can  still be inferred
from the  annihilation   of $\bar
D3$-branes  with 3-fluxes~\cite{KPV02} in the adverse case where
this process occurs before the $\bar D3 $-branes tunneling
to a metastable supersymmetry breaking vacuum.  The condition that
moduli stabilization dominates over brane-flux contributions sets the
upper bound~\cite{brownwolfe09},  $ \ln ( \mu / \sqrt {\a '}) << {2
  \calv ^{1/3} /( 3 M)} (\e ^{2/3} / \sqrt {\a '} ) $.

Useful constraints on $\mu $   are  provided  by the
holographic gauge mediation model~\cite{benini09} in  which 
the deformed   non-supersymmetric  Klebanov-Strassler
background   (in the large radius limit) is used   as a
hidden sector  with   a  spontaneously broken supersymmetry set
by the scale parameter $\cals $~\cite{dewolfe08}.  The observable sector
is the grand unified $SU(5)$ gauge theory on a $n_f\ D7$-brane stack
($n_f =5$) localized near the ultraviolet boundary of the throat.  The
massive gauginos of $D7$-branes in the throat with broken
supersymmetry mediate soft contributions to other brane modes via
their coupling with pairs of messenger meson modes $\chi ^{(m)} \sim
\bar Q O _m Q $.  The bound on   the gauginos mass for
low energy supersymmetry breaking, where $\a (\L _\mu )$ is the
$ n_f \ D7$-brane running gauge coupling constant at the scale $\L
_\mu \simeq \mu ^{2/3 } /\a ' $, yields  the condition~\cite{benini09}
\bea && m_\l \simeq {\a (\L
  _\mu ) n_f \over 4 \pi } {\cals \a ^{'3} \over \mu ^2} {1\over (4
  \pi g_s N )^{1/2} } \simeq 10^{+1} \times {\cals \a ^{'3} \over \mu
  ^{2}} \times 10^{-2} > O(100) \ GeV \ \Longrightarrow \ {\cals \a
  ^{'3} \over \mu ^2} > O(1) \ TeV , \eea
which translates in the  case  of an intermediate scale
breaking of supersymmetry, $ \cals ^{1/4} \sim 10^9 \ GeV $,  into
the  conditions on $\mu $ and $L$,
\bea && \mu ^{2/3} \leq {10 ^{11} \ GeV \over m_s
  ^2} \ \Longrightarrow \ L= {\mu \over \e } \leq 10^{-21/2} {\calv
  \over (w_s (\pi g_s M)^{1/2} ) ^{3/2} } ({\cals \over 10^{36} \ GeV
  ^4 })^{3/2} .\eea

The scale evolution  towards the  ultraviolet of the $ n_f \ D7$-brane
running gauge coupling  in this model~\cite{benini09}  provides
an additional  constraint on $\mu $.   As the floating  energy
scale  exceeds the mesons mass, $ \L > \L _\mu \sim \mu ^{2/3 } /\a ' $,
$\a (\L )$  picks up positive logarithmic contributions from 1-loops of virtual
pairs of messengers that could produce a Landau p\^ole (LP) $\a (\L )
\to \infty $ invalidating the model. The flow of $x (\L ) = 2 \pi /
\a ( \L ) $ towards the ultraviolet
is expected to proceed through pairs of
self-similar Seiberg dualities of the Klebanow-Witten gauge theory, $
SU(N) \to SU(N+M) \to SU(N +2M) $ in steps $ \d t =\d \ln Q = 2\pi / (3
g_s M) \d k ,\ [t= \ln Q ]$    with  the $D7$-brane gauge theory beta function
$ \b _k  (x) = - 3  k  M/ 2 ,\ [k \in Z _+ ] $.
The integrated   increase of the $ SU(N)$ gauge group rank
from $N _\mu = k _\mu M \to N _{uv} = k _{uv } M $  induces the
growth of the $ SU(n_f)$ gauge theory coupling constant,
\bea && \D x = 2 \pi ( \a ^{-1} (\L _{uv}) - \a ^{-1} (\L _\mu ) ) = -
\sum _{k_\mu } ^{k _{uv}} \d t \b _k \simeq - {\pi \over 2g_s} (k^2
_{uv} - k^2_\mu ) \ \Longrightarrow \ ( \a ^{-1} (\L _\mu ) - \a ^{-1}
(\L _{uv} ) ) \simeq {1\over 4 g_s} ( k^2 _{uv} - k^2_\mu ) . \eea
For  an  initial coupling constant at the GUT
value, $ \a (\L _\mu ) \simeq 1/ 25 $, the Landau p\^ole $ \a (\L
_{LP} ) \simeq \infty $ occurs at $ 1/ \a (\L _\mu ) \simeq (k^2 _{LP}
- k^2_\mu ) / (4 g_s) $, corresponding  (for $g_s \simeq 1/4 $) to
$k _{LP} \leq 5 $.  The resulting  upper   bound on  the ratio of mass scales
(for $ 8\pi / 3 M \sim 1 $), 
\bea && \ln {\L _{LP} \over \L   _\mu }
= \D t = t _{LP} - t_\mu = {2\pi \over 3 g_s M} (k_{LP} -
k_{\mu } ) = {8\pi \a ^{-1} (\L _\mu ) \over 3 M (k_{LP} + k_\mu ) }
\simeq  {25 \over k_{LP}} \geq  5  \ \Longrightarrow \
\L _\mu / \L _{LP} \leq e ^{-5} \simeq 10 ^{-2} ,\eea
implies  a  problematic narrow window  of scales from $\L _\mu \to \L _{uv} $.
For    the  values $ w_s \sim 10 ^{-4} ,\
\L _\mu = \mu ^{2/3} m_s ^2,\ \L _{LP} = m_s :\   \L
_\mu /\L _{LP} < 10^{-2} \ \Longrightarrow \ L< 10^{-3} w_s^{-3/2}
/\calv ^{1/4} $.
As discussed  in~\cite{benini09},  one can widen  this bound
via the  conifold orbifolding.  For the orbifold $ \calc _6 /Z_Q $, the flux
parameter gets  reduced, $ M\to M/Q $, hence  $ \b _k \to \b _k/ Q $,
but the scale evolution $\d t / \d k= 2\pi /(3 g_s M) $ is
unaffected. The modified  bound  for $ Q=5$ reads,
$1/ \a (\L _\mu ) \simeq (k^2 _{LP}
- k^2_\mu ) / (4 g_s Q) \ \Longrightarrow \ \L _\mu / \L _{LP} \leq e
^{-5 Q} \simeq 10 ^{-10} $. 

We   conclude  from the    above    comparisons 
that the conditions on   $L = \mu  /\e $
favour values $ L= O(1)$  in hadronic physics applications
and   larger values $L > O(10^{2})$ in particle  physics applications
which   feature a sensitive dependence on the     string theory and
compactification  parameters $\a ' ,\ \e  ,\ \calv $.

\subsubsection{Mass spectra   of scalar and vector  mesons} 
\label{sec2sub3II}

\begin{figure}[b]
\begin{subfigure}[b]{0.49\textwidth} 
\includegraphics[width=0.9\textwidth]{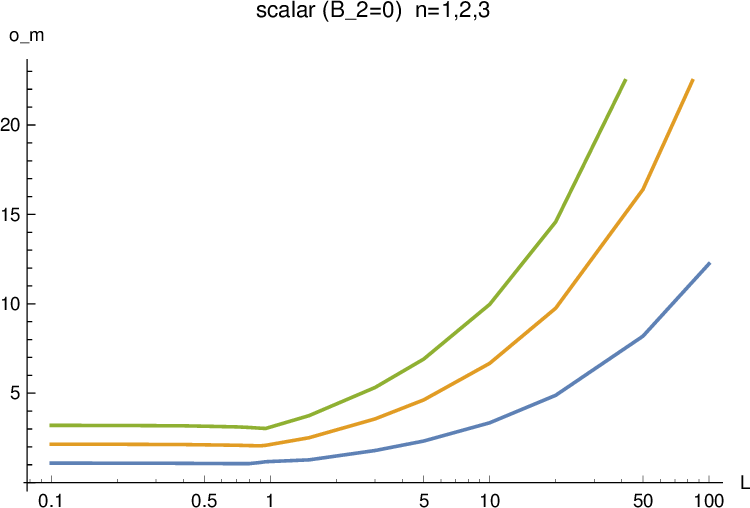}
\caption{\it Scalar modes for $ B^{cl} _2 = 0$. } \end{subfigure}
\begin{subfigure}[b]{0.49\textwidth} 
\includegraphics[width=0.9\textwidth]{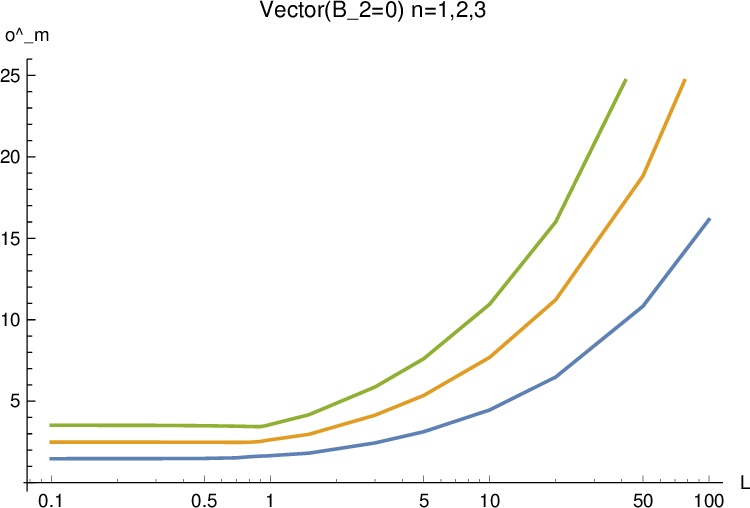}
\caption{\it Vector modes for $ B ^{cl} _2 = 0$.}  \end{subfigure}
\begin{subfigure}[b]{0.49\textwidth} 
\includegraphics[width=0.9\textwidth]{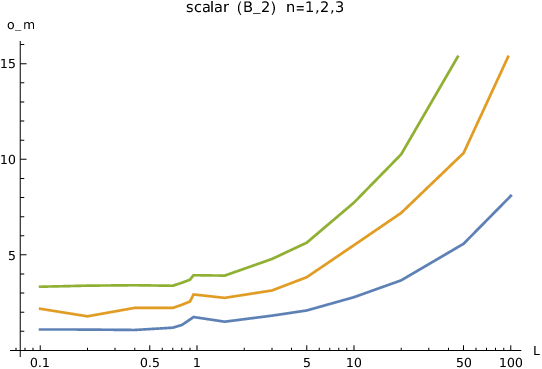}
  \caption{\it Scalar modes for $ B^{cl} _2 \ne 0$.} \end{subfigure}
\begin{subfigure}[b]{0.49\textwidth} 
\includegraphics[width=0.9\textwidth]{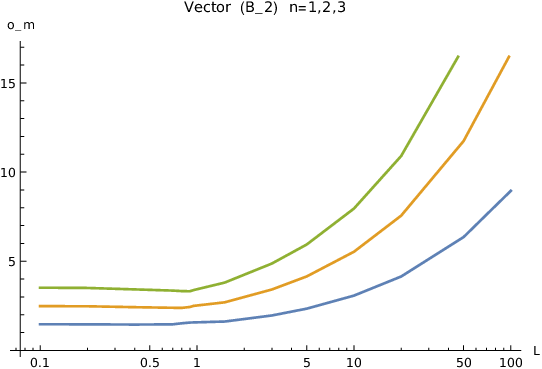}
\caption{\it Vector modes for $ B^{cl} _2 \ne 0$.}  \end{subfigure}
\caption{\it \label{GRChiD7} The scalar $ \chi ^{(m)} $ (left) and the
  vector $ A _\mu ^{(m)} $ (right) mesons masses $\hat \o _m $ are
  plotted as a function of the parameter $ L = \mu / \e \in [0 \to 100
  ]$.  The curves from bottom to top refer to the ground state and
  first two radially excited modes $ n=1,2,3$.  The upper panels
  $(a),\ (b)$ refer to the case $ B_2 ^{cl} =0$ (ignoring the NSNS
  field)  and those in the lower panels $(c),\ (d)$ to the case $ B_2
  ^{cl} \ne 0$ for the full background solution.  We set $ g_s = 1$
  in numerical calculations.}
\end{figure}

Our   applications were all   developed within  the
semi-classical approach using `Mathematica' numerical
tools.    The predictions for the mass spectra of the radial
excitations of scalar and vector mesons are
listed in Table~\ref{SPECTRUM} and displayed as a function
of real $ L = \sqrt {L^2} \in [ 0, \ 100 ] $ in Fig.~\ref{GRChiD7}.
It is   clearly  seen  that scalar modes are significantly lighter than vector
modes. The masses are nearly constant for $ L\in (0, 1) $ with a weak
inflection point at $L=1$ and a slow growth until $ L= O(10)$ beyond
which they follow the asymptotic power law regimes $\hat \o _{1}
\simeq (0.8 \ - \ 0.6 )\  L^{2/3}$ for scalars and $\hat \o _{1} \simeq
(1. \ - \ 0.7 )\  L^{2/3}$ for vectors.  The masses grow with the radial
quantum number as $ \hat \o _m \propto n $, as expected for low
curvature throats and previous results for $ \caln = 2 $
backgrounds~\cite{erdmenger07}. This contrasts with the growth law $
\hat \o _m \propto n ^{1/2} $ in strongly coupled gauge
theories~\cite{karch06} and the mixed law $ \hat \o _n \sim (a + b n )
$ in analyses for multidimensional fields
spaces~\cite{berg06,dymkov06,benna07,dymkovyov08}.  The mass ratios $
[\hat \o _{2 } / \hat \o _{1} ,\ \hat \o _{3} / \hat \o _{1} ] $ are
approximately independent of $L$, ranging inside $ [2, 3] $ for
scalars and $ [1.7 , 2.4] $ for vectors.

The   mass gap,   defined as the
ratio of  the mesons mass  $\o _{mes} \simeq \mu ^{2/3} m_s^2 / (g_s M) $
(at $L >>1$) to the  open string mass (in flat spacetime) $ m_Q
\simeq \mu ^{2/3} T ({F1}) \sim \mu ^{2/3} / (2\pi \a ') $,
sets typically at $ \o _{mes}/ m_Q \sim 1 / (g_s M)$. Note this is smaller than
the ratio in the undeformed conifold case~\cite{leviyang05}, $ M_{mes}
\sim m_q /\sqrt {g_s N} $.  If one used   instead the $F1$-string tension
near the conifold apex, $T ({F1}) \simeq w_s m_s ^2 \calv ^{-1/3}$,
a substantially larger mass gap  wold result, $\o _{mes}/ m_Q (F1)
\simeq (\e ^{2/3 } m_s )^{-2} \simeq \calv ^{1/3} / (g_s M w _s ^2 )
$.

For completeness, we compare in the table below our semi-classical
predictions for the reduced masses of (ground and first radial
excitation) scalar and vector mesons $\hat \o _m ( \chi ) ,\ \hat \o
_m ( A_\mu ), \ [m=1,2] $ with those of other authors which made use
of the shooting technique.  Four cases $ {\bf A,\ B,\ C,\ D }$ are
considered in the successive columns.  The results refer to $\o _m /
m_{gb} = 2^{1/3} \hat \o _m $ in cases ${\bf A,\ B}$ and to $\l _m =
\o _m L^2_{eff} /\mu ^{2/3} = ( {81 \ln ( r_\mu / r_0 )\hat \o _m^2 /
  ( 8 L^{4/3}) } )^ {1/2} \to 5.13 \hat \o _m /L^{2/3} $ in cases
${\bf C,\ D}$. In all cases but case ${\bf B}$, we set $ B_2^{cl} =0
$. The comparisons reflect on the numerical evaluation scheme used but
also on the parameter choices, since the previous studies considered
the limit $\mu =0 $ in ${\bf A}$ and the limit $\e =0$ in ${\bf C
  ,\ D}$.  The results show some disparity but they roughly agree
within a factor 2.

  
\begin{center} \begin{tabular}{|c|cc|c||cc |c|}\hline
 & $ {\bf A} (\mu \simeq 0 )  \to $ & $ L \simeq 0.1 $~\cite{kuperstein04}  &
 $ {\bf B} (L=1.5) $~\cite{cotroneDK10}  &$ {\bf C} (\mu >> 1 )
 \to $ & $ L=50$~\cite{benini11}  & $ {\bf D}
 (\mu = 10, \e =0)  \to L=100 $~\cite{benini09}  \\ \hline &
 $2^{1\over 3} \hat \o _m ( \chi )$ & $2^{1\over 3} \hat \o _m (
 A_\mu ) $&$2^{1\over 3} \hat \o _m ( A_\mu ) $&$ \l _m (\chi ) $ &
 $ \l _m (A_\mu ) $ & $ \l _m (A_\mu ) $ \\ Shooting & $ (3.38,
 4.92) $ & $(4.32, 5.8 ) $&$ (1.62, 3.04 ) $&$ (1.89, 4.39) $ & $
 (2.56, 4.97) $&$ (1.89, 3.73) $ \\ JWKB & $(1.37, 2.70) $ & $
 (1.85, 3.13) $&$ (2.03, 3.39) $&$ (3.08, 6.16) $ &$ (4.09, 7.11) $
& $ (3.83,\ 6.69 ) $ \\ \hline  \end{tabular}\end{center}

\begin{table}[t]
\caption{\it \label{SPECTRUM} List of masses $ \hat \o ^2 _{1, 2, 3} $
  of the scalar and vector ground state and two first radially excited
  mesons at a set of real positive values of $ L = \mu / \e \in [ 0,
    \ 100 ] $.}
\begin{tabular}{|cc|ccccccc||ccccccc|} \hline  
$L$ & & 0.1& 0.2& 0.4& 0.7& 0.8& 0.9 & 0.95 & 1.5 & 3& 5& 10& 20& 50&
  100 \\ \hline $ \chi ^{(m)} $ & $ \hat \o _1^2 $ & 1.187& 1.181&
  1.161& 1.124& 1.131& 1.273& 1.367 & 1.620& 3.221 & 5.408 & 11.19 &
  23.84& 67.07& 149.6 \\ & $ \hat \o _2^2 $ & 4.610& 4.595& 4.535&
  4.367 & 4.290 & 4.241& 4.350 & 6.307 & 12.70 & 21.40& 44.51& 95.17&
  268.7 & 600.2 \\ & $ \hat \o _3^2 $ &10.25 & 10.22 &10.09 &9.72 &
  9.516 &9.26 &9.170 &14.03 & 28.31 &47.73 &99.41 &212.7 & 600.76 &
  1366.3 \\ \hline \hline $ A_\mu ^{(m)} $ & $ \hat \o _1^2 $ & 2.161
  & 2.146 & 2.206 & 2.316 & 2.519 & 2.648& 2.673 & 3.276 & 5.938 &
  9.75& 19.90& 42.01& 117.2& 260.5 \\ & $ \hat \o _2^2 $ & 6.19 & 6.13
  & 6.13 & 6.10 &6.12 & 6.37 & 6.68 & 8.807 & 17.14 & 28.64 & 59.18
  &125.9 & 354.2 &790.4 \\ & $ \hat \o _3^2 $ & 12.39 & 12.27 &12.20
  &11.99 &11.86 & 11.80 & 12.20 & 17.40 & 34.4 & 57.87 & 120.0 &
  256.17 &722.3 & 1614.0 \\ \hline
\end{tabular}  \end{table} 

The results in Fig.~\ref{GRChiD7} and Table~\ref{tabNSNS}  show that
 the classical field $ B_2 ^{cl}$  has a negligible effect at $L < 1$
 but  contributes  significantly at $L > 5 $, where masses  are
 reduced by a factor $\sqrt 2$  while
 mass splittings between radial excitations  are  enhanced.  We
have also evaluated the mass spectra in the  region of the
parameter space $ L^2 < 0 $.
The results in Table~\ref{tabL2N} show minor changes relative to those
for $L^2 > 0 $ (in Table~\ref{tabNSNS}). The variation in $ \vert L
\vert < 1 $ is weaker but the growth $\hat \o _m \sim \vert L \vert
^{2/3 } $ is stronger for $ \vert L \vert >>1 $ and more efective for
$\chi ^{m} $ relative to $ A_\mu ^{m} .$

\begin{table}[t]
  \caption{\it \label{tabNSNS} Scalar and vector mesons masses of the
    ground state and first radially excited mesons including the
    contributions fom NSNS $ B_2 ^{cl} $ solution evaluated for a list
    of real values of $L^2 >0$.}
\begin{tabular}{|cc|ccccccc||ccccccc|} \hline
$L$ & & 0.1& 0.2& 0.4& 0.7& 0.8& 0.9 & 0.95 & 1.5 & 3& 5& 10& 20& 50&
  100 \\ \hline $ \chi ^{(m)} $ & $ \hat \o _1^2 $ & 1.20 & 1.19 &
  1.14 & 1.41 & 1.77 & 2.60 & 3.05 & 2.26 & 3.31 & 4.37 & 7.76 & 13.4
  &31.1 & 65.4 \\ & $ \hat \o _2^2 $ & 4.75 & 3.18 & 4.95 & 4.95 &
  5.70 & 6.53 & 8.57 & 7.56 & 9.83 & 14.6 & 30.3 & 51.7 & 106.6 &
  245.5 \\ \hline \hline $ A_\mu ^{(m)} $ & $ \hat \o _1^2 $ & 2.15 &
  2.14 & 2.10 & 2.14 & 2.32 & 2.44 & 2.47 & 2.62 & 3.84 & 5.48 & 9.45
  & 17.18 & 40.41 & 80.09 \\ & $ \hat \o _2^2 $ & 6.17 & 6.13 &5.88 &
  5.71 & 5.69 & 5.93 & 6.245 & 7.278 & 11.68 & 17.21 & 30.62 & 57.12 &
  137.8& 277.5 \\ \hline
\end{tabular}  \end{table}

\begin{table}[t]
  \caption{\it \label{tabL2N} Scalar and vector mesons masses of
    ground state and first radially excited modes with NSNS $ B_2
    ^{cl} $ solution for a list of real values of the ratio parameter
    $L ^2 = \mu ^2 /\e ^2 < 0 $.}
\begin{tabular}{|cc|ccccccc||ccccccc|} \hline
$-i L $ & & 0.1& 0.2& 0.4& 0.7& 0.8& 0.9 & 0.95 & 1.5 & 3& 5& 10& 20&
  50& 100 \\ \hline $ \chi ^{(m)} $ & $ \hat \o _1^2 $ & 1.21 & 1.22 &
  1.25 & 1.35 & 1.39 & 1.40 & 1.42 & 2.83 & 4.51 & 6.58 & 11.5 & 21.2
  & 50.3 & 100.2 \\ & $ \hat \o _2^2 $ & 4.77 & 4.43 & 5.36& 5.86 &
  5.84 & 5.85 & 5.96 & 11.75 & 19.17 &28.5 & 51.2 & 96.2 &233.3 &
  470.3 \\ \hline \hline $ A_\mu ^{(m)} $ & $ \hat \o _1^2 $ & 2.16 &
  2.17 & 2.26 &2.35 & 2.40 & 2.46 & 2.49 & 2.86 & 4.00 & 5.58 & 9.50 &
  17.2 & 40.4 &80.0 \\ & $ \hat \o _2^2 $ & 6.20 & 6.25 & 6.54 & 6.83
  & 7.01 &7.20 & 7.30 & 8.51 & 12.3 & 17.5 & 30.8 & 57.2 & 137.8 &
  277.5\\ \hline
\end{tabular}\end{table}

\begin{table}[b] 
  \caption{\it \label{HADRONS} Comparison of ratios for the ground
    state and first or second radially excited modes between mesons
    and hadrons for $ L =[0.1, 0.2, 10, 100] $ assigned to $[(u, d ),
      s, c, b ]$ quark flavours.  No data is available for the second
    radial modes of the $\eta _c,\ \eta _b $ hadrons.}
   \begin{tabular}{|c|c cccc||c cccc|} \hline  
     $ 1 ^{--} $&$ {\o _2 \over \o _1 } $&$ \rho $&$ \phi $&$ J/\Psi
     $&$ \Upsilon $&$ {\o _3 \over \o _1 } $&$ \rho $&$ \phi $&$
     J/\Psi $&$ \Upsilon $ \\ \hline Hadrons & &$ 1.88 $&$ 1.65 $&$
     1.19 $&$ 1.06$& & $ 2.73 $&$ 2.12 $&$ 1.21 $&$ 1.08 $ \\ Mesons &
     & $ 1.69 $&$ 1.69 $&$ 1.72 $&$ 1.74 $& & $ 2.39 $&$ 2.39 $&$ 2.45
     $&$ 2.49 $ \\ \hline $0^{-+}$& $ {\o _2\over \o _1 }$&$\pi $&$
     \eta $&$ \eta _c $&$ \eta_b $& $ {\o _3 \over \o _1 } $&$ \pi $&$
     \eta $ &$\eta _c $&$ \eta_b $ \\ Hadrons & & $1.39 $&$ 1.35 $&$
     1.22 $&$ 1.04$ & & $ - $& $ 1.47 $&$ - $&$ - $ \\ Mesons & & $
     1.97 $&$ 1.97 $&$1.99 $&$ 2.0 $& & $ 2.94 $&$ 2.94 $&$ 2.98 $&$
     3.02 $ \\ \hline
\end{tabular}  \end{table}

Before  addressing  the holography
correspondence~\cite{cotroneDK10} to the QCD  (Quantum Chromodynamics)
hadrons, it is useful  to    recall that
the space inversion and charge conjugation parities  assigned   to 
meson modes  are linked to the choice of boundary conditions 
near the origin $\rho =0 ,\ \O =0$.
The transformations under $P,\ C$ of the 4-d fields
$ f_m (\rho ) \chi ^{(m)} (x) $ and $a _m
(\rho ) a_\mu ^{(m)} (x) $    follow from  the induced action of
these symmetry operators on the (transversal) spatial and
(longitudinal) gauge coordinates $\chi \sim X^{8+i9} ,\ A_\mu \sim
X_\mu ] $ in the $D7$-brane theory, $ P:\ [\chi \to - \chi ,\ A_\mu
    \to - A_\mu ] $ and $C:\ [\chi \sim \chi ,\ A_\mu \to - A_\mu ^T] $,
as discussed  in~\cite{Sakaito04,kupsonn08,benna07,pufu10}.
One can  choose, for instance, radial and angular 
wave functions  $ f_m (\rho ,\O ) ,\ a _m (\rho
,\O ) $    that are even or odd  under the reflections,
$ \rho \to -\rho ,\ \O \to - \O $.
The angular dependence (on $\O $) can be ignored for singlet
modes. For radial wave functions $ f_m \sim \hat f_m ,\ a_m \sim \hat
a_m $ even under $\rho \to - \rho $,  hence 
obeying the boundary conditions at $\rho =0:\ f_m (0) \ne 0
,\ \dh _\rho f_m (0) = 0 $ and $a_m (0) \ne 0 ,\ \dh _\rho a_m (0) = 0 $,
the assigned $ J^{PC}$ quantum numbers are $ 0^{-+} $ for
scalar mesons and $ 1 ^{--} $ for vector mesons.

It is safe to restrict  our  study to  $S$-wave quarkonia and
their radial excitations, ignoring the $
0^{++}$ scalar hadrons which exhibit large decay widths attributed to
mixing with glueballs.  We focus on the natural parity vector and
pseudoscalar hadrons of spacetime quantum numbers $J^{PC} = 1 ^{--
},\ 0^{-+} $, and their radial excitations and restrict to the
electric charge neutral members of $ SU(3)_{fl}$ multiplets.  We also
discard the $\pi ^0 (140) $-meson, since spontaneously broken chiral
symmetry is definitely absent in the present model.  We thus consider
the masses (in MeV units) of (1) the ground and radially excited
vector mesons of $ u, \ d $ flavours for $\rho :\ [770, 1450 , 2100 ]
$ and $\o: \ [782, 1420, 1650]$ and of the mixed $(u,d, s)$ mixed
flavours for $\phi : \ [ 1020, 1680 , 2170] $ and (2) the excited
pseudoscalar mesons of $(u, d) $ flavours, $\pi :\ [1300, 1800] $ and
of $(u, d, s ) $ mixed flavours, $ \eta :\ [958, 1295, 1405, 1475 ] $.
We include in our test the heavy flavours hadrons of same quantum
numbers.

The    comparison  of holographic mesons to QCD hadrons reveals two
discouraging features   given that scalar mesons  are lighter than vector
mesons and the masses for mesons with light quark flavours change
little in the relevant parameter range $ L_i = \mu _i /\e < 1$. With
the choice $ L _i \simeq (m_{q_i} / \L )^{3/2 } = [0.1 , 0.2, 10, 100]
$ for the  $ (q, s, c, b)$ flavours, fitting the
predicted vector meson ground state mass $\o _1 \equiv \L _{ir} \hat
\o_1 /(g_s M) ,\ [ \hat \o_1 = \sqrt {2.16} ] $ to the $\rho
^0$-hadron mass determines  $\L _{ir} / (g_s M) = 524 $ MeV. The
resulting mass predictions for the ground state mesons $ [(770), 767,
  2336, 8454] $ MeV compare poorly to those of the lightest $ 1 ^{--}$
hadrons, $ m [\rho, \phi , J/\psi , \Upsilon ] = [(770), 1015, 3097,
  9460] $.  The predicted scalar meson mass $ \o _1 (0 ^{-+}) = 629$
MeV also misses the observed value by a factor 2.  We turn next to the
ratios of mesons masses where one expects a weaker sensitivity to 
the parameters.   The predicted ratios of the first and second
radially excited meson modes relative to the ground state mode are
compared to those of quarkonia hadrons in Table~\ref{HADRONS}.  In
spite of the improved agreement, we fail to account for the  uniform 
decrease   of hadronic mass ratios  upon   going  from light
to heavy  flavours. 
 
Two  main drawbacks of supersymmetric
holography   models for hadrons  are the presence of
fermionic superpartners  and the  absence of a
spontaneous breaking of chiral symmetry $ U (N_f) _L \times U (N_f) _R
\to U (N_f) _V $ (excluding light Nambu-Goldstone  pseudoscalar mesons).
To estimate the uncertainties one could   examine   from the gauge theory side
how    soft supersymmetry breaking effects
affect the transition to the non-supersymmetric  theory.
Unfortunately, the contributions from F-term components of spurion
superfields   at small energy scales  are understood
moderately  for the gauginos and
scalars  mixing~\cite{Aharony95,evans97}   and  poorly for  the
scalars masses.  On the other hand,  relaxing  the
supersymmetry constraints on the supergravity side (holomorphic
4-cycles with anti-self dual, primitive gauge fields, $\calf = -\star
_{S _4} \calf,\ \calf \wedge J =0$)~\cite{kupsonn08}  entails
starting  anew   from  $U$-shaped embedding of
$D7 -\bar D7$-branes in the ultraviolet
joining together towards the infrared~\cite{Sakaito04}  which also  require
activating     gauge   and dilaton-axion
fields~\cite{kruczenski03,sakaisonn03,erdghorokukirsch07,kupsonn08,dymarkupson09}. 
We here consider  the   non-supersymmetric 
deformed Klebanov-Strassler background   embedding  a
smeared distribution of $ p (D3 - \bar D3) $-branes
near the conifold apex  which is     dual  to the  $ SU(N-p
+M)\times SU(N-p)$  gauge theory in a metastable supersymmetry breaking
vacuum~\cite{dewolfe08}.  The dilaton,  2-form  and
energy-momentum tensor  fields $ O_+ \sim e
^{-\phi } ,\ O_- \sim e ^{-\phi } \int _{S^2} (B_2 + i C_2), \ \T
_{\mu \nu } $  acquire finite VEVs in the  deformed solution of size
set by the   warped vacuum energy scale, $\cals = {p\over N} (w_s m_s )^4 $.
The    corrections to   the vector mesons masses  in this
background  were actually  evaluated in~\cite{benini09}
\bea && \o _m ^2 \to \o
_m ^2 (1 - {\cals \a ^{'4} r_m \over \mu ^{8/3} } ) = \o _m ^2 (1- {p
  \calv ^{2/3} r _m \over N (g_s M)^2 L^{8/3 } } ) , \eea where the
constant coefficients $ r_m = \d \b _m / \l _m ^2 $ take numerical
values of $ O(10) $ increasing with the radial excitation.  The
negative sign correction reduces the masses of light relative to heavy
flavour mesons, owing to the factor $ \mu ^{- 8/3} $, while the factor
$ r_m = O(10) $ reduce the ratios $ \o^2 _2 /\o _1^2 \propto (r_2- r_1
) ^{1/2} $ upon going from light to heavy modes.  Both features have
the potential of removing the disagreements found in the above
comparison  using Table~\ref{HADRONS}.

\subsubsection{Interactions of scalar mesons}  
\label{sec2sub3III}

\begin{figure}[t]
\begin{subfigure}{0.4\textwidth} 
\includegraphics[width=0.99\textwidth]{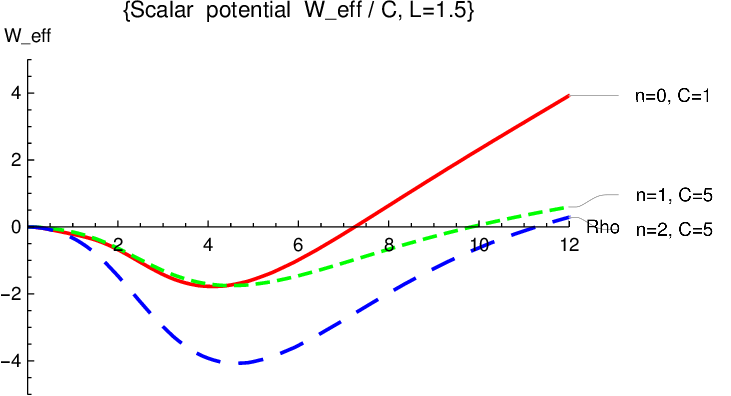}
  \caption{\it Potentials $ W _{eff} (\rho ) $ of the scalar mesons
    ground and first two radial excitations $ n=(0, 1,2) $ at $L=1.5$,
    with the indicated constant rescaling factors $C=  (1,5, 5)$.}
\end{subfigure} 
\begin{subfigure}{0.4\textwidth} 
\includegraphics[width=0.99\textwidth]{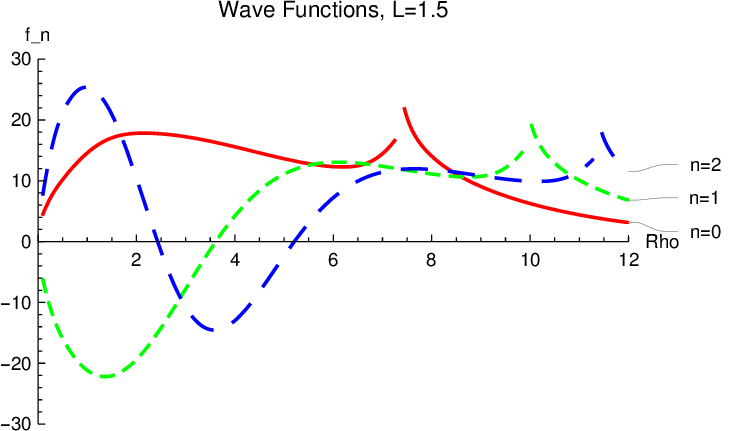}
\caption{\it Wave functions $f_n  (\rho  ) $ (in JWKB  approximation)
  of ground and   first two radial excitations $ n=(0, 1,2)$ at $L=1.5$}
\end{subfigure}
\begin{subfigure}{0.4\textwidth} 
\includegraphics[width=0.99\textwidth]{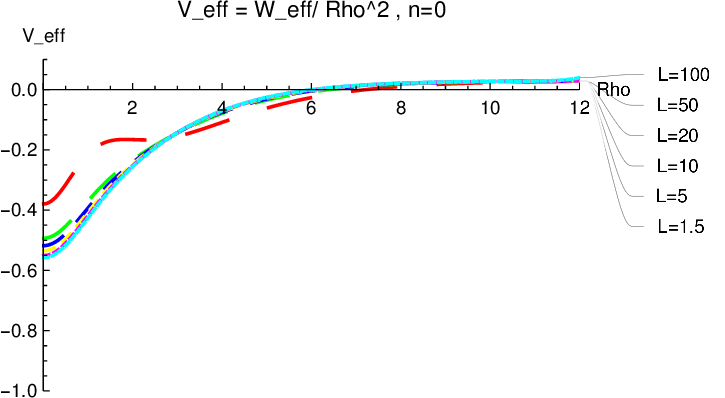}
\caption{\it Potentials $ \hat V_{eff} (\rho )= W_{eff} (\rho ) /\rho
  ^2 $ of ground state mode at $L=[1.5, 5, 10, 20, 50, 100]$.}
\end{subfigure}
\caption{\it \label{GR3wcofXD7} Plots versus $\rho $ of
  the effective potentials  $W_{eff} (\rho ),\  \hat V_{eff} (\rho ) =
  W_{eff} (\rho ) /\rho  ^2 $    and wave functions $f _n (\rho ) $
  (in units of $\e =1,\   g_s M \a ' =1$) of   the scalar meson mode fields
  $\chi ^{(n)} (x)  $  at fixed   values of $ L= \mu /\e $ in the case  $
  B_{2} ^{cl}  = 0$.}
\end{figure}

\begin{figure}[b]
\begin{subfigure}{0.4\textwidth} 
\includegraphics[width=0.99\textwidth]{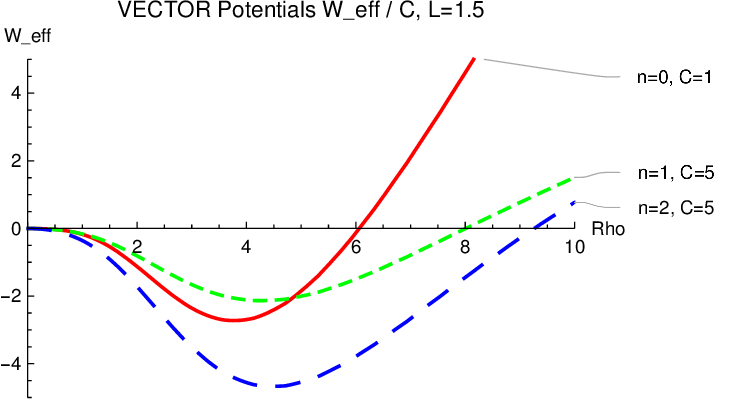}
\caption{\it Effective potentials $ W_{eff} (\rho ) / C $
  of ground and first two  radial excitations $ n=(0,1,2) $ at $L=1.5$
with the rescaling  factor $C= (1,5,5 )$.}
\end{subfigure} 
\begin{subfigure}{0.4\textwidth} 
\includegraphics[width=0.99\textwidth]{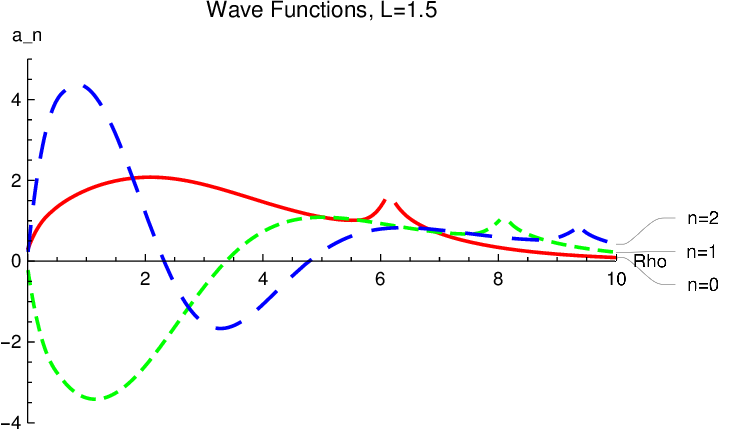}
\caption{\it Wave functions $a_n  (\rho ) $  
  of ground    and first two radial excitations $ n=(0,1,2) $ at $L=1.5 .$}
\end{subfigure}
\begin{subfigure}{0.4\textwidth} 
\includegraphics[width=0.99\textwidth]{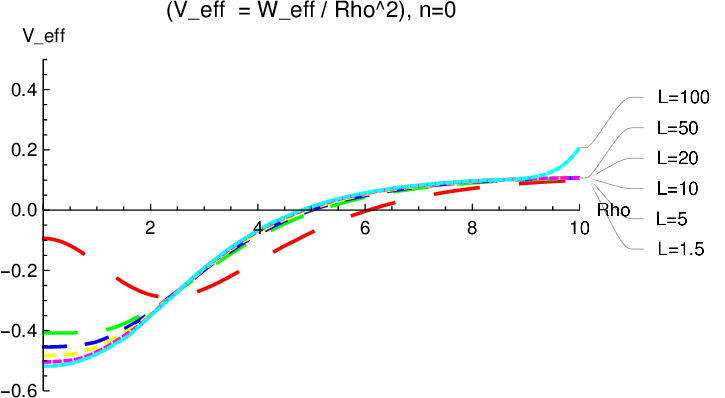}
\caption{\it Potentials $ \hat V_{eff} (\rho ) =W_{eff} (\rho ) /\rho
  ^2 $ of ground state mode at $L=[1.5,5, 10, 20, 50, 100]$.}
\end{subfigure}
\caption{\it \label{GRVVVD7} Plots versus $\rho $ of the effective potentials
 $W_{eff} (\rho ),\  \hat V_{eff} (\rho ) =W_{eff} (\rho ) /\rho
  ^2 $    and wave functions $a_n  (\rho ) $ (in units of $\e =1,\
  g_s M \a ' =1$)  of    the vector fields modes  $A_\mu ^{(n)}  $
 at  fixed   values of  $ L= \mu /\e $  in the case  $ B^{cl} _{2} = 0$.}
\end{figure}

The tree level couplings of modes are  obtained by expanding the
$D7$-brane action in powers of field fluctuations and evaluating
coupling constants in terms of overlap integrals of  wave
functions.  It is useful to first have a look on the radial profiles
of the effective potentials and wave functions which are displayed in
Fig.~\ref{GR3wcofXD7} for scalars and Fig.~\ref{GRVVVD7} for vectors.
The plots of the potentials $W_{eff} (\rho )$ in panels $(a) $ feature
'well' regions getting deeper and broader with increasing radial
quantum number $ n$ (or mass)  with  monotonic  growth  beyond the
classical turning points $\rho _0 $.   The potentials for vector modes
are deeper and narrower than those of scalar modes.

The semi-classical wave functions $\hat f _m (\rho ) $ and $\hat a_m (\rho
) $ in panels $(b) $ are peaked at the 4-cycle tip and decrease with
long tails beyond the turning points. The previously
announced boundary conditions
near $\rho =0$ are visible on these plots.  One sees the expected
oscillatory behaviour at large $\rho $ for radially excited modes with
exponentially decreasing tails beyond the turning points.  The vector
modes have smaller amplitudes.  The radial profiles of the effective
potentials $ \hat V_{eff} =W_{eff} /\rho ^2 $, displayed in panels
$(c) $, shows that the $L$-dependence significantly saturates beyond $
L \sim 10$.  The smooth dependence of the predicted masses on $ L$ is
a consequence of the slow variation of the parameters $ X $ and $\tau
_{min}$ and of the effective potentials turning points $\rho _0 $, as
illustrated in the following table.

\begin{center} 
\begin{tabular}{c|c|c|c|c} \hline
$L$ & $ \tau _{min}$ & $ X $ & $ \rho _0 (\chi ) $ & $ \rho _0 ( A_\mu
  )$ \\ \hline $ 0. \to 0.9 $ & 0 & $ 0.1 \to 2.0 $ & $ 6.7 \to 9.6 $
  & $ 5.4 \to 8.2 $ \\ $ 1.5 \to 100 $ & $ 1.9 \to 10 $ & $ 1.3 \to
  1.0 $ & $ 7.2 \to 6.0 $ & $ 6.1 \to 4.8 $ \\ \hline
\end{tabular}
\end{center} \vskip 0.3 cm

For imaginary  parameter values  $ L= i (0.1 \to 0.9) $  the
variations  are $ \tau _{min} \simeq (0.2 \to 1),\ X \simeq (0.1 \to 1.7).$
We also note that  the predictions depend weakly
on the auxiliary ultraviolet cutoff parameter values as long as one
chooses $\tau _{uv}< 25 $ or $\rho _{uv} < 15 $.
For example, increasing $\rho _{uv} = 10 \to 12 $ changes the mass at $
L=100 $ from $ \o _m ^2 = 152 \to 149 $.  The $\tau
$-profiles are similar to the $\rho $-profiles except for the
successive thresholds at $\tau _{min} = \tau (\rho =0)$ which grow
with increasing $L $.


The glueball-meson coupling constants $\l ^ A _\D $ of dimension $\D =
5, 6$ and the mesons self couplings $ \l ^S _\D $ of dimension $\D =
2, 3, 4 $ are described by the Lagrangians
\bea && L_{EFF} = \l ^A _{mnp} h ^{(m) \mu
  \nu } \dh _\mu \chi ^{(n)} \dh _\nu \bar \chi ^{(p)} + \l ^A _{mnlp}
h ^{(m) \mu \nu } \dh _\mu \chi ^{(n)} \dh _\nu \bar \chi ^{(p)} \chi
^{(l)} \cr && +\l ^S _{mn} \chi ^{(m)} \bar \chi ^{(n)} + \l ^S _{mnp}
\chi ^{(m)} \bar \chi ^{(n)} \chi ^{(p)} + \l ^S _{mnpq} \chi ^{(m)}
\bar \chi ^{(n)} \chi ^{(p)} \bar \chi ^{(q)} , \eea using
Eqs.~(\ref{eq.grmes2}), (\ref{eq.selfmes2}) and~(\ref{eq.grAmu}).  We
associate to $\l ^ A _\D ,\ \l ^ S _\D $ the reduced 
dimensionless coupling constants
$ \hat \l ^A _{\D } = \l ^A _{\D } / \calr ^A _\D  ,\  \hat \l ^S _{\D } =
\l ^S _{\D }  / \calr ^S_\D $ by factoring out the dependence on parameters
in the coefficients $ \calr ^A _\D ,\ \calr ^S _\D $ (which differ
from one coupling type to the other and may include powers of $L$).
Our numerical predictions for the reduced coupling constants $\hat \l
_\D ^{A, S}$ are listed in Table~\ref{COUPLINGS} for 3 values of $
L=[0.5 , 1.5, 3]$ where  the  expressions  for the coefficients $\calr _\D $
are displayed in the second line.  We see that the gravitational couplings
are strongly suppressed for $ L > 3 $, owing to the negligible overlap
due to the widened distance between glueballs and mesons.
The reduced coupling constants are seen to be $ O(1)$ with the
exception of $A_6$.  Inside the interval $ L\in [0, 3] $ the
gravitons-mesons coupling constants decrease by a factor 5, while the
derivative mesons couplings $A_6$ and the non-derivative reduced
couplings $S_2, S_3, S_4 $ increase by a factor $ < 2 $.  However,
upon taking the $L $-dependence in $ \calr $ into account, one finds
that also $\l ^S _{3,4}$ decrease slowly with increasing $L > 1$.

The term $ S_2 \simeq \l ^S _2 \vert \chi \vert ^2 $ corresponds to a
mode independent tree level contribution to the scalar meson masses
coming on top to the squared masses $\o _m ^2$.  The mass shifts, $\d
m ^2_\chi =\calr _2 ^S \hat \l ^S _2 \simeq ( M_\star w )^2
/(g_s M \calv ^{1/3} ) $ are typically smaller that the masses $ \o _m
^2 = m_s ^4 \e ^{4/3} \hat \o _m ^2 /(g_s M )^4 \simeq ( M_\star w )^2
\hat \o _m ^2 /(g_s M \calv ^{1/3} ) $ and are strongly suppressed at
large $ L$  since  $\hat \o _m ^2 \propto L^{4/3} .$

\begin{table} [t]
  \caption{\it \label{COUPLINGS} Reduced coupling constants
 $ \hat \l _\D = ( \hat \l  ^{A_5^n} ,\ \hat \l  ^{A_6} ,\ \hat \l  ^{S_n} ) $
for  the dimension $\D =5$ operators  $ A ^{(n)} _5(h \dh \bar \chi \dh \chi) $ 
 in the ground and first two   radial modes $ n=(0, 1, 2) $,
 the dimension $\D =6$ operator $ A_6 (h \chi  \dh \bar \chi \dh \chi) $and the dimension $\D =   2, 3 , 4 $ operators $S_2 (\chi ^2) ,\ S_3 (\chi ^3), \ S_4 (\chi ^4)$ in the ground state mode. The ratios of the physical  coupling
 constants  to the reduced ones,  $\calr _\D ^A = [\l ^{A} _{5,6} / \hat \l
    ^{A} _{5,6} ,\ \calr _\D ^S = \l ^{S} _{2,3,4} / \hat \l ^{S}
   _{2.3 ,4} ] $, including the dependence on  free parameters,
 appear in the second line entry.  The next lines
  entries display the reduced coupling constants at the values $
  L = \mu /\e  = (0.5,\ 1.5,\ 3)$.}
  \begin{tabular}{|c|c|c||ccc|} \hline  
 & $A_5^0\ \ \ \ \ \ \ A_5^1\ \ \ \ \ \ \ \ \ \ A_5^2 $&$ A_6 $ &$ S_2
  $&$ S_3 $&$ S_4 $ \\ \hline
$ \calr _\D = {\l _\D \over \hat \l _\D } $ & ${ 0.787 \ 10^4 \rho
    ^{1/4} \over (g_s M^3)^{1/2} M_\star w } $ & $ { 0.899 \ 10^{6}
    (\rho / g_s ^5 )^{1/2} \over M ^3 L (M_\star w )^2 } $ & $ {0.754
    (M_\star w)^2 \over g_s M \rho^{1/2} } $ & $ {42.9 M_\star w \over
    g_s^3 M ^{ 5/2 } L \rho ^{1/4} } $ & $ {612. \over g_s^5 M^4 L^2 }
  $ \\ \hline L= 0.5 & $ 0.171 \ \ \ \ \ \ 0.181 \ \ \ \ \ \ 0.182 $
  &28.78 & 0.954 & 8.741 & 2.052 \\ L= 1.5 & $0.106 \ \ \ \ \ \ 0.101
  \ \ \ \ \ \ 0.112 $ & 47.98 &0.873 & 3.029 & 0.619 \\ L= 3 & $0.043
  \ \ \ \ \ \ 0.053 \ \ \ \ \ \ 0.045 $ &62.31 & 3.044 &11.88 & 3.519
  \\ \hline
\end{tabular}
\end{table} 

\section{Fermionic   sector of  $D7$-branes} 
\label{sec3}

We examine in this section the fermionic modes on $D7$-branes wrapped
over the Kuperstein 4-cycle $\S _4$ of the Klebanov-Strassler
background.  The towers of 4-d spinor fields from the Kaluza-Klein
decomposition of the 10-d bispinor field $\T (X) $ on $ M_4 \times
\calc _6 $ are examined for the (linearized) wave equations derived
in~\cite{marche08}.  The Dirac operator in the simplified action in
Eq.~(\ref{eq.branesdp}) (dropping the contributions from classical
form felds except the $ F_5 $-form) is utilized in a truncated version
ignoring the twisting effects from the  normal  bundle and the extrinsic
curvature of $\S _4$~\cite{sorokin99,bandoskin06}.
Motivated by the similarity of the present model to that of
non-compact conic manifolds of $G_2$ holonomy~\cite{hartnoll02},
we also address the possible existence of normalizable massless fermion modes.

\subsection{Dimensional reduction of spinor fields}  
\label{sec3sub1}

The embedding of the brane world volume $ M_8 \subset M_{10} = M_4
\times \calc _6 = M_4 \times \S_4 \times N_2 $  involves the consecutive
splittings of dimensions $10\to 4+6$ and $6\to 4+2$ for the curved and flat
coordinates of the $M_{10}$ spacetime, \bea && X^M = (X^{\mu } , X^m )
= ( X^{\mu } , X^{r} , X ^{u} ),\ [ M=0, \cdots ,9 ,\
\mu = 0, \cdots , 3, \ m = 1 , \cdots , 6, \ r = 1 , \cdots , 4 , \ u
=1, 2 ] \cr && X^{\ua } = (X^{\underline \mu } , X^{\underline m } ) =
(X^{\underline \mu } , X^{a} , X ^{i } ) ,\ [\ua =0, \cdots ,9 ,\ a =
  1 , \cdots , 4 , \ i =1, 2 ]\label{eq.indBRA} \eea where the
coordinates $X^{r} ,\ X^{a} ,\ [r,\ a = 1,\cdots , 4] $ indices are
ordered as $(\rho , \hat h_3 , \hat h_1, \hat h_2 ) .$ Similar
splittings occur for the curved and flat (constant) Dirac gamma
matrices $ \G ^M$ and $ \G ^A$ acting in the $32 $ dimensional spinor
of the tangent spacetime group $SO(9,1)$. The Dirac matrices
representations adapted to the brane consist of direct products of
Dirac matrices acting on the $4 $ and $ 8 $ spinors of the $ SO(1,3)
\times SO(6) $ group and on the $ 4,\ 4,\ 2 $ spinors of the $ SO(1,3)
\times SO(4) \times SO(2)$ group,
\bea && \G^\mu = \g ^\mu \otimes 1_4 \otimes 1_2,\ \G^a = \g
_{(4)}\otimes \tilde \g ^a \otimes 1_2,\ \G^r = \g _{(4)}\otimes
\tilde \g _{(4)} \otimes \hat \g ^r, \ [\G _{(4)} = i \G ^{\underline
    {0123}} =\g _{(4)} \otimes 1_{8} ,\cr && \G _{(6)} = -i \G
  ^{\underline {456789}} =1 _{(4)} \otimes \tilde \g _{(6)} ,\ \G _{\S
    _4} = \G ^{\underline {4567}} = 1 _{(4)} \otimes \tilde \g _{(4)}
  , \ \G _{N_2} = -i \G^{\underline {89}} = 1 _{(4)} \otimes 1 _{(4)}
  \otimes \tau _3 ,\ \G _{(8)} = - i \G ^{01\cdots 67} ]
.  \label{eq.DCmatrx} \eea The 10-d chirality matrix $\G _{(10)}
,\ [\G _{(10)} ^2 =1] $ factors into the chirality matrices of the
submanifolds, $ M_4 ,\ \calc _6 $ and $\S _4 , \ N_2 $,
$\G _{(10)} = \G _{(4)} \G _{(6)} = \G _{(4)} \G _{\S _4} \G _{N_2}
=\g _{(4)} \otimes \tilde \g _{(6)} = \g _{(4)} \otimes \tilde \g
_{(4)} \otimes \hat \g _{(2)} $.  We also specialize below to the
matrix representations, \bea && \g ^\mu = \pmatrix{0 & -\bar \s ^\mu
  \cr \s ^\mu & 0 } = \pmatrix{ 0 & [-1, \s _{xyz} ] \cr [1, \s _{xyz}
  ] & 0 },\ \tilde \g ^a = \pmatrix{ 0 & [-i, \s _{zxy} ] \cr [i, \s
    _{zxy} ] & 0 } ,\ \hat \g ^r = (\tau _1 ,\tau _2 ) ,\cr && [\mu
  =0,\cdots , 3,\ a=1,\cdots , 4,\ B = \G^2 \G ^4 \G ^7 \G ^9 =
  \pmatrix{ 0 & -\s _2 \cr \s _2 & 0 }\otimes \pmatrix{-i\s _2 &0 \cr
    0 & -i\s _2} \otimes \pmatrix{0 & -i \cr i & 0 } , \cr && \g
  _{(4)} = diag(1 _2, -1 _2 ),\ \tilde \g _{(6)} = diag(1 _8, -1
  _8),\ \tilde \g _{(4)} = diag(1 _2, -1 _2 ),\ \hat \g _{(2)} =
  diag(1, -1 ) ] .  \label{eq.DCmatry} \eea Note that our choice of
the $SO(6)$ gamma matrices representations differs
from~\cite{marche08}
but coincides with~\cite{conlon08}.  The fermionic brane action of
quadratic order in the bi-spinor fields, $\T = (\t _1 \ \t _2 )^T $,
involves the extended Dirac operator in $ M_4,\ \S_4$ including the
classical 5-form~\cite{marche08}, \bea && \g ^{\a \b } ( \G _\a D_\b +
{i\over 8} \G _\a \Fslash _5 \G _\b \otimes \s _2 ) = \G ^\mu \cald
_\mu + \G ^a \cald _a ,\ [ \cald _\mu = \nabla _\mu -{1\over 4} \G
  _\mu \dslash _{\S _4} (\ln h (\tau ) ) P_+^{O3} ,\cr && \cald _a =
  \nabla _a + {1\over 8} \dh _a (\ln h (\tau ) ) - {1\over 4} \dslash
  _{\S _4} (\ln h (\tau ) ) \G _a P_+^{O3} , \ P_\pm ^{O3} = {1 \over
    2} (1 \pm \G _{(6)} \otimes \s _2 ) ] . \eea

The quantities adapted to the unwarped metric (independent of the warp
profile $ h(\tau )$) are introduced via the replacements \bea &&
\Dslash _{M_4} = \G ^\mu \cald _\mu \to h^{1/4} (\tau ) \Dslash _{M_4}
,\ \Dslash _{\calc _6 } = \G ^m \cald _m \to h^{-1/4} (\tau ) \Dslash
_{\calc _6 }, \eea where the tilde symbols were suppressed, for
simplicity.  The projections on $\S _4$ of the covariant derivatives
of $\calc _6 $  comprise both the Lorentz spin connection $\o_a $ and
the gauge connection $ A_a $, with $ \nabla _{\calc _6 } = \G^a \nabla
_a ,\ [\nabla _a = \dh _a + {1\over 4} \o _a ^{\hat a\hat b} \G _{\hat
    a\hat b} + A_a ] $   but  ignore the topological twisting terms from
the R-symmetry group $SO(2)$~\cite{bandoskin06} to be discussed at the
end of this subsection.  Note that the pull-back transformations of
bulk fields do not affect scalar quantities such as $\dslash \phi (X)
,\ \dslash h(\tau ),\ \Fslash $ and that the cross terms $(d\rho + i \hat
h_3 ) d \bar \chi + H.\ c.  $ in the conifold metric $ g^{a b} \G _a
D_b \to g ^{\rho \chi } \G _\rho \dh _\chi $ drop out in the kinetic
action.  The fermionic $D7$-brane \bea && \d S_F ^{(2)} (D7) = i \tau
_7 \int d ^7 \xi \sqrt {-\g } e^{\phi } \bar \T P_- ^{D7} [h^{1/4}
  \Dslash _{M_4} + h^{-1/4} \Dslash _{\calc _6 } + \dslash _{\S _4}
  (\ln h (\tau ) ) ( {1\over 8} - {1\over 2} P_+^{O3} ) ] \T , \cr &&
[P_- ^{D7} = (1- \G _{(8)} \otimes \s _2 )/2 ,\ \G _{(8)} = -\G _{(4)}
  \G _{\S_4 }] \label{eq.actionD8} \eea includes the projection operator $ P_- ^{D7} $  ensuring $\kappa $-symmetry.  The Kaluza-Klein
ansatz for the component Dirac spinors involves direct products of the
4-d and 6-d Weyl spinors, $\t _{i,4}^{(m)} (x) $ and $\t ^{(m)}_{i, 6
} (y) $ of same chiralities $\pm 1 $, \bea && \t _i (X) = \sum _m [\t
  _{i,4}^{(m)} (x) \otimes \t ^{(m)}_{i, 6 } (y) + (B_4 \t
  ^{(m)}_{i,4} )^\star \otimes (B_6\t ^{(m)}_{i,6})^\star ]
, \label{eq.KKsumf} \eea
where the linear combinations ensure that the 10-d Majorana-Weyl
spinors $\t _{i} (X) $  satisfy the reality condition which was
expressed in terms of the factorized form of the charge conjugation
matrix, $ B= B_4 \otimes B_6 ,\ [B_4 \g ^\mu B_4^\star = \g ^{\mu
    \star } ,\ B_6 \tilde \g ^m B_6^\star = - \tilde \g ^{m \star }]$
in Eq.~(\ref{eq.DCmatry}) for the gamma matrices.  The summation over
the  modes   index label $ m$   involves   products of (same chirality) 4-d and
6-d Dirac spinors  to  which   are  added  the  products  of  opposite
chirality spinors.   Consider specifically the basis  $
\l _{1,2,3,4} = [(+++), (+--) , (--+), (-+-) ] $  of  the
$+$ chirality spinors in $ M_{10}:\ \t
_i \in 32 ,\ [\G _{10} \t _i = \t _i ]$    formed    from
direct products of   the $\pm $-chiralities spinors
$ 4, \ 4, \ 2 $ of  $ M_4 \times \S _4 \times N_2 $. 
The projection on Majorana-Weyl spinors
selects the  basis of spinors, $\l_1 +\l _4 = \l_1 + (B \l_1 )^\star $
and $ \l_2 +\l _3 = \l_2 + (B \l_2 )^\star $, of chiralities $ +1 $
and $-1 $ in $\S _4$, but mixed chiralities in $M_4,\ N_2 $, as
seen on their  representations in terms of the  $2, 2, 1 $   Weyl spinors
$\xi ,\ \eta ,\ \zeta $ of $ M_4, \ \S _4 ,\ N_2 $,
\bea && \l _1 + \l _4 = {\xi \choose 0 }
 {\eta \choose 0 } {\zeta \choose 0 } +
{0  \choose -\s _2 \xi ^\star } {-i\s _2 \eta ^\star \choose 0} {0
\choose -i \zeta ^\star } ,\cr && \l _2 + \l _3 = {\xi \choose 0 }
{0 \choose \eta } {0 \choose \zeta } + {0 \choose -\s _2 \xi ^\star
} {0 \choose -i\s _2 \eta ^\star } {i \zeta ^\star \choose 0} .\eea
Substitution of Eq.~(\ref{eq.KKsumf}) into Eq.~(\ref{eq.actionD8})
yields sums over pairs of square and cross matrix elements, $
\overline{(\t + (B\t )^\star )} \Dslash (\t + (B\t )^\star ) = ( S_1 +
S_2 )+ ( C_1 + C_2 ) $, which by virtue of the relations $
\overline{(B\t )^\star } \Dslash (B\t )^\star = - \eta \bar \t \Dslash
\t ,\ \bar \t \Dslash (B\t )^\star = - (\t ^T C \Dslash \t )^\star
,\ (B\t )^\star \Dslash \bar \t = \t ^T C \Dslash \t ,\ [B^T = C \G^0
  ,\ B^\star = \G _0 C] $ and the identities $\G _0 \Dslash \G _0 =
\Dslash ^\dagger,\ \G_a ^\dagger = \G _0 \G _a \G _0 $ satisfy $ S_{1}
+ S_2 = 2 S_{1} $ and $C_1 + C_2 = C_1 - C_1 ^\star = 0$.


The calculations are facilitated by selecting a suitable gauge for the
local fermionic $\kappa $-symmetry.  For the simple choice, $ \T \equiv
(\t _1 , \t _2) = (\t ,0)$, the wave equations for modes of mass $\o
_m $, subject to the mass shell conditions, $ \G _{(4)} \Dslash _{M_4
} \T _4 =-\o _m (B_4\T _4 (x ) )^\star $, are derived from the
stationarity property of the action in the form \bea && \G _{(4)}
[\Dslash _{\calc _6 } - {1\over 8} \dslash _{\S _4} ( \ln h ) (1 + 2
  \G _{\S _4} )] \t ^{(m)} _{6} (y) = h ^{1/2} (y) \o _m (B_6 \t
^{(m)} _{6} (y) )^\star , \eea where the $\G _{(4)} $ factor ensures
that the external and internal space Dirac operators commute.
The 6-d massless ($\o _m =0 $)  zero modes of
chirality $\S _4 = \pm 1$ have the explicit dependence on the warp profile
$h^{3/8} (\tau ) $ and $h^{-1/8} (\tau )$,
as verified from the wave equatioons \bea && [\Dslash
  _{\calc _6 } - {1 \over 8} {3\choose -1 } \tilde \dslash _{\S _4} (
  \ln h ) ] \t ^{(0)} _{6,\pm } (y) =0 \ \Longrightarrow \ \Dslash
_{\calc _6 } \psi _\pm =0 , \ [\t _{6, +} ^{(0)} (y)= h^{3/8} \psi _+
  (y) ,\ \t _{6, -} ^{(0)} (y) =h^{-1/8} \psi _- (y) ] .\eea Note that
if the classical $F_5$-term were ignored, all  modes would have the
same warp profile factor, $ \t _{6, \pm } = h^{1/8} \psi _\pm $.  The
kinetic energy action for zero modes becomes \bea && \d S _F ^{(2)}
(D7)= i \tau _7 \int _{M_4}d^4 x \sqrt {-\tilde g_4} e ^{\phi } \sum
_{r} \int _{\S _4} V_0 (\rho ) [(\bar \t ^{(r)} _{4,+} \tilde \Dslash
  _4 \t ^{(r)}_{4,+}) h (\tau ) (\psi ^\dagger _+ \psi _+ (\rho ) )+
  (\bar \t _{4 , -} ^{(r)} \tilde \Dslash _4 \t _{4 , -} ^{(r)}) (
  \psi ^\dagger _-\psi _- (\rho ) )]. \label{eq.red1} \eea


The correspondence between fermionic and bosonic superpartner modes is
more transparent in the alternative $\kappa$-symmetry gauge fixing
condition~\cite{marche08,bandoskin06}, $P_- ^{Dp} \T =0
\ \Longrightarrow \ \T = ({\t , \ - i \G _ {(8)} \t })^T $.  Using the
relations $P_\pm ^{O3} \T _{6, \pm } = \T _{6, \pm } ,\ P_\mp ^{O3} \T
_{6, \pm } =0 $ in the extended Dirac equation $ [\Dslash _{M_4} +
\Dslash _{\calc _6} + \dslash _{\S_4} (\ln h ) (1/8 - P_+^{O3} /2 ) ] \T
=0 ,$ yields the wave equations, \bea && -h^{1/2}
(\tau ) \o _m(B_6\T _6 ^{(m)} )^\star + \G _{(4)} ( \Dslash_{\calc _6
} -{1\over 8} \dslash _{\S _4 } (\ln h ) (1 +2 \G _{\S _4 } \otimes \s
_2 ) ) \T _6 ^{(m)} =0 ,\ [\T ^{(m)}_{6} = (\t _{1, 6} ^{(m)} \ \t
  _{2, 6} ^{(m)}) ] .\label{eq.fer0} \eea The zero modes of $\S
_4$-chirality $\pm 1$ have the same warp profile dependence, $\T
^0 _{6, \pm } = h^{[3/8, -1/8]} \Psi _\pm $, and the
correspondence to the bosonic superpartner modes of same warp profile
dependence associates $ \Psi _+ $ to gaugino and modulino and $ \Psi
_- $ to wilsonini.  Recall that for compact Calabi-Yau
manifolds~\cite{marche08}, the ($ SO(2)_R$ singlet) gaugino wave
function are built from the covariantly constant spinor $\eta ^{CY}$.
If $\S _4$ is a complex manifold of K\"ahler metric, $ \g _{A\bar B} d
W^A d W^{\bar B} $ in a suitable basis of holomorphic coordinates $
W^A$, one can then build  the modulini and wilsonini wave functions
by acting with linear combinations of Dirac
matrices~\cite{gsw}, $ \eta _m = m_{AB} \G^{W^A W^B} \eta ^{CY} ,
\ \eta _W = W_A \G ^{W^A } \eta ^{CY} $. The   independent
harmonic 2- and 1-forms $m _{AB} ,\ W_A $ on $\S_4 $ (of Betti numbers
$b_2 (\S _4) = h^{2,0} $ and $b_1 (\S _4) = 2 h^{1,0}$)  also  serve to
build  up the  wave functions of bosonic superpartner modes,
$\chi \sim X^{8,9} $ and $ A _{\hat
  a }\sim X^a $.  The wave functions for gaugino and modulini $\t _{6,
  +}\sim (\l _{2} ,\ \l _{3} ) $ and wilsonini $ \t _{6, -}\sim (\l
_{1} ,\ \l _{4}) $ are of form~\cite{marche08}, \bea && \t _{6, +} = {
  h^{3/8}\over \sqrt 2 } \bigg [ [\pmatrix {i \eta ^{CY} \cr \eta
      ^{CY} } + \cdots ],\ [ \pmatrix {i \eta _m \cr \eta _m } +
    \cdots ] \bigg ] ,\ \ \t _{6, -}= {h^{-1/8} \over \sqrt 2 }
[\pmatrix { -i \eta _W \cr \eta _W } + \cdots ] , \eea where the
central dots refer to the terms projecting on Majorana spinors.  
While the above  properties   need not  apply to non-compact manifolds,
normalizable massless fermions  might  still  exist   in the conifold
case.   This is found to be the case
for Killing spinors in supergravity backgrounds~\cite{arean06} and
also for the massless axino~\cite{argurio06} superpartner of the axion
mode~\cite{GHK04} induced through the spontaneous breaking of the 
baryon charge symmetry $ U(1)_b $  in Klebanov-Strassler background.
For completeness, we also  recall that fermionic zero modes
arise typically in 4-d gauge
theories with instantons~\cite{acharya06} or magnetic
backgrounds~\cite{marche08,conlon08,abe15} and in theories with extra
dimensions with $D$-branes intersecting at points
$D$-branes~\cite{marche10}, or instanton (Euclidean) branes wrapped on
divisors of the compactification
manifold~\cite{witten.inst96,kalloshpolo05,bergshoeff05}.  In theories
with anomalous global symmetries the numbers of massless fermions are
determined via the Atiyah-Singer index or the holomorphic (arithmetic)
index,  $ \chi (D _{6, 4})= \sum _{p} (-1)^p h ^{(0,p)} $.

Before analyzing  further  the   issue    of fermionic zero modes, we review
briefly the  geometrical approach~\cite{bandoskin06}  to
embed  superbranes  in curved manifolds. (Generalized
versions exist for superspacetimes and superworld
volumes~\cite{sorokin99}).  One assumes   here a linear relationship between
the bulk spacetime vielbeins $e ^{\ua } = e ^{\ua }_M d X^M$ and the
tangential and normal   vielbeins of the brane world volume
$\e ^{a}= \e ^{a} _m d \xi ^m ,\ \e ^{i} = \e ^{i} _m d \xi ^m $,
involving the local Lorentz transformations $ u ^{\ua } _{\ub } (\xi )
\in SO(1,9)$, \bea && \e ^a = u ^a _{\ub } e ^{\ub } ,\ \e ^i = u ^i
_{\ub } e ^{\ub } , \ [u _{\ua } ^{\ub } = (u _{\ua } ^{b }, u _{\ua }
  ^{j} ) = (u ^{\ub } _{a}, u ^{\ub } _{i} ) =\pmatrix{u _{a } ^{b } &
    u _{a } ^{j} \cr u _{i} ^{b } & u _{i} ^{j} } ,\ \e ^a \cdot \e ^i
  =0 ] \label{eq.embLH} \eea which align $(p+1)$ of the bulk tangent
space basis onto the world volume tangent space basis.  Similar
relations hold between the bulk gamma matrices and the
gamma  matrices, $\G^a = \G ^{\ua } u _{\ua } ^a ,\ \G^i = \G ^{\ua } u _{\ua
} ^i$  along tangential and normal directions.
Imposing the embedding equations $\e^i \equiv u ^i _{\ub } e
^{\ub } =0$, which imply that the $(9-p)$ normal vielbeins are
annihilated by the Lorentz transformation aligning the $ \e ^a $ along
the world volume, makes $u$ become auxiliary (non-dynamical fields).
The 10-d metric and the world volume embedding $\dh _m X^M $ determine
the tangential sub-block $ u ^{\ua } _b = e ^{\ua } _M \dh _m X^M \e
_b ^m $ (reducing to  $ e ^{\ua } _i \dh _m X^i \e _b ^m $ in the static gauge)
while the orthogonality conditions $u _{\ua } ^i u _{\ub , i} = \eta
_{\ua \ub} - u _{\ua } ^a u _{\ub } ^{b} \eta _{a b }$ determine the normal 
sub-blocks.  With these definitions and a suitable choice for the
gamma matrices, one can express after some work the covariant
derivatives $D_m = \dh _m + \O _m $ involving the bulk spin connection
$\O _m $ in terms of tangential and normal induced spin connections $
\o _m ^{ab} ,\ A _m ^{ij} $ for the reduced spinors of $ SO(1,p) $ and
$SO(9-p)$, \bea && D_ x \equiv (e ^{\uc } _m \dh _{\uc } - {1\over 4}
\O ^{\ua \ub } _m \G _{\ua \ub } ) = \cald _m - {1\over 4} u ^{\uc \ua
} \dh _m u _{\uc} ^{\ub } \G _{\ua \ub } - {1\over 2} K_m ^{b i} \G
_{b i } ,\ [\cald _m = e ^{\uc } _m \dh _{\uc } - {1\over 4} \O ^{a b
  } _m \G _{ab} - {1\over 4} A ^{ij} _m \G _{ij } ,\cr && \o ^{bc} = u
  _{\ub }^b \O ^{\ub \uc } u _{\uc }^c - u ^{\ua b} d u _{\ua } ^c,
  \ A ^{ij} = u _{\ub }^i \O ^{\ub \uc } u _{\uc }^j -u^{\ua i}d u
  _{\ua }^j,\ K ^{i} _b = - D (e ^{\ua } _b ) u_{\ua } ^{i} = - (d(e
  ^{\ua } _b ) + u_{b} ^{\ub } \O _{\ub } ^{\ua } ) u_{\ua } ^{i}
] \label{eq.covD1} \eea where $ K ^i = d \xi ^b K^i _b$ denotes the
world volume extrinsic curvature (second fundamental form).

\subsection{Fermionic zero  modes}  
\label{sec3sub2}

We discuss in this subsection the existence of  massless
fermions on $D7$-branes in the Klebanov-Strassler background, within  the
simplified description using  the Dirac operator restriction on $\S _4$,
ignoring  twisting terms, \bea && \tilde \Dslash \vert _{\S _4} = \G ^m (\dh _m
+ {1\over 4} \o ^{ \hat a \hat b} _m \G _{ \hat a \hat b} ) ,\ [\G _m
  = e _m ^{\hat a } \G _{\hat a },\ \G ^m = e ^m _{\hat a } \G ^{\hat
    a } ,\ \G _{\hat a } = \d _{\hat a \hat b}\G ^{\hat b } ,\ \{ \G
  _{\hat a } ,\G _{\hat b} \} = 2\d _{\hat a \hat b} ] \eea where the
indices $ m$ and $\hat a $ refer to curved and flat directions and the
spin connection is evaluated from Eq.(\ref{eq.spinc0}) in the basis of
vielbeins defined in Eq.~(\ref{eq.spincp0}).  Thanks to the $ SO(3)$
isometry of Kuperstein 4-cycle, the general formulas for the Dirac
operator in~Eqs.(\ref{eq.spinc1}) and~(\ref{eq.spinc2}) of
Appendix~\ref{appD7subFe1} simplify to \bea && \tilde \Dslash \vert
_{\S _4} = {1\over \eta ^{\hat \rho } } \G ^{\hat \rho } (\dh _\rho +
\ud \sum _{\hat k } \dh _\rho (\ln \eta ^{\hat k} ) + X(\rho ) \G _{\S
  _4 } + \eta ^{\hat \rho } \G _{\hat \rho } \G ^{\hat k } e ^m_{\hat
  k } \dh _{m } ) = \eta _{\hat \rho }\G ^{\hat \rho } ( \dh _\rho +
\ud \dh _\rho \ln D + X(\rho ) \G _{\S _4 } + \eta ^{\hat \rho } \eta
_{\hat k } \G ^{\hat \rho \hat k } \nabla _{\hat h _k} ) ,\cr && [ X (
  \rho ) = { (\eta ^{\hat \rho })^2 \over 8 D } ({N_1 \over \eta
    ^{\hat 1} } + {N_2 \over \eta ^{\hat 2} } + {N_3 \over \eta ^{\hat
      3} } ) , \ D = \eta ^{\hat 1} \eta ^{\hat 2} \eta ^{\hat 3} ,
  \ N_1 = - \eta ^{\hat 1} + \eta ^{\hat 2} + \eta ^{\hat 3} ,\ N_2 =
  \eta ^{\hat 1} - \eta ^{\hat 2} + \eta ^{\hat 3} ,\ N_3 = \eta
  ^{\hat 1} + \eta
  ^{\hat 2} - \eta ^{\hat 3} ] .  \label{eq.spincp4} \eea
The  successive    contributions 
refer to  the radial derivative  term,  two potential terms (proportional to
 the unit matrix and chirality matrix $ \G _{\S _4 } = \G ^{\hat \rho
   \hat 1 \hat 2 \hat 3} $) and  the angular term $\Dslash (S^3 ) $.
 The latter term drops out for singlet
 modes, satisfying the conditions  $ e _{\hat k } ^m\dh _{\a ^m} \psi
 (\rho , \a ) =0$, to which we restrict hereafter.  In the gamma
 matrices representation of $\S _4$,
 \bea && \G ^{\hat a } = \g _{(4)} \times
 \tilde \g ^{\hat a} ,\ \tilde \g _{(4)} = \tilde \g ^{ \hat \rho \hat
   1 \hat 2 \hat 3 } = \pmatrix{1& 0 \cr 0 & -1} ,\ [\hat a = (\hat
   \rho , \hat k ) ,\ \tilde \g ^{\hat \rho } =\pmatrix{0 & -i \cr i &
     0} , \ \tilde \g ^{\hat 1, \hat 2, \hat 3 } = \pmatrix{0& \s _{z,
       x, y} \cr \s _{z, x, y} & 0 } ] \label{eq.spinc3}\eea
quoted from  Eq.~(\ref{eq.DCmatrx}),   the
 chirality matrix diagonal $ \G _{\S _4 } = diag (1 _2,- 1_2 ) $ and
 the 4-spinors split up into a decoupled pair of 2-spinors of $\pm 1 $
 chirality, $\psi = (\psi _+,\ \psi _-) $.  The solutions $ \psi _\pm
 (\rho ) $ of the decoupled radial wave equations, \bea && (\dh _\rho
 + \ud \sum _{\hat k } \dh _\rho (\ln \eta ^{\hat k} ) \pm X(\rho ))
 \psi _\pm (\rho ) =0 \ \Longrightarrow \ \psi _\pm (\rho ) = C _\pm
 \psi _0 (\rho ) e ^ {\mp \int ^{\rho } _{\rho _C} d \rho ' X ( \rho
   ')} \xi _\pm , \ [\psi _0 (\rho ) = {1 \over \prod _{\hat k } (\xi
     ^{\hat k})^{1/2} } ] \eea
 are expressed in terms of the 2-d
 constant spinors $\xi _\pm $ are 2-d constant spinors and the
 constant factors $C _\pm $, along with the boundary values $ \rho _C$
 in the  definite integrals, are set through the wave functions
 normalizations. For modes of canonical kinetic actions in
 Eq.~(\ref{eq.red1}), the normalization  conditions,   $ 1= C ^{2}_+
 \int d\rho \sqrt { -\g ^{(0)} } h(\tau ) \vert \psi _+ \vert ^2,\ 1=
 C ^{2}_- \int d\rho \sqrt { -\g ^{(0)} } \vert \psi _- \vert ^2$,
 determine $ C$ as a function of $ \rho _C$.  The existence of
 normalizable zero modes is thus dictated by the behaviour of radial
 wave functions at small and large distances.

\begin{figure}[t]\begin{subfigure}{0.45\textwidth} 
\includegraphics[width=0.99\textwidth]{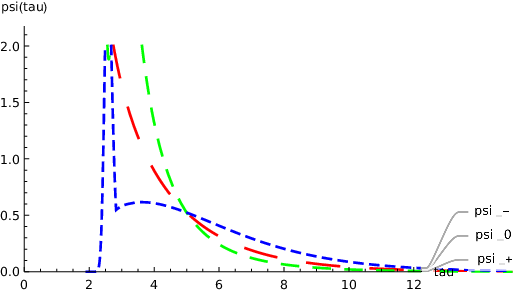}  
\end{subfigure} 
\begin{subfigure}{0.45\textwidth} 
\includegraphics[width=0.99\textwidth]{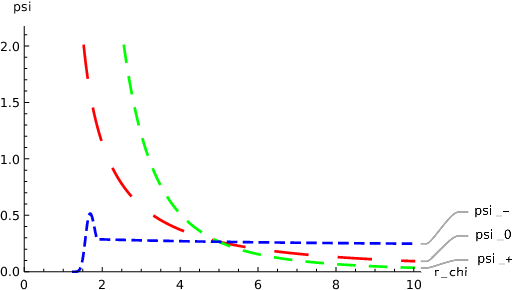} 
\end{subfigure}
\caption{\it \label{wzmD7}  Plots of the fermionic zero mode wave functions
  (in units $\e =1, \ g_s M  \a ' =1 $)
on Kuperstein 4-cycle    as a function of   the
deformed conifold radial variable $\tau $  for
$L= 1.5 ,\ [\tau  \geq \tau _{min} = 2.58] $ (left panel) and
the undeformed conifold radial variable $r _\chi $
for $\mu = 1.5 ,\ [r _\chi  \geq (r _\chi )_{min}=
  1.65 ] $ (right panel). 
The (red) curves in long dashes refer to the  prefactor $ \psi _0(\tau ) $ and
the (green and blue) curves in shorter dashes
refer to  the unnormalized radial wave functions, $\psi
_{\pm } (\rho ) = C_{\pm } \psi _0 e ^{\mp \int _{\rho _C} ^\rho d
  \rho X(\rho ) } $  of  chiralities $\pm 1 $.  The
values of the normalization constant $ C_\pm $  depend on   the
choice  of the arbitrary  lower  limit $ \rho _{B} $  affecting
 the  wave functions  and the   normalization integrals.} 
\end{figure}

We display in Fig.~\ref{wzmD7} the solutions for the zero modes radial
wave functions.  The zeroth-order factor $\psi _0 (\tau ) $ is regular at infinity but diverges at threshold.  Both solutions $\psi _\pm $
slope to zero at large $\tau $, with $\psi _+ $ converging much
more rapidly  than $\psi _-$.
Near the threshold  ($ \tau = \tau _{min} \simeq 2.58 $ for
$L=1.5$),   $\psi _- $ vanishes with a rapid slope and $\psi _+ $
diverges.  Since the normalization integrals diverge for both
$\psi _+ $ and $\psi _-$, whether the warp profile is taken into account or not,
it follows that    the existence  of zero modes  is ruled out.
Only if one imposed an ultraviolet radial cutoff $\tau _{uv}$ could  the
normalizable mode $\psi _- $ be present.


Similar conclusions are reached in the undeformed conifold case.  The
radial Dirac operator (for singlet modes) \bea && \Dslash
= \eta _{\hat \rho } \G^{\hat \rho } (\dh _\rho + \ud \dh _\rho (\ln D
) + X(\rho ) \G _{\S _4} ) ,\ [D ( \rho )= \eta ^{\hat 1} \eta ^{\hat
    2} \eta ^{\hat 3} = { \vert \mu \vert ^2 \over 16 \sqrt 3} ( (2
  r_\chi ^3 -1) ( r_\chi^{3} -2) ) ^{1/2} ,\cr && \cr && {X( \rho )
    \over d r_\chi / d \rho } = {(2 r_\chi^3 -1)^{1/2} \over \sqrt 6
    r_\chi ^{5/2}} +{1\over 12 r_\chi } [ 3 + { 2\sqrt 6 (r_\chi ^3
      +1) (2 r_\chi^3 -1)^{1/2} \over (r^3_\chi -2)^{3/2} }]
]  \label{eq.spinc4} \eea
admits   the zero modes solutions  \bea && \psi
_\pm (r_\chi ) = C_\pm \psi _0 (r_\chi ) e ^{ \mp \int ^{\rho } _{\rho _C} d \rho
  X( \rho )}  ,\ [\psi _0 (r_\chi ) = {4 \cdot 3 ^{1/4} \over \vert \mu \vert (
    (r_\chi ^3 -2 ) (2 r_\chi ^3 -1) )^{1/4} } ] \label{eqsinc8} \eea
where the indefinite integral in the exponential is expressed by the
analytic formula, \bea && \int ^{r_\chi } d \rho X(\rho ) = {1\over
  36} [-6 \sqrt 6 ({ 2 r_\chi^3 -1 \over r_\chi^3 -2} )^{1/2} -4 \sqrt
  6 ({2 r_\chi^3 -1 \over r_\chi^{3} })^{1/2} \cr && + 8 \sqrt 3 (
  Arcsinh ( \sqrt {2\over 3} (r_\chi^3 -2 )^{1/2} ) + Arcsinh
  ((2r_\chi^3 -1)^{1/2} ) ) +2 \sqrt 3 Arctanh(({r_\chi^3 -2 \over 2
    (2 r_\chi^3 -1) } )^{1/2}) + 9 \log r_\chi ] .\eea The series
expansions near threshold $z \equiv (r - 2^{1/3}) \to 0 $ and the
boundary $ r_\chi \to \infty $ of the various factors in the
integrands of the normalization integrals, $ N (\psi _\pm ) = \int
_{\rho _C} ^\rho d\rho \sqrt {-\g ^{(0)} } [\psi _+ ^2 h(r) ,\ \psi _-
  ^2] $ are displayed in the table below for the parameter value $ \mu
=1.5$. The values of the constant coefficients $ C_\mp $ are
correlated with the value chosen for the radial integral lower bound
$\rho _C$.


\vskip 0.3 cm
\begin{center} 
  \begin{tabular}{|c|c|c|c|c|c|c|} \hline \centering
& Limit $ {dr _\chi \over d \rho } $ & $ \psi _0 $ & $e ^{\int ^ {r }
      d \rho X(\rho ) } $ & $\psi _\mp $ & $ N(\psi ^2_-) $ & $ N(\psi
    ^2_+ h(r) ) $ \\ \hline $z \to 0 $&$ ({2 ^{1/6} / \sqrt 3 }) z
    ^{1/2} $&$ {1.8 z ^{-1/4}} $&$ e^ {\mp 0.32 z ^{-1/2} } $&$ 1.8
    z^{-1/4} e ^{\mp 0.32 z^{-1/2} }$ & & \\ \hline $r_\chi \to \infty
    $&$ r_\chi /3 $&$ 2.95 r _\chi ^{-3/2} $&$ r _\chi ^{(1/4 +
      2/\sqrt 3 )} $&$ \pmatrix{C_- r_\chi ^{-5/4 + 2/\sqrt 3 } \cr
      C_+ r _\chi ^{-7/4 - 2/\sqrt 3 } } $&$ r _\chi ^{3/2 + 4/\sqrt 3
    } $&$ r _\chi ^{-7/2 - 4/\sqrt 3 } $ \\ \hline
\end{tabular}
\end{center}
\vskip 0.3 cm

The normalization integral for $\psi _+ $ converges at large radius
but diverges near threshold, while that for $\psi _- $ diverges at
large radius but converges near threshold, as verified from the
limited expansions above.  The radial wave functions displayed in
Fig.\ref{wzmD7} are qualitatively similar to those in the deformed
conifold case.  The solution $\psi _- $ vanishes at threshold $
r^{min} _\chi \simeq (2\mu ^2 ) ^{1/3} \simeq 1.65 $, and jumps
rapidly to a plateau that slopes very slowly to zero.  This is
explained by  our  choice for  the arbitrary parameter $\rho _C$
which contributes a large constant  part to the  wave
function,  $\psi _ ( r_\chi )$ that is compensated by  the
normalization  factor $ C_- $.  We thus
conclude that  also  the undeformed conifold does not  support singlet zero modes
but could allow   one $\psi _- $ if an ultraviolet cutoff  were imposed. 

\section{Summary  and conclusions} 
\label{sect4}

We examined in this work the Kaluza-Klein theory for $D7$-branes
embedded in Kuperstein 4-cycle of Klebanov-Strassler background.  The
mass spectra and wave functions of ground state and radially excited
meson modes, singlet under the base manifold isometry group, were
evaluated using semi-classical tools.  The reduced masses $\hat \o _m
$ (in same units $ \e ^{2/3} (g_s M \a ' ) /\e ^{2/3} $ as for
glueball masses $\hat E _m $) exhibit a nearly uniform dependence on
the ratio $ L= \mu /\e \in R $, being nearly constant in $ L\in [0,1]
$ and growing monotonically with the power law $\hat \o _m \sim
L(^{2/3} $ at $L > 5$.  Similar results hold for $ L ^2 <0 $ with no
qualitative differences are expected for complex values of $L$.
The classical 2-form  $ B_2 ^{cl}$  gives  contributions
resulting  in lighter modes changing the power law growth
from $ \o _m ^2 \approx 0.5 \ L^{4/3} \ \to \o _m ^2 \approx (0.4 -
0.3) \ L^{4/3}.$ The meson modes are more strongly bound than
glueball modes with typical ratios $\hat \o ^2 _n / \hat E ^2_n \simeq
1/3 $ for $ L\in [0,1] $. Scalar modes are always lighter than vectors
modes.  The mass splittings between radial excitations depend weakly
on $ L$.  The mass splittings for radial excitations are independent
of $ L$.

The coupling constants of interactions between meson and glueball
modes were evaluated from the overlap integrals of meson and
glueball wave functions.  The gravitational couplings of meson modes
decrease rapidly beyond $ L> 3 $ while their self couplings increase
slowly with $L$.  The existence of normalizable fermionic massless
modes is ruled out unless one imposes an ultraviolet cutoff on radial
distances.

Our test of   duality identifying the holographic meson  modes to the
hadronic quarkonia of light $u,\ d ,\ s $ and heavy $ c, \ b$ flavour
was  inconclusive partly due to the uncertain gauge-gravity duality
correspondence relations between parameters.  Improvements are
expected if supersymmetry breaking effects are taken into account.

\begin{appendix} 
\label{appendices}  
\section{Embedding  of $D7$-branes  in  deformed conifold} 
\label{appwdcsubD7}

We review in this appendix useful tools for computing the reduced
$D7$-brane action on Kuperstein submanifold~\cite{kuperstein04} $\S
_4:\ w _4 \equiv {1\over 2i} Tr (W) = \mu $ of the deformed conifold
in Eq.~(\ref{eq.conufol}), where the complex parameter $\mu \in C ,\ [
  [\mu ] = L^{2/3}]$ measures the minimal distance from its apex to
the conifold apex.  Helpful details complementing the discussion in
the text are provided in the  following four subsections.  We start
with geometrical properties of the 4-cycle embedding, discuss in turn
the reduced bosonic  and  fermionic actions of $D7$-branes
and finally  conclude with intermediate results in the undeformed
conifold limit.

\subsection{Geometrical    properties of  4-cycle embedding} 
\label{appD7sub0}

The metric in the spacetime $M_4 \times \calc _6$ is expressed in a
convenient form by introducing the basis of cotangent space frame
vectors, $ e ^{\hat \mu } ,\ ( e ^{\hat \rho } , \ e ^{\hat k } ,\ e
^{\hat \chi } ),\ [\mu =0, 1,2,3,\ k=1,2,3]$ and the dual orthogonal
basis of tangent space frame vectors $e _{\hat\mu } ,\ ( e _{\hat \rho
} , \ e _{\hat k } ,\ e _{\hat \chi } ) $, such that the sets of
components of these two bases, $ ( e^{\hat \mu }_\nu ,\ e ^{\hat \rho
} _{\rho } ,\ e ^{\hat k} _m ) $ and $ (e_{\hat \mu }^\nu ,\ e _{\hat
  \rho } ^{\rho } ,\ e _{\hat k} ^m ) $ describe mutually inverse
matrices.  (The present notation for covariant and contravariant
vectors is also used for differential forms and derivatives.)  The
warped metric consists of 4-d and 6-d parts, $ds ^2 _{10} = h^{-1/2}
(\tau ) d \tilde s_4^2 + h^{1/2} (\tau ) d \tilde s_6^2 ,$ defined as
\bea && d \tilde s ^2_4 = \tilde g _{\mu \nu} d x^\mu d x^\nu = \eta
_{\hat \mu \hat \nu} e^{\hat \mu } e^{\hat \nu } = \eta _{\hat \mu
  \hat \nu} e^{\hat \mu } _\mu e^{\hat \nu } _\nu d x^\mu d x^\nu ,
\cr && d \tilde s_6^2 = \tilde g_{\rho \rho } d \rho ^2 + \tilde g
_{\hat h_i \hat h_j }(\rho ) \hat h_i \hat h_j + (\tilde g_{\chi \bar
  \chi } d \chi d \bar \chi + \tilde g_{\rho \bar \chi } d \rho d \bar
\chi + \tilde g_{\hat h_i \bar \chi } \hat h_i d \bar \chi + H.\ c. )
= (e^{\hat \rho } )^2 + (e^{\hat h_i } )^2 + 2 e ^{\hat \chi } (e ^{
  {\hat\chi } } ) ^\star .  \label{eq.metrans} \eea
It  proves convenient to separate out the radial variable dependence of the frame
vectors along the angular directions of the 4-cycle manifold $\S _4$,
$ e ^{\hat k} = \eta ^{\hat k} (\rho ) h ^{\hat k} , \ e _{\hat k} =
\eta _{\hat k}(\rho ) h _{\hat k} $, where $ h ^{\hat k} _m ,\ h_{\hat
  k} ^m $ denote the components of base manifold $ S^3/Z_2$ vielbeins,
depending only on the angle variables $ \a ^m $.  The non-vanishing
components of the $\calc _6 $ metric
in Eq.~(\ref{eqINDm}) take the equivalent forms
\bea && \tilde g _{\mu \nu } =\eta _{\mu \nu } ,\ \tilde g _{\rho \rho
} \equiv (\eta ^{\hat \rho } )^2 =\tilde g _{\hat h_3 \hat h_3 }
\equiv (\eta ^{\hat 3 })^2 = {F_1\over 2} \vert \eta _\chi \vert ^2
A_1,\ \tilde g _{\hat h _1 \hat h_1} \equiv (\eta ^{\hat 1 })^2=
{F_1\over 4} \vert \eta _\chi \vert ^2 (1 + C _\rho ) , \cr && \tilde
g _{\hat h _2 \hat h_2} \equiv (\eta ^{\hat 2})^2 = {F_1\over 4} \vert
\eta _\chi \vert ^2 (-1 + C _\rho ), \ \tilde g _{\rho \chi } =
(\tilde g _{\rho \bar \chi }) ^\star = i \tilde g _{\hat h_3 \chi } =
-i (\tilde g _{\hat h_3 \bar \chi })^\star = \ud F_1 \bar \eta _\chi
S_\rho A_2 ,\ \tilde g _{\chi \bar \chi } = \tilde g _{\bar \chi \chi
} = F_1 A_3 , \cr &&
\cr && [A_1 =(C _\rho + 2 \vert \eta _\chi \vert ^2 R S ^2 _\rho)
  ,\ A_2 = X (1 + 2 \vert \eta _\chi \vert ^2 R ({\bar X \over X} + C
  _\rho ) ) ,\ A_3= 1 + X \bar X ( C _\rho + 2 \vert \eta _\chi \vert
  ^2 R \vert 1 + C _\rho {\bar X \over X} \vert ^2 )
] \label{eqmetS4}\eea where $C _\rho = \cosh \rho ,\ S _\rho = \sinh
\rho $.  The cross terms $ \bar \eta _\chi d \bar \chi (d \rho - i
\hat h_3) + H.\ c. $ signal that $\calc _6$ is not a direct product
manifold of $\S _4 $ times its complementary sub-manifold $ N_2 $ in
$\calc _6$.  It is  still possible to express the metric as a diagonal
quadratic form in the basis of vielbein vectors, \bea && e^{\hat \rho
} = (\tilde g _{\rho \rho } - 2 {\vert \tilde g _{\rho \chi } \vert ^2
  \over \tilde g _{\chi \bar \chi } } ) ^{1/2} d \rho ,\ e^{\hat h_3 }
= (\tilde g _{\hat h_3 \hat h_3 } - 2 {\vert \tilde g _{h_3 \chi }
  \vert ^2 \over \tilde g _{\chi \bar \chi } } ) ^{1/2} \hat h_3 ,\cr
&& e ^{\hat \chi } = \tilde g _{\chi \bar \chi }^{1/2} ( d \chi +
{\tilde g _{\rho \bar \chi } \over \tilde g _{\chi \bar \chi } } (d
\rho - i \hat h_3) ) , \ e^{\hat h_1 } = \tilde g _{h_1 h_1 }
^{1/2}\hat h_1 ,\ e^{\hat h_2} = \tilde g _{h_2 h_2 } ^{1/2} \hat h_2
.\eea in same notations as Eq.~(\ref{eq.metrans}).  The complex
vielbein $ e^\chi $ splits up into two real vielbeins, $ e ^{\hat \phi
  _1} =\tilde g _{\chi \bar \chi }^{1/2} ( d \phi _1+ ({\tilde g
  _{\rho \bar \chi } / \tilde g _{\chi \bar \chi } }) d \rho ) ,\ e
^{\hat \phi _2} =\tilde g _{\chi \bar \chi }^{1/2} ( d \phi _2-
{\tilde g _{\rho \bar \chi } / \tilde g _{\chi \bar \chi } } \hat h_3
) , \ [e ^{\hat \chi } e ^{\bar {\hat\chi } } = ( e ^{\hat \phi _1}
  )^2 + ( e ^{\hat \phi _2} )^2 ] ,$ corresponding to the pair of real
moduli fields in $ \chi = \phi _1 + i\phi _2 ,\ \bar \chi = \phi _1 -
i\phi _2 $.  The restriction $ \calc _6 \to \S _4 $ leads to a metric
and scalar Laplacian for $\S _4 $ of diagonal form in the coordinates
$ ( \rho ,\ \hat h_i )$ consistently with the underlying $SO(3)$
isometry group, \bea && d \tilde s^2(\calc _6) \vert _{\S _4} =
(e^{\hat \rho } _\rho ) ^2 d \rho ^2 + \d _{ij} e^{\hat i } _m e^{\hat
  j} _n d \a ^m d \a ^n = (e^{\hat \rho } _\rho ) ^2 d \rho ^2 +\eta
_{\hat k} ^2 (\rho ) \d _{mn} h _{\hat k} ^m k_{\hat k} ^n ,
\ \ \nabla ^2 \vert _{\S _4} = e ^\rho _{\hat \rho } \dh _{\rho } e
^\rho _{\hat \rho } \dh _{\rho } + e_{\hat i } ^m {\dh \over \dh {\a
    ^m } } e_{\hat i} ^n \dh _{ \a ^n} , \cr && [e ^{\hat \rho } = e
  ^{\hat \rho } _\rho d \rho ,\ e^{\hat k} = e^{\hat k}_l \hat h_l =
  e^{\hat k}_m d \a ^m = \eta ^{\hat k} (\rho ) h^{\hat k} _m( \a ) d
  \a ^m ,\ e _{\hat \rho } = e _{\hat \rho } ^\rho \dh _ \rho ,\ e
  _{\hat k} = e _{\hat k} ^m \nabla _{m} = e _{\hat k} ^m {\dh _{\a ^m
} } ] . \label{eq.mtrcS4} \eea The components of the dual bases of
frame vectors $ h ^{\hat k} _m ,\ h_{\hat k} ^m $ are expressed via
the (row-column) matrix notation for the matrices $ A,\ B$ in terms of
the Euler angle coordinates $\a ^m = (\t , \phi , \g /2 \in [0, 2 \pi
] )$, describing the (Lens space) orbifold $S^3 / Z_2$ as the Hopf map
onto the fibre $ S^1 (\g /2 )$ over $ S^2 (\t ,\phi ) $, \bea && h
^{\hat k} _m = A _{k m} \equiv \pmatrix{ -2 c_\g & -2 s_\g s_\t & 0
  \cr -2 s_\g & 2 c_\g s_\t & 0 \cr 0 & 2 c_\t & 1},\ h_{\hat k} ^m =
B _{m k} = \ud \pmatrix{- c _\g & - s _\g & 0 \cr -{s _\g \over s _\t
  } & {c _\g \over s _\t} & 0 \cr {2 c _\t s _\g \over s _\t} & -{2 c
    _\t c _\g \over s _\t } & 1},
\label{eq.cof2}\eea 
where $ [c_\g , \ s_\g ]= [\cos {\g / 2} ,\ \sin {\g / 2} ],\ [c_\t ,
  \ s_\t ] = [\cos \t , \ \sin \t ]$.  The resulting formulas for the
metric tensor, \bea && \tilde g ^0_{mn} = \sum _{\hat k} h ^{\hat k}
_m k^{\hat k} _n = \pmatrix{ 4 & 0 & 0 \cr 0& 4 & 2 c _\t \cr 0 & 2 c
  _\t & 1} ,\ \tilde g ^{0, mn} = \sum _{\hat k} h _{\hat k} ^m
k_{\hat k} ^n= \pmatrix{ 1& 0 & 0 \cr 0& 1/s^2_\t & -2 c _\t /s^2_\t
  \cr 0 & 2 c _\t/s^2_\t & 1/s^2_\t } \cr && \Longrightarrow \ d s^2 (
S^3 / Z_2)=\tilde g^o_{mn } d \a ^m d \a ^n = 4 ( d\t ^2 + s_\t ^2 d
\phi ^2 + ( {d \g \over 2} + c _\t d \phi )^2 ) , \eea give the
expected results for the volume integral of the (unit radius)
orbifold~\cite{gubser98}, $ Vol( S^3 / Z_2 ) =2^{-3} \int \hat h _1
\wedge \hat h _2 \wedge \hat h _3= 8 \pi ^2 = Vol( S^3 ) /2 $. The
geometry of $\S _4$ at large radial distances $1 << \rho \sim \hat r /
\mu ^{2/3} $ is that of a real cone over a squashed $ S^3 / Z_2$ (in
agreement with Eq.~(\ref{eqsinc3})), while that near $ \rho \to 0 $ is
that of a real cone over the direct product of $ S^2 $ times a
collapsed $ S^1$, \bea && \lim _{ \rho \to \infty } d\tilde s ^2
(\calc _6 )\vert _{\S _4} = {3 \over 2} d \hat r^2 + {\hat r^2 \over
  6} (\hat h_{1}^2 +\hat h_{2}^2 + {4 \over 3} \hat h_{3}^2) ,
\ \ \lim _{ \rho \to 0 } d s^2 (\S _4) = {\vert \eta _\chi \vert ^2
  \over 2} F_1 ^2 (\tau ) ( d \rho ^2 + \hat h_3^2 + \hat h_2^2 +
         {\rho ^2 \over 4 } \hat h_1^2 ) .\eea


We  now consider  the harmonics of the base manifold $S^3 =
SO(4)/SO(3)$.  The basis of scalar harmonics $ \caly ^{{l/ 2}, {l/ 2}
} _{m_l, m_r} (\a ),\ [j \equiv l/2 =0, 1/2 ,\dots ,\ (m_l,\ m_r) \in
  [-j,\cdots , j] $,  running over the irreducible representations of $ SO(4)
  \sim SU(2)_l \times SU(2)_r$,  consists of eigenfunctions of the $S^3
  \sim SU(2) $ scalar Laplacian $- \nabla ^2 (S^3) \to l(l+2) = 4 j
  (j+1) ,\ [j= l/2 ] $ of equal   angular momentas  $j =
  l/2,\ r= l/2 $  of the left and right  acting subgroups  $SU(2) _{l,r} $.
The left magnetic quantum number  $ m_l $ labels the
states of $SU(2)_l $ while the right  one $ m_r $
labels the representations part of the harmonic basis, with
the  multiplicities  $ D_3(l) = (l+1 )^2 = (2j+2)^2 $.  The harmonic basis
in the Euler angles parameterization,
$Y ^j _{m,n} (\a _E)  ,\ [\a _E = (\t , \phi , \g  )] $ and 
 that in the hyperspherical angles parameterization, $Y _{j, L, M} (\a _S ) ,\
 [\a _S= (\chi , \ \t , \phi  ) ]$ 
 involving    Wigner   functions~\cite{vanNieu12} and
 products of Jacobi polynomials times spherical
 harmonics~\cite{higuchi87}, respectively  are  linearly
related  via the Clebsch-Gordan coefficients~\cite{lechtmar20}
(enforcing  the conditions $ L \leq 2 j ,\ M= m+r$),  
  \bea &&
\caly  ^j _{m,r} (\a _E)   = \sum _{L, M} <L,M\vert j m, jr >  Y_{j, l, M}  (\a _S ),
\cr &&   \bigg [\caly ^{j} _{m,r} (\a _E) =  D ^{{l\over 2}, {l\over 2}} _{m, r}
  (\t, \phi , \g  ) ,\ [j\equiv  {l\over 2} = 0,\ud ,\dots ,\
(m ,\ r ) \in (-j ,\cdots , + j)] \cr &&   Y _{j, L, M} (\a _S ) =
\phi _{jL} (\chi ) Y _{LM} (\t ,\phi ) ,\ [j=0, 1/2 , 1, \cdots ,
  \ L=0,1,\cdots , 2j ]  \bigg ].    \eea
The   allowed functions   on $S^3/Z_2$,  invariant  under the action
of $Z_2$,  are  restricted  for   the  Euler angles
harmonics  by the projection  condition $ r = 0\ mod\  1$. 

The harmonic decomposition for the gauge field components $ A_\mu $
along $ M_4$ utilizes the scalar harmonics while that for the
longitudinal and transversal components along $\S _4$, defined by the
respective conditions, $ \e ^{ijk} \nabla _j A_k ^\parallel =0
,\ \nabla ^{ k} A_k ^\perp =0 $, utilizes the bases of scalar and
vector harmonics,  \bea && A_k ^\parallel = \nabla _{ k } \caly
^{{l\over 2}, {l\over 2} } (\a ),\ [E_l= - l(l+2) +2 ,\ D_l =(l+1)^2 ]
\cr && A_k ^\perp = \caly ^{l, \pm } _{m,r} = [\caly ^{{l+ 1 \over 2}
    , {l- 1 \over 2} } _{m, r} (\a ) ,\ \caly ^{{l - 1 \over 2} , {l+
      1 \over 2} } _{m, r} (\a ) ],\ [ E_l=- l(l+2)+1, \ D_l =2 l(l+2)
] \eea with the above indicated values of the eigenvalues and
degeneracies~\cite{rubinez85} $E_l$ and $D_l$.

The spinor harmonics of $S^3$ can be constructed
recursively~\cite{higcampo95} in terms of the basis of $2\times 2 $
Dirac matrices, $ \G ^1= - \s _2, \G ^2 = \s _1 ,\ \G ^3 = -i \G ^1\G
^2 = \s _3$.  The Dirac operator harmonic eigenfunctions of angular
momentum $ n $, $0= (\Dslash (S^3) \mp i (n + 3/2 ) ) \psi _{\pm , n l
  l_1} (\chi , \O _2) $, are given by linear combinations of $S^2$
spinor harmonics $\chi ^{(\pm )} _{l l_1} (\O _2) $, eigenfunctions of
the Dirac operator of angular momentum $ l $, $ (\G^3 \Dslash _{S^2}
\mp (l +1) ) \chi ^{(\pm )} _{lm} (\O _2) =0 $. The latter are in turn
expressed in terms of $ S^1$ spinor harmonics, eigenfunctions of Dirac
operator of angular momentum $ l_1$, $(\Dslash _{S^1} \mp i (l_1 +1/2)
) \chi ^{(\pm )} _{l_1} (\phi ) =0 $, \bea && \psi _{\pm , nl l_1}
(\chi , \O _2)= \phi _{n l} (\chi ) \chi ^{(-)} _{\pm , l l_1 } (\O _2
) \pm i \psi _{n l} (\chi ) \chi ^{(+)} _{\pm , l l_1 } (\O _2 ) ,\cr
&& [\chi _{\pm , l l_1 } ^{(-)} (\O _2 ) = \pmatrix{\phi _{l l_1 } (\t
       )\cr \pm i \psi _{ l l_1 } (\t ) } \chi ^{(-)} _{ l_1 }(\phi )
     ,\ \chi _{\pm , l l_1 } ^{(+)} = \pmatrix{i\psi _{l l_1 } (\t
       )\cr \pm \phi _{l l_1 } (\t ) } \chi ^{(+)} _{l_1 } (\phi )
     ,\ \chi ^{(\pm )} _{l_1} (\phi ) = e ^{\pm i (l+1/2)\phi } ] \eea
where the $\pm $ lower indices in $\psi _{\pm ll _1 } $ and $\chi_{\pm
  ll _1 } $ label the independent solutions while the uper indices
$^{(\pm )} $ refer to the $ S^2 $ chiralities. The integer angular
momenta for the harmonics of $ S^3,\ S^2,\ S^1$ satisfy $n \geq l \geq
l_1 ,\ [n =0, 1,\cdots ] $ with degeneracies $D_3 (n) = (n+1)(n
+2),\ D_{2}(l) = 2 (l+1),\ D_1 (l_1) =1$ and the scalar harmonic
functions $\phi _{nl} ,\ \psi _{nl} $ are expressed in terms of Jacobi
polynomials.

\subsection{Reduced bosonic action} 
\label{appD7subB1}

The induced metric on the $D7$-brane world volume of intrinsic
coordinates $\xi ^\a $ is the pull-back transform of the target
spacetime metric through the directional derivatives of the
coordinates $X^M (\xi ) $, defining the world volume embedding in
spacetime,
\bea && d s^2 (D7) = \g _{\a \b } d \xi ^\a d \xi ^\b ,\
 [\g _{\a \b } = {\dh X^M \over \dh {\xi ^\a } } {\dh X^N \over
     \dh {\xi ^\b } } g_{MN} ].  \label{eq.metS0} \eea
 The invariance under  coordinates diffeomorphisms allows choosing
 the static gauge in which the intrinsic coordinates are
 identified to subset of spacetime coordinates tangential to the
 world volume, $\xi ^\a = X^\a (\xi )=[\rho, \ (\hat h _3, \hat h
   _1, \hat h _2 ) ] $.  The remaining coordinates along
 transversal directions $ X^u (\xi ) $ are interpreted as brane
 moduli fields $\chi (\xi ) $ and are assigned classical values
 (VEVs) describing the brane embedding in spacetime. The induced
 metric \bea && d s^2 (D7) = \g _{\mu \nu } d x^{\mu \nu } +
 \g_{\rho \rho } d\rho ^2 + 2 \g _{h_i h_j } \hat h_i \hat h_j + 2
 \g _{\rho h_i } \hat h_i d\rho , \eea can also be directly
 evaluated by substituting the decomposition $ d \chi =\dh _{\mu }
 \chi d x ^\mu + \dh _{\rho } \chi d \rho + \dh _{h_i } \chi h_i $
 into Eq.~(\ref{eq.metrans}) for the spacetime metric.  We
 consider for definiteness the induced warped metric.  This
 consists of two additive parts, $\g _{\a \b} (\chi ) \equiv \g
 ^{(0)} _{\a \b} + \a _{\a \b} = g _{\a \b} + \a _{\a \b} $, the
 first coinciding with the spacetime metric restriction to the
 brane world volume and the second involving directional
 derivatives of the scalar moduli fields $ \dh _{\mu , \rho , \hat
   h_i} \chi $.  Both parts are functions of $\chi $ and $\mu
 _\chi $, the first $g _{\a \b } $ being of diagonal form in the
 present case and the second $\a _{\a \b} $ of non-diagonal form,
 \bea && \g _{\mu \nu } = \g ^{(0)} _{\mu \nu } + \a_{\mu \nu } =
 h^{-1/2} \tilde g _{\mu \nu } + 2 h^{1/2} \tilde g _{\chi \bar
   \chi } \dh _\mu \chi \dh _\nu \bar \chi ,\cr && \g _{\a \nu } =
 \g _{\nu \a } = \a_{\nu \a } = h^{1/2} (\tilde g _{\a \chi } \dh
 _\nu \chi + \tilde g _{\a \bar \chi } \dh _\nu \bar \chi + \tilde
 g _{\chi \bar \chi } (\dh _\nu \chi \dh _\a \bar \chi +\dh _\nu
 \bar \chi \dh _\a \chi ) ,\ [\a = \rho ,\ \hat h_3] , \cr && \g
 _{\rho \rho } = \g ^{(0)} _{\rho \rho } +\a_{\rho \rho } =
 h^{1/2} (\tilde g_{\rho \rho } + 2 \tilde g_{\rho \chi } \dh
 _\rho \chi + 2 \tilde g_{\rho \bar \chi } \dh _\rho \bar \chi +2
 \tilde g _{\chi \bar \chi } \dh _\rho \chi \dh _\rho \bar \chi )
 ,\cr && \g _{\rho h_i } = h ^{1/2} \tilde g _{\rho h_i } +
 h^{1/2} (2 \tilde g _{h_i \chi } \dh _\rho \chi +2 \tilde g _{h_i
   \bar \chi } \dh _\rho \bar \chi ) ,\cr && \g _{ \hat h_i \hat
   h_j} = h^{1/2} (\tilde g _{ \hat h_i \hat h_j} +2 \tilde g
 _{\chi h_i } \dh _{h_j} \chi + 2\tilde g _{\bar \chi h_j} \dh
 _{h_j} \bar \chi + 2\tilde g _{\chi \bar \chi } \dh _{h_i } \chi
 \dh _{h_j} \bar \chi ) ,\ [ \tilde g _{\rho h_i } =0,\ \tilde g
   _{ \hat h_i \hat h_j} =0] . \label{eq.metS2} \eea


The linear and quadratic order contributions in $ d \chi $ and $\eta
_\chi $ are given by the explicit formulas for the warped metric
components \bea && \a _{\mu \nu } = 2 F_1 h^{1/2} (\tau ) A_3 \dh _\mu
\chi \dh _\mu \bar \chi ,\ \a _{\rho \rho } = F_1 h^{1/2} (2 A_3 \dh
_\rho \bar \chi \dh _\rho \chi + \bar A_2 \eta _\chi S_\rho \dh _\rho
\bar \chi + A_2 S_\rho \bar \eta _\chi \dh _\rho \chi ) ,\cr && \a
_{h_3 h_3 } = F_1 h^{1/2} (2 A_3 \dh _{h_3} \chi \dh _{h_3} \bar \chi
+ i \bar A_2 \eta _\chi S_\rho \dh _{h_3} \bar \chi - i A_2 S_\rho
\bar \eta _\chi \dh _{h_3} \chi ),\cr && \a _{h_2 h_2} = 2 A_3 F_1
h^{1/2} \dh _{h_1} \chi \dh _{h_1} \bar \chi ,\ \a _{h_1 h_1} =2 A_3
F_1 h^{1/2} \dh _{h_2} \chi \dh _{h_2} \bar \chi , \cr && \a _{\mu
  \rho } = {1\over 2} F_1 h^{1/2} (2 \bar A_3 \dh _\rho \bar \chi \dh
_\mu \chi + 2 A_3 \dh _\mu \bar \chi \dh _\rho \chi + \bar A_2 S_\rho
\dh _\mu \bar \chi \eta _\chi + A_2 S_\rho \bar \eta _\chi \dh _\mu
\chi ) ,\cr && \a _{\mu h_3 } = {i\over 2} F_1 h^{1/2} S_\rho (\bar
A_2 \dh _\mu \bar \chi \eta _\chi - A_2 \bar \eta _\chi \dh _\mu \chi
), \ \a _{\mu h_1 } = A_3 F_1 h^{1/2} \dh _{h_1} \bar \chi \dh _\mu
\chi ,\ \a _{\mu h_2 } = A_3 F_1 h^{1/2} \dh _{h_2} \bar \chi \dh _\mu
\chi , \cr && \a _{\rho h_3 } = {i\over 2} F_1 h^{1/2} S_\rho (\bar
A_2 \dh _\rho \bar \chi \eta _\chi - A_2 \bar \eta _\chi \dh _\rho
\chi ), \ \a _{\rho h_1 } = A_3 F_1 h^{1/2} \dh _{h_1} \bar \chi \dh
_\rho \chi ,\ \a _{\rho h_2} = A_3 F_1 h^{1/2} \dh _{h_2} \bar \chi
\dh _\rho \chi , \cr && \a _{h_3 h_1 } = A_3 F_1 h^{1/2} \dh _{h_3}
\chi \dh _{h_1} \bar \chi - {i\over 2} A_2 h^{1/2} S_\rho \bar \eta
_\chi F_1 \dh _{h_1} \chi , \ \a _{h_1 h_2 }= A_3 F_1 h^{1/2} \dh
_{h_1} \chi \dh _{h_2} \bar \chi,\cr && \a _{h_3 h_2 } = A_3 F_1
h^{1/2} \dh _{h_3} \chi \dh _{h_2} \bar \chi -{i\over 2} A_2 F_1
h^{1/2} S_\rho \bar \eta _\chi \dh _{h_2} \chi , \cr && [ \eta _\chi =
  {\mu _\chi \over X} , \ A_1=C_\rho + 2 \eta _\chi \bar \eta _\chi
  S_\rho ^2 R ,\ A_2= X + 2 \bar \mu _\chi \eta _\chi (1 + C_\rho {X
    \over \bar X} ) R ,\ A_3 = 1+ C_\rho X\bar X + 2 \mu _\chi \bar
  \mu _\chi R \vert 1+ C_\rho { X \over \bar X } \vert ^2 ]
. \label{eq.metS3} \eea

The action for the Kaluza-Klein modes of $ \chi $ depends on the 4-d
Minkowski spacetime derivatives $ \dh _\mu \chi $ as well as the
radial and angle derivatives, $ \dh _\rho \chi $ and $ \dh _{ h_i }
\chi $.  For the singlet modes under the 4-cycle isometry group (as
opposed to charged modes) only the radial derivatives are relevant.
The square root of the induced metric determinant is evaluated as a
power expansion in the fields derivatives by means of the familiar
formula \bea && Det ^{1/2} (- (\g ^{(0)} + \a ) ) = Det ^{1/2}(-\g
^{(0)} ) (1 + \ud Tr ( \g ^{(0) -1 } \a ) -{1\over 4} Tr ( \g ^{(0) -1
} \a \g ^{(0) -1 } \a ) + {1\over 8} (Tr ( \g ^{(0) -1 } \a ))^2
+\cdots ) . \label{eq.expdet} \eea


The $D7$-brane bosonic action up to $O ((d \chi )^4)$ is written
below  with  the linear, quadratic, cubic and quartic order (square
and cross) terms in $ d \chi $ appearing in succession,
\bea && S _B (D7) = - \mu _7 \int d^4 x
\sqrt {g _4 ^{(0)} } \int d\rho \sqrt {-\g
  ^{(0)} } e ^{\phi } \int \hat h_1 \wedge \hat h_2 \wedge \hat h_3
\cr && \times \bigg [ 1
+ \bigg ( {S_\rho \over \vert \eta _\chi \vert ^2 T_3 } [ A_2 \bar
  \eta _\chi ( \dh _{\rho } \chi - i \dh _{h_3 } \chi ) - F_1 h A_3
  \bar A_2 \eta _\chi (\dh _\mu \bar \chi ) ^2 ( \dh _{\rho } \chi + i
  \dh _{h_3 } \chi ) ] + H.\ c.  \bigg ) \cr &&
+ F_1 h (\tau ) (A_3 - {\vert A_2 \vert ^2 S_\rho ^2 \over T_3 } )
\tilde g^{\mu \nu } \dh _{\mu } \bar \chi \dh _{\nu } \chi + {2 \over
  \vert \eta _\chi \vert ^2 T_3 } (A_3 - {\vert A_2\vert ^2 S_\rho ^2
  \over T_3 } ) ( \dh _{\rho } \bar \chi \dh _{\rho } \chi + \dh _{h_3
} \bar \chi \dh _{h_3 } \chi ) \cr && + {4\over (1 + C_\rho ) \vert
  \eta _\chi \vert ^2 } (A_3 - {\vert A_2 \vert ^2 S_\rho ^2 \over T_3
} ) \dh _{h_1 } \bar \chi \dh _{h_1 } \chi + {4\over (-1 + C_\rho )
  \vert \eta _\chi \vert ^2 } (A_3 - {\vert A_2\vert ^2 S_\rho ^2
  \over T_3 } ) \dh _{h_2} \bar \chi \dh _{h_2 } \chi
 + ( { 2i \vert A_2 \vert ^2 S_\rho ^2 \over \vert \eta _\chi
   \vert ^2 T_3 ^2 } \dh _\rho \chi \dh _{h_3} \bar \chi + H.\ c. )
\cr &&  + { 1 \over 2 \vert \eta _\chi \vert ^4 T_3 ^2 } ((\dh _{\rho } \chi
)^2 + (\dh _{h_3 } \chi )^2 )  \bigg ( ( {4 A_2 A_3 \bar \eta _\chi S_\rho }
  (- \dh _\rho \chi + i \dh _{h_3 } \chi ) + H.\ c . )  - 4 A_3 ^2 (
  (\dh _\rho \bar \chi )^2 + (\dh _{ h_3 } \bar \chi )^2 )\bigg ) \cr &&
- { 1 \over \vert \eta _\chi \vert ^4 } (\dh _{h_1} \chi )^2 [ { 4 A_2
    A_3 \bar \eta _\chi S_\rho \over (1+C_{\rho }) T_3} \dh _\rho \bar
  \chi + {4 A_3 ^2 \over (1+C_\rho ) T_3} (\dh _{\rho } \bar \chi )^2
  + {8 A_3 ^2 \over (1+C_\rho )^2 } (\dh _{ h_1} \bar \chi )^2 ] \cr
&& - {1 \over \vert \eta _\chi \vert ^4 } (\dh _{h_2} \chi )^2 [ { 4
    A_2 A_3 \bar \eta _\chi S_\rho \over (- 1+C_{\rho }) T_3} \dh
  _\rho \bar \chi + {4 A_3 ^2 \over (-1+C_\rho ) T_3} (\dh _{\rho }
  \bar \chi )^2 + {8 A_3 ^2 \over (-1+C_\rho )^2 } (\dh _{ h_2} \bar
  \chi )^2 ]\bigg ] . \label{app.indactP}\eea
The cross angular terms, $ (\dh _{h_i } \bar \chi \dh _{h_j } \chi )
,\ [i\ne j]$ are absent owing the symmetry under the isometry group
$SO(3) \sim SU(2) / Z_2 $.  Note also the absence at quadratic order
of complex terms mixing the real and imaginary field components, $
(\dh _{\mu } \chi )^2, \ (\dh _{\rho } \chi )^2,\ (\dh _{h_j } \chi )
^2 $.

The pull-backs of the K\"ahler, NSNS 2-form and complex 3-form fields
$J,\ B_2,\ G_3 $ in Einstein frame are conveniently evaluated by
substituting the expressions in
Eq.~(\ref{eq.conufol})  for the complex coordinates $ (w_a) \in
C^4 $  in the $SO(4)$ invariant expressions of the Klebanov-Strassler
classical solutions~\cite{herzog01}, \bea && \bullet \ J= i {K(\tau )
  \over \e ^{2/3} } [\d ^{ab} - {1\over \e ^2 \sinh ^2 \tau } (\cosh
  \tau - {2 \over 3 K ^3(\tau )} ) \bar w^a w^b ] d w _a \wedge d \bar
w_b , \cr && \bullet \ B_2 ^{cl} = i b(\tau ) \e _{abcd } w ^a \bar
w^b d w ^c \wedge d \bar w^d ,\ [b(\tau ) = {g_s M \a ' \over 2 \vert
    \e \vert ^4 } { (-1 + \tau \coth \tau) \over \sinh ^2\tau } ] \cr
&& \bullet \ G_3 = G_{(2,1)} = F_3 - i e ^{-\phi } H_3 = g_1 (\tau )
\e _{abcd} w ^ a \bar w^b d w ^c \wedge d \bar w^d \wedge \bar w_e d
w^e + g_2 (\tau ) \e _{abcd} w ^a \bar w^b d w ^c \wedge d w ^d \wedge
w_e d \bar w^e ,\cr && [g_1 (\tau ) = {M\a ' \over 2 \e ^6 \sinh ^4
    \tau } {\sinh 2 \tau - 2 \tau \over \sinh \tau },\ g_2 (\tau ) =
  {M\a ' \over 2 \e ^6 \sinh ^4 \tau } 2 ( 1 -\tau \coth \tau)
] . \label{PULLJBC} \eea
The resulting formulas for the differential
forms restriction to $\S _4$ read \bea && \bullet \ J\vert _{\S _4} =
{ \vert \eta _\chi \vert ^2 \over 2 \e ^{2/3} } K(\tau) [(\cosh \rho -
  {\vert \eta _\chi \vert ^2 \over \e ^2 \sinh ^2 \tau } (\cosh \tau -
  {2 \over 3 K ^3(\tau )} ) \sinh ^2 \rho ) \ d \rho \wedge \hat h_3 +
  \ud \sinh \rho \ \hat h_1 \wedge \hat h_2 ] \cr && = { \vert \eta
  _\chi \vert ^2 \over 2 \e ^{2/3} } K(\tau) [ K_2 (\tau ) \ d \rho
  \wedge \hat h_3 + \ud \sinh \rho \ \hat h_1 \wedge \hat h_2 ], \cr
&& \bullet \ B_2 ^{cl}\vert _{\S _4} = \vert \mu _\chi ^2 Y \vert ^2
b(\tau ) [{i\over 2} (Y -\bar Y )\sinh ^2 {\rho \over 2} (\cosh {\rho
    \over 2} d \rho \wedge \hat h_1 + \sinh {\rho \over 2} \hat h_3
  \wedge \hat h_2) \cr && - {1\over 2} (Y +\bar Y ) \cosh ^2 {\rho
    \over 2} (\sinh {\rho \over 2} d \rho \wedge \hat h_2- \cosh {\rho
    \over 2}\hat h_3 \wedge \hat h_1 ) ] ,\cr && \bullet \ G_3\vert
_{\S _4} = {\vert \mu _\chi ^2 Y \vert ^2 \over 4}\sinh \rho [ g_1
  (\tau ) \vert \mu _\chi \vert ^2 \vert Y \vert ^2 [ (Y -\bar Y)
    \sinh ^2 {\rho \over 2 } ( \cosh {\rho \over 2} d\rho \wedge \hat
    h_1 + \sinh {\rho \over 2 } \hat h_3 \wedge \hat h_2 ) \cr && + i
    (Y +\bar Y) \cosh ^2 {\rho \over 2 } ( \sinh {\rho \over 2} d\rho
    \wedge \hat h_2 - \cosh {\rho \over 2 } \hat h_3 \wedge \hat h_1 )
  ] \wedge ( d\rho + i \hat h_3 ) \cr && + g_2 (\tau )\mu _\chi ^2 Y^2
  (Y + \bar Y \cosh \rho ) [ \cosh {\rho \over 2} ( d\rho + i \hat
    h_3) \wedge \hat h_1 + i \sinh {\rho \over 2 } ( d\rho + i \hat
    h_3) \wedge \hat h_2 ] \wedge ( d\rho - i \hat h_3 ) ]
. \label{eq.pullJBC} \eea We see that $ \hat J \equiv J \vert _{\S _4}
/2 $ and $\calf = B_2 ^{cl} \vert _{\S _4} + F_2 $ obey (for $F_2=0$
and the value $\t =0 $ of the arbitrary parameter $\t $) the
conditions set by the $\kappa $-symmetry~\cite{marino99}, \bea && \ud
(\hat J\wedge \hat J ) \vert _{\S _4} = vol (\S _4),\ ( \hat J \wedge
\calf )\vert _{ \S _4} = \ud \tan \t ( \hat J \wedge \calf - \calf
\wedge \calf ) \vert _{ \S _4} , \eea as expected~\cite{benini09} in
the case of a calibrated (holomorphic, supersymmetry preserving)
4-cycle.


The pull-backs of the background solutions also include extra
contributions depending on $ d \chi $.  The K\"ahler 2-form and the
3-form solutions acquire the correction terms, \bea && \d J = i K
(\tau ) \e ^{-2/3} [{1\over 2} (- \bar \eta _\chi X S_\rho ( d \rho -i
  \hat h_3 ) \wedge d \chi + \eta _\chi \bar X S_\rho ( d \rho + i
  \hat h_3 ) \wedge d \bar \chi )
  + (1 + X\bar X C_\rho ) d \chi \wedge d \bar \chi ] \cr &&
+ i {K'(\tau  ) \e ^{-8/3} \over \sinh (\tau ) } [- \ud (\mu_\chi \bar \mu_\chi ^2
  (1 + C_\rho ) S_\rho (d \rho -i \hat h_3 ) \wedge d \chi - H.\ c. )
  + \vert \mu _\chi \vert ^2 (1 + C_\rho )^2 d \chi \wedge d \bar \chi
] ,\cr && \d G_3 = g_1 (\tau ) [i \vert \mu _\chi \vert ^4 \cosh ^2
  {\rho \over 2} (\sinh {\rho \over 2} d \rho \wedge \hat h_2 - \cosh
  {\rho \over 2}\hat h_3 \wedge \hat h_1 ) ] \wedge ( (\bar \mu_\chi +
C_\rho \bar \eta_\chi ) d \chi + {\vert \eta _\chi \vert ^2 \over 2}
S_\rho ( d \rho + i \hat h_3) ) \cr && + g_2 (\tau ) [-\ud \vert \eta
  _\chi \vert^2 \mu_\chi (X - Y) {S_\rho ^2 \over \sinh {\rho \over 2}
  } \hat h_1 \wedge d \chi -i \vert \eta _\chi \vert^2 \mu _\chi (X -
  Y) \sinh {\rho \over 2} S^2_\rho \hat h_2 \wedge d \chi ] \wedge
((\mu_\chi + C_\rho \eta_\chi ) d \chi + {\vert \eta _\chi \vert
  ^2\over 2} S_\rho ( d \rho - i \hat h_3) ) . \eea The additional
terms in $ \d B _{\a \b } = B_{\a \b } - B_{\a \b }\vert _{\S _4} = 2
B _{\a \chi } \dh _\b \chi + 2 B _{\a \bar \chi } \dh _\b \bar \chi +
2 B _{\chi \bar \chi } \dh _\a \chi \dh _\b \bar \chi $ are found to
vanish in the limit $\mu >> \e $ (in which $X \simeq 1/ Y \to 1 $) and
hence will not be quoted.

We now  explain how to include the
contributions from the classical fields $B_2 ^{cl} ,\ F _2 ^{cl} $
components along $ \S _4$ present in   (Einstein
frame) Klebanov-Strassler solution.
The dependence   enters   through the (unwarped) effective metric
$ \tilde q _{\a \b } = \tilde \g _{\a \b } + e ^{-\phi /2} h^{-1/2}
\calf _{\a \b } ,\ [ \calf _{\a \b }= B _{\a \b }+ F_{\a \b }] $ where
the index $\a =  (\mu =0, 1, 2, 3, \ a= 1,2,3, 4) $ labels directions
along $M_4 $ and $\S _4$ in  the ordering convention   adopted   for the
tangent space basis of $\S _4 :\ (d \rho ,\ \hat h_3, \ \hat
h_1,\ \hat h_2) $.  The matrix $ \tilde q $ is block diagonal with one
sub-block $ \tilde g _{\mu \nu }$ in $M_4 $ while the other sub-block
$ \tilde q = \tilde \g + b,\ [b = e ^{-\phi /2} h^{-1/2} B _2^{cl} ] $
along $\S _4 $ is the sum of a symmetric (diagonal) matrix $\tilde \g
_{ab} $ and an antisymmetric (anti-diagonal) matrix $ b _{ab} $, \bea
&& \tilde \g =diag ( \g _1, \g_2, \g_3, \g_4), \ b= \overline{diag}
(b_1, b_2, -b_2, - b_1),\ [\g_{a=1,\cdots , 4} = \ud F_1 \vert \eta
  \vert ^2 [ A_1, A_1 , \ud (1 + C_\rho ) ,\ud (-1 + C_\rho ) ], \cr
  && b_{[1,2]} = - e ^{-\phi /2} h ^{-1/2} b _0 (\tau ) [\sinh {\rho
      \over 2} ,\ -\cosh {\rho \over 2} ] ,\ b_0 (\tau ) = {\vert \eta
    _\chi \vert ^4 \vert X \vert ^2 \over 4 } (1 + C_\rho ) b(\tau )
  ,\ b (\tau ) = {g_s M \a ' \over 2\e ^4 } {( -1 + \tau \coth \tau )
    \over \sinh ^2 \tau } ] . \label{eq.termsTXB1} \eea In the
convenient matrix notation inthe tangent space of $\S _4$, $\tilde q $
and its inverse $ \tilde q^{-1} $ are both given by a sum of diagonal
and anti-diagonal (antisymmetric) matrices,
\bea && \tilde q = \tilde \g + b =\pmatrix {\g_1& 0& 0& b_1 \cr 0 &
  \g_2& b_2& 0 \cr 0& -b_2& \g_3& 0 \cr -b_1& 0& 0& \g_4 }
\ \Longrightarrow \ \ \tilde q ^{-1} = ( \tilde \g + b ) ^{-1}
=\pmatrix {g_1& 0& 0& -\b_1 \cr 0 & g_2& -\b_2& 0 \cr 0& \b_2& g_3& 0
  \cr \b_1& 0& 0& g_4 },\cr && [g_{a} \vert _{a=1,4} = {1\over \g _i
    R_i } ,\ R_1 = R_4 = 1 + {b_1^2 \over \g _1 \g _4} ,\ R_2 = R_3 =
  1 + {b_2^2 \over \g _2 \g _3} ,\ \b _i \vert _{i=1,2} = {1\over b _i
    \rho _i } ,\ \rho _1 = 1 + {\g _1 \g _4\over b_1^2 } ,\ \rho _2 =
  1 + {\g _2 \g _3 \over b_2^2},\cr && \tilde q \equiv Det (\tilde q)
  = (b_2 ^2 +\g _2 \g _3) (b_1 ^2 + \g _1 \g _4) \tilde \g +b + b_1 ^2
  \g _2 \g _3 + b_2 ^2 \g _1 \g _4 ,\cr && \tilde \g \equiv Det (\tilde \g
  ) = \g _1 \g _2 \g _3 \g _4 ,\ b \equiv Det (b ) = (Pf(b))^2 = (b_1
  b_2)^2 ].  \label{eq.termsB1PP} \eea


Substituting the expressions in Eq.~(\ref{eq.termsTXB1}) for the
matrices entries, $ \g _a,\ b_i $ and $ g _a,\ \b_i ,\ [a=1,\cdots
  ,4,\ i=1,2 ] $ specifies the parameters, $ R_1= R_2= 1+ {\sqrt b
  \over \sqrt {\tilde \g } } , \ \rho _1= \rho _2= 1+ {\sqrt {\tilde
    \g } \over \sqrt b }$, hence yielding the simplified formulas for
the matrix determinant, $ \sqrt {\tilde q} = \sqrt {\tilde \g } +
{\sqrt b } ,$ and for the inverse matrix entries, \bea && g_a = {1
  \over R_S \g _a } \ \Longrightarrow \ \tilde q^{\rho \rho } ={1
  \over R_S \g _1} ,\ \tilde q^{\hat h_3 \hat h_3 }= {1 \over R_S \g
  _2} ,\ \tilde q^{\hat h_1\hat h_1}= {1 \over R_S \g _3},\ \tilde
q^{\hat h_2\hat h_2}= {1 \over R_S \g _4}, \ [ R_S ={\sqrt {\tilde q}
    \over \sqrt {\tilde \g } } ] \cr && \b _i = - {1 \over R_A b_i }
\ \Longrightarrow \ \tilde q ^{[\rho \hat h_2] } =- {1\over R_A b_1 }
, \ \tilde q ^{[\hat h_3 \hat h_1] } = - {1 \over R_A b_2} ,\ [R_A
  ={\sqrt {\tilde q } \over \sqrt {b} } ] . \label{eq.termsTXB3} \eea
The effective metric and its inverse along $ M_4 \oplus \S _4 $ can be
represented in a general (basis independent) way by block diagonal
matrices $\tilde q _{\a \b } = [\tilde \g _{\mu \nu } , \tilde q _{a b
} ] ,\ \tilde q ^{\a \b } = [\tilde \g ^{\mu \nu } , \tilde q ^{a b }
] $, satisfying the relations~\cite{benini09}, \bea && Det ( \tilde
q)= ( \sqrt {Det (\tilde\g )} + h ^{-1} \sqrt { Det (B) })^2 ,
\ \tilde q ^{ (\a \b )} = { \sqrt {\g } \over \sqrt {\tilde q} } \g
^{\a \b } = {\g ^{ \a \b } \over  R_S },\ \tilde q ^{ [\a \b ]} = - {\sqrt {b
  } \over \sqrt {-\tilde q } } B ^{ \a \b } = -{B ^{\a \b } \over  R_A },\cr
&& [\sqrt {\tilde \g } = \ud (\ud F_1 \vert \eta \vert ^2 )^2 A_1
     \sinh \rho , \ \sqrt { Det (B) } = Pf(B) = \ud b_0^2 \sinh \rho
     ,\ R_S = 1+{\sqrt {b} \over \sqrt {\tilde \g }} ,\ R_A =1 +
           {\sqrt {\tilde \g } \over \sqrt { b}} ] \label{eq.termsB2}
\eea where the indices enclosed between pairs of parentheses or
brackets are to be symmetrized or antisymmetrized.  One finds for a
constant dilaton profile, \bea && R_S = 1 + {Pf (B) \over g_s h (\tau
  ) V_0 (\rho ) } = 1 + {S_\rho b_0 ^2 (\tau ) \over g_s h (\tau )V
  (\rho ) } = 1 + {2 ^{4/3} \vert L \vert ^4 g_s } { \cosh ^4(\rho /2)
  \over I (\tau ) K^2 (\tau ) T_3 (\rho ) } ({-1 + \tau \coth \tau
  \over \sinh ^2 \tau }) ^2 ].  \label{eq.termpB2} \eea

The effective metric determinant in Eq.~(\ref{eqbracts}) for the $\chi
$-field action, \bea && V_0 ^2 (\rho )= -Det (\tilde q )_{M_8} = - Det
(\tilde \g )_{M_8} \times Det (\tilde q ) _{\S _4} , \ [(\tilde q
  )_{M_8} = diag ( (\tilde \g )_{M_4} , \ (\tilde q ) _{\S _4}
  ),\ (\tilde \g )_{\S _4} = ( \tilde \g + b ) \vert _{\S _4} ]\eea
shows that the contributions from $ B_2^{cl}$ may be included by
replacing $ V_0 (\rho ) =\sqrt {-\tilde \g ^{(0)} } \to \sqrt {-\tilde
  q } = R_S \sqrt {-\tilde \g ^{(0)} } $ and adding the linear and
quadratic order corrections in $ \dh _{\mu , \rho } \chi $ from $
Tr(\tilde q^{-1} \a ) $ and $ Tr(\tilde q ^{-1} \a \tilde q ^{-1} \a )
.$ In Eq.~(\ref{eq.fol4}), for instance, the coefficient $ (T_1 - T_2
^2 / T_3) $ of $ \vert \dh _{\mu } \chi \vert ^2 $ is replaced by $
(T_1 - T_2 ^2 / (R_S T_3) )$ and the coefficient $ (T_1 - T_2 ^2 /
T_3) / T_3 $ of $ \vert \dh _{\rho } \chi \vert ^2 $ is replaced by $
R_S^{-1} (T_1 - T_2 ^2 / (T_3 R_S) ) / T_3 $, noting that the first
and second terms in these coefficients arise from linear and quadratic
order corrections.  The wave equations for $\chi $ modes are modified
likewise by replacing in Eq.~(\ref{eq.fol4}), \bea && Q (\rho ) \to
F_1 h (\tau ) (A_3 - {\vert A_2 S_\rho \vert ^2 \over A_1 R_S } )
,\ P(\rho ) \to (A_3 - {\vert A_2 S_\rho \vert ^2 \over A_1 R_S } ) /
A_1 . \label{eq.recipeNSNS} \eea

The contributions from the classical 2-form gauge fluxes are
conveniently included by noting the formal similarity of $F_2$ with $
B_2$.  For instance, the $(1,1)$-form solution derived
in~\cite{benini11}, $ F_{II}= -i \l \e_{ijk} \bar w_i d w_j \wedge d
\bar w_k + H.\ c. $ is of same form as the background solution for the
NSNS field $ B_2^{cl}$, so its contribution is found by replacing $
b(\tau ) \to \l / \mu _\chi $.  The comparison with
Eq.~(\ref{PULLJBC}) yields the explicit expression for the gauge flux
2-form in $\S _4$ \bea && F_{II}= (f_1 d \rho \wedge \hat h_2 + f_2
\hat h_3 \wedge \hat h_1 ) ,\ \ [f_{[1,2]} = \l \vert \mu _\chi \vert
  ^2 \bar \mu _\chi \cosh ^2 {\rho \over 2} [-\sinh {\rho \over 2},
    \ \cosh {\rho \over 2}] ] \label{eqgaugeflux} \eea which is
represented in the antisymmetric part of the effective metric tensor
by the anti-diagonal matrix, $F_{II} = \overline{diag} (f_1, f_2,
-f_2, -f_1)$.

\subsection{Reduced Dirac  operator}
\label{appD7subFe1}

We here examine the Dirac operator on the 4-cycle $\S _4 \subset \calc
_6$ restricted to the part of the covariant derivative depending on
the spin connection along the brane longitudinal directions, $\Dslash
\vert _{\S _4} =(\dslash + {1\over 4 } \o ^{\hat a\hat b} \G _{\hat
  a\hat b} ) $.  The solution of the flat torsion condition, $0= D_\mu
e ^{\hat a } _\nu = (D_\mu e )^{\hat a } _\nu - \G _{\mu \nu }^\l e_\l
^{\hat a } ,\ [(D_\mu e )^{\hat a } _\nu= 
  ) ^{\hat a } _\nu \dh _\mu e ^{\hat a } _\nu +\o _\mu ^{\hat a \hat
    b } e_{\nu , \hat b} ] $ provides the  explicit  formula for the
spin connection in terms of the tangent frame vectors of $\S _4$,
, \bea && \o _\mu ^{\hat a \hat b} = - e ^{\nu , \hat
  b} ( \dh _\mu e ^{\hat a } _\nu - \G _{\mu \nu }^\l e_\l ^{\hat a }
) = \ud [ e ^{\nu \hat a } (\dh_\mu e_\nu ^{\hat b } -\dh_\nu e_\mu
  ^{\hat b } ) - e ^{\nu \hat b } (\dh_\mu e_\nu ^{\hat a } - \dh_\nu
  e_\mu ^{\hat a } ) - e^{\l \hat a } e ^{\s \hat b} e _{\mu \hat c}
  (\dh_\l e_\s ^{\hat c} -\dh_\s e_{\l } ^{\hat c} )
] \label{eq.spinc0} \eea
where  $ e ^{\hat a } _\mu (\rho , \hat h _k ) $
carry vector  indices   along curved and flat space directions,
$\mu = ( \rho, \ m ) $ and $\hat a = ( \hat \rho , \hat k
)$.  In the static gauge, $\g _{\a \b } \to g_{\a \b } $, the non
vanishing components of the spin connection are evaluated from the general
formulas, \bea && \o ^{\hat \rho \hat k} _\rho =0 , \ \o ^{\hat j \hat
  k} _\rho = \ud H ^{\hat j \hat k } (\eta ^{\hat j} \eta ^{\hat k}
\dh _\rho \eta ^{\hat j} \dh _\rho \eta ^{\hat k} ) = 2 (A_{\hat j
  \hat k} - A_{\hat k \hat j}),\cr && \o ^{\hat \rho \hat k} _m = -
    {1\over 2} e ^{\rho \hat \rho } (\dh _\rho e^{ \hat k} _m + e
    _{m\hat c} \dh _\rho e ^{\hat c} _n e ^{n\hat k} ) = - {1 \over 2}
    (e ^{\rho \hat \rho }h^{ \hat k} _m \dh _\rho \eta ^{ \hat k} + C
    _{m \hat k} ) \cr &&
 \o ^{\hat j \hat k} _m = e ^{n \hat j} \dh _{ [m } e ^{\hat k} _ {n ]
 } - e ^{n \hat k} \dh _{ [m } e ^{\hat j} _ {n ] } - e ^{n \hat j } e
 ^{p \hat k } e_{m \hat c } \dh _{ [n } e ^{\hat c} _ {p ] } = \ud X
 _{m, \hat j \hat k} ,\cr && [e^{n \hat j}=\eta ^{\hat j} (\rho ) h^{n
     \hat j} (\a ) ,\ A _{\hat j \hat k} = {1\over 4} H^{\hat j \hat
     k} \eta ^{\hat j} \dh _\rho \eta ^{\hat k} ,\ C _{ m \hat k} = e
   ^{\rho \hat \rho }\eta ^{\hat k}\eta _{\hat j} \dh _\rho \eta ^{
     \hat j} h _{m \hat l} H^{\hat l \hat k}, \ H ^{\hat j \hat k } =
   h ^{\hat j} _n h ^{\hat k n} =\tilde g ^{nn'} h ^{\hat j} _n h
   ^{\hat k } _{n'} , \cr && X _{m, \hat j \hat k} = 2 [ {\eta ^{\hat
         k} \over \eta ^{\hat j}} h ^{n }_{\hat j} \dh _{ [m } h
       ^{\hat k} _ {n ] } - \eta ^{\hat j} \eta ^{\hat k} h ^{n
     }_{\hat k} \dh _{ [m } h ^{\hat j} _ {n ] } - { \eta ^{\hat l 2}
       \over \eta ^{\hat j} \eta ^{\hat k} } h _{\hat j}^n h_{\hat
       k}^p h_{m} ^{\hat l } \dh _{ [n } h ^{\hat l} _ {p ] } ] ]
 . \label{eq.spinc1} \eea

The gamma matrices identities, $ \G _{ \hat \rho } \G _{\hat l} \G _{
  \hat \rho \hat k} = - \G _{\hat l} \G _{\hat k} ,\ \G _{\hat l} \G
_{ \hat j \hat k} = i \G ^{\hat \rho} \tilde \g _{(4)} \e _{\hat \rho
  \hat j\hat k \hat l} + \d _{\hat j\hat l } \G _{\hat k} - \d _{\hat
  k\hat l } \G _{\hat j} $, yield the simplified form of the Dirac
operator, \bea && \tilde \Dslash \vert _{\S _4} \equiv \G ^ {\rho }
(\dh _\rho + \o _\rho ) + \G ^ {m} (\dh _m + \o _m ) \cr && = e ^{\rho
}_{\hat \rho } \G ^{\hat \rho } (\dh _\rho + {1 \over 2} \eta _{\hat
  k} \dh _\rho \eta ^{\hat k} + \G _{\hat j\hat k} A_{\hat j\hat k} )
+ e ^{m }_{\hat l } \G ^{\hat l } (\dh _m - {1 \over 4} \G _{\hat \rho
  \hat k} C _{m \hat k} + {1 \over 8} \G _{\hat j\hat k} X _{m,\hat j
  \hat k} ) \cr && = e ^{\rho }_{\hat \rho } \g _{(4)} \otimes \tilde
\g ^{\hat \rho } [\dh _\rho + {1 \over 2} \dh _\rho \ln (\eta ^{\hat
    1}\eta ^{\hat 2} \eta ^{\hat 3}) - i \e _{\hat \rho \hat j\hat k
    \hat l} \g _{(4)} \otimes \tilde \g ^{\hat l } \tilde \g _{(4)}
  A_{\hat j\hat k} \cr && + \eta ^{\hat \rho } e ^m _{\hat l } (\tilde
  \g _{\hat \rho \hat l} e ^m _{\hat l} \dh _m + {1 \over 4} C
  _{m,\hat k} \tilde \g _{\hat l \hat k} + {1 \over 8} X _{m,\hat j
    \hat k} (i\e _{\hat \rho \hat j\hat k \hat l} \tilde \g _{(4)} +
  \d _{\hat j\hat l } \tilde \g _{\hat \rho \hat k} - \d _{\hat k\hat
    l } \tilde \g _{\hat \rho \hat j} ) )]. \label{eq.spinc2}\eea

 Applying these formulas to the deformed conifold case, using
 Eq.~(\ref{eq.cof2}) and the expressions implied by the $ SO(3)$
 isometry of $\S _4$, \bea && e_m ^{\hat l } =\eta ^{\hat l } h_m
 ^{\hat l } ,\ e^m _{\hat l } =\eta _{\hat l } h^m _{\hat l } ,\ g
 ^{nn'} = \sum _{\hat l } e^{n} _{\hat l } e^{n'} _{\hat l } ,\ H
 ^{\hat j \hat k } = \d _{\hat j\hat k } \eta _{\hat j} ^2 ,\ \sum
 _{m} h_{m}^{\hat l } h^{m} _{\hat l ' } = \d _{\hat l \hat l' } , \cr
 && g _{\hat j \hat k } = \d _{\hat j\hat k } \eta ^{\hat j 2} ,\ g
 ^{\hat j \hat k } = \d _{\hat j\hat k } \eta _{\hat j}^ 2 ,\ h^{
   m\hat j} = \eta _{\hat j}^ 2 h^m_{\hat j} ,\ h_{ m\hat j} = \eta
 ^{\hat j 2} h_m^{\hat j} , \label{eq.spincp0} \eea one finds that $ C
 _{ m \hat k} $ and $A_{\hat j \hat k} = - A_{\hat k \hat j} $ vanish
 identically while $ e _{\hat l} ^m X _{m ,\hat j \hat k} $ is
 independent of angles.  The resulting Dirac operator \bea && \tilde
 \Dslash \vert _{\S _4} ={1\over \eta ^{\hat \rho } } \G ^{\hat \rho }
 (\dh _\rho + \ud \sum _{\hat k } \dh _\rho (\ln \eta ^{\hat k} ) +
 {\eta ^{\hat \rho }\over 8 \eta ^{\hat l }} h^m_{\hat l}
 X_{m,\hat j \hat k} \G _{\hat \rho } \G _{\hat l}\G _{ \hat j
   \hat k} + \eta ^{\hat \rho } \G _{\hat \rho } \G ^{\hat k }
 e^m_{\hat k} \dh _{m } ) , \label{eq.spincp2}\eea includes on
 side of the first radial derivative term, angle independent
 curvature contributions in the second and third terms (which
 commmute with the chirality matrix $ \G _{\S _4 } $)  and the
 base manifold Dirac operator $ \G ^{\hat k } e^m_{\hat k} \dh
 _{m } = \G ^{\hat k } \nabla _{\hat h _{k} } = \Dslash (S^3
 /Z_2) $ in the fourth term.


\subsection{Undeformed  conifold  limit} 
\label{appD7subsing}

The useful intermediate results  in the undeformed
conifold case are obtained from the deformed conifold results
by taking the limit $\e \to 0, \ \tau \to \infty $ at fixed radial
conic variable $r$ and $\mu $, absorbing the dependence on $\e $
through the change of radial variable $\tau \to r$ and of the parameter
$L \to \mu = L \e $.  The correspondence between the 4-cycle radial
variable $\rho $ and the conifold radial variable $r$ is given by \bea
&& \hat r ^3 \equiv (2/3)^{3/2} r ^3 \equiv \vert \mu _\chi \vert ^2 r
_\chi ^3 = \vert \mu _\chi \vert ^2 (1+\cosh \rho ) ,\ {d r _\chi
  \over d \rho } = {1\over 3} ({r _\chi ^3 -2 \over r _\chi } )^{1/2}
, \label{eqsinc1} \eea where we introduced the useful rescaled and
dimensionless radial variables, $\hat r $ and $ r_\chi $. Note that
the change of variable $\rho \to r _\chi $ picks up the factor ${\dh
  \rho \over dr _\chi }$
which diverges at the 4-cycle apex,  $\hat r _{min} \equiv
(2/3)^{1/2} r _{min} \equiv \vert \mu ^2 \vert ^{1/3} (r _\chi )_{min}
= \vert 2 \mu ^2 \vert ^{1/3} .$
The coefficient functions $ F_{1, 2}$ and the warp profile  have  the
limiting forms, \bea && F_1 \simeq (2 \e ) ^{-2/3} e ^{-\tau /3}
\simeq {3 ^{1/2} 2 ^{-3/2} \over r } = {1 \over 2 \vert \mu _\chi
  \vert ^{2/3} r _\chi } ,\ R = {F_2 \over F_1} \simeq - {e ^{-\tau }
  \over 3 \e ^2} \simeq - {1 \over 6 \vert \mu _\chi \vert ^2 (1
  +\cosh \rho ) } = - {3 ^{1/2} 2 ^{-5/2} \over r^3 } = - {1 \over 6
  \vert \mu _\chi \vert ^{2} r _\chi ^3 } ,\cr && \sqrt {-\g ^{(0)} }
\simeq {\vert \mu _\chi \vert ^{4} \over 64 r^2 } S_\rho (1 + 2 C_\rho
) = {\vert \mu _\chi \vert ^{8/3} \over 96 r_\chi ^2 } ( r _\chi ^3 (
r _\chi ^3 -2) ) ^{1/2} (2r _\chi ^3 -1) ,\ h (r)  =  { L^4_{eff}
  \over r ^4} ,\ [ L_{eff} ^4 = {81 \over 8 } (g_s M \a ' )^2 \ln {r
    \over r _{ir} } ] \label{eqsinc2} \eea
where the  effective curvature radius $ L_{eff} (r) $ was
expressed in terms of the infrared cutoff radius $r_{ir}$ at which $
L_{eff} (r_{ir} ) =0 $.  The conifold parameterization as a foliation
by Kuperstein 4-cycle $\S _4$  is described by the metric \bea && d
\tilde s^2 (\calc _6) = 2^{-5/2} 3^{1/2} {\vert \mu _\chi \vert ^2
  \over r } [ {(1+ 2 \cosh \rho ) \over 3} ( d \rho ^2 + \hat h_3 ^2)
  + \cosh ^2 (\rho /2)\hat h_1 ^2 + \sinh ^2 (\rho /2)\hat h_2 ^2 \cr
  && + {4\over 3} (1+ \cosh \rho ) d \chi d \bar \chi + ({2\over 3}
  \mu _\chi \sinh \rho ( d \rho + i \hat h_ 3 ) d \bar \chi +
  H.\ c. )] \cr && = {1\over 8 \hat r} [ \hat r^3 \hat h_1 ^2+ ( \hat
  r^3 - 2 \vert \mu _\chi \vert ^2 ) \hat h_2 ^2 +{2\over 3} ( 2 \hat
  r^3 - \vert \mu _\chi \vert ^2 ) \hat h_3 ^2 + 6{\hat r (2 \hat r ^3
    - \vert \mu _\chi \vert ^2 ) \over (\hat r ^3 - 2 \vert \mu _\chi
    \vert ^2 ) } d \hat r ^2 \cr && + {8\over 3} \hat r^3 d \chi d
  \bar \chi + ( {4\over 3} ( \hat r^3( \hat r^3 - 2 \vert \mu _\chi
  \vert ^2 ) )^{1/2} (3 ({\hat r \over \hat r^3 - 2 \vert \mu _\chi
    \vert ^2 })^{1/2} d \hat r + i \hat h_3 ) d \bar \chi + H.\ c. )
] . \label{metriconfp} \eea
We quote for convenience the metric tensor
components, \bea && \tilde g _{\rho \rho } = \tilde g _{\hat h_3 \hat
  h_3 } = 2^{-5/2} 3 ^{1/2} {\vert \eta _\chi \vert ^2 \over 3 r} (1+
2 \cosh \rho ) = {\vert \mu _\chi \vert ^2 \over 12 \hat r} ( {2 \hat
  r ^3 \over \vert \mu _\chi \vert ^2 } -1 ) ,\cr && [\tilde g _{\hat
    h_1 \hat h_1}, \tilde g _{\hat h_2 \hat h_2} ] = {2^{-7/2} 3
  ^{1/2} \vert \eta _\chi \vert ^2 \over r} [\pm 1+ \cosh \rho ] =
{\vert \mu _\chi \vert ^2 \over 8 \hat r} [ { \hat r ^3 \over \vert
    \mu _\chi \vert ^2 } -2 ,\ { \hat r ^3 \over \vert \mu _\chi \vert
    ^2 } ] ,\cr && \tilde g _{\chi \bar \chi } = {\hat r^2 \over
  6},\ \tilde g _{\hat r \bar \chi } = {\hat r^2 \over 2},\ \tilde g
_{\hat h_3 \bar \chi } = - {i\over 6} (\hat r (\hat r ^3 - 2 \vert \mu
_\chi \vert ^2 ) )^{1/2} . \label{eqsinc4} \eea

The conifold metric restriction to $\S_4$ is given by \bea && d \tilde
s^2 (\calc _6) \vert _{\S _4} = {\vert \mu_\chi \vert ^{4/3} \over 8 r
  _\chi } ( r _\chi ^3 \hat h_1 ^2 + (r _\chi ^3 -2) \hat h_2^2 +
{2\over 3} (2r _\chi ^3 -1) (\hat h_3^2 + 9 ({r _\chi \over r_\chi^3 -
  2} ) d r_\chi^2 ) ) , \cr && = {3\over 4} [{2 \hat r^3 - \vert \mu
    \vert ^2 \over \hat r^3 - 2 \vert \mu \vert ^2 } d \hat r ^2 +
  {\hat r^2 \over 6 } \hat h_1^2 + {\hat r^3 - 2\vert \mu \vert ^2
    \over 6 \hat r} \hat h_2 ^2 + {2 \hat r^3 - \vert \mu \vert ^2
    \over 9 \hat r} \hat h_3 ^2 ] .  \label{eqsinc3} \eea

The reduced $ D7$-brane action for the scalar modes descending from
the open moduli field $\chi $ is given by \bea && S _B(D7) =- {\mu _7}
\int d^4 {x} d \rho \int \hat h_1 \wedge \hat h_2 \wedge \hat h_3
\sqrt {-\g ^{(0)} } e ^{\phi } [1 +{A_2 S_\rho \over A_1} ({\dh _\rho
    \bar \chi \over \bar \eta _\chi } +{\dh _\rho \chi \over \eta
    _\chi } ) \cr && + {2^{1/2} \over 3^{1/2} r } h (r) {(1 + C_\rho )
    \over 1 + 2C_\rho } \tilde g ^{\mu \nu } \dh _\mu \chi \dh _\nu
  \bar \chi + {4\over \vert \mu _\chi \vert ^2 } ({1 + C_\rho \over 1
    + 2C_\rho } + {1 \over (1 + 2C_\rho )^2 } ) \dh _\rho \chi \dh
  _\rho \bar \chi + \cdots ] \cr && = - {\mu _7 \over g_s ^2 } \int
d^4 x \int \hat h_1 \wedge \hat h_2 \wedge \hat h_3 e ^{\phi } \int d
\rho \sqrt {-\g ^{(0)} } [ 1 + 2 {(r_\chi ^3 (r_\chi ^3 -2 ) )^{1/2}
    \over ( 2r_\chi ^3 -1) } ({\dh _\rho \chi \over \eta _\chi } +
  H.\ c.)  \cr && + {1 \over 3 \vert \mu _\chi \vert ^{2/3}} h (r)
  r_\chi ^2 \bigg (1 - {2 (r_\chi ^3 -2 ) \over ( 2r_\chi ^3 -1) }
  \bigg ) \dh _\mu \chi \dh ^\mu \bar \chi + { 4 r_\chi ^3 \over 3
    \vert \mu _\chi \vert ^{2} ( 2r_\chi ^3 -1) } \bigg (1 - {(r_\chi
    ^3 -2 ) \over ( 2r_\chi ^3 -1) } \bigg ) \dh _\rho \chi \dh _\rho
  \bar \chi ].\label{eqsinc6} \eea
The  first terms (in the last line above)
inside the two pairs large parentheses
in front of $\vert \dh _{\mu , \rho } \chi \vert ^2 $, 
reproduce  the results of~\cite{benini09}  while  
the second terms   arising   from additional contributions
to the Born-Infeld  determinant  are  absent there. 
The pull-back of the  NSNS classical solution on $\S _4$ 
\bea && B_2 ^{cl} =  e ^{-\phi /2} h ^{-1/2}  \vert \eta _\chi
\vert ^4  b (r) \cosh ^2 {\rho \over 2} [- \sinh {\rho
    \over 2} d \rho \wedge \hat h_2 + \cosh {\rho \over 2} \hat h_3
  \wedge \hat h_1 ],\ [b(r) = {81\over 16} {g_s M \a ' \over r^6} \ln
  {r \over r ' _{ir } } ] \label{eqsincB2} \eea
agrees with~\cite{benini09}  and involves an  infrared
radius  distinct   from the warp profile radius in
Eq.~(\ref{eqsinc2}), $ \ln r '_{ir} = {1\over 3} - \ln (2 ^{5/6} 3
^{-1/2} \e ^{-2/3} ) = {1\over 4 } + \ln r_{ir} $.
The  contributions from the classical NSNS solution $ B_2 ^{cl} $ and the
gauge 2-form solution~\cite{benini11} $F_2 ^{cl}= {P \over r^6} \Re
(i\bar \mu \e _{ijk} w_i d w_j \wedge d w_k ) $ in
Eq.~(\ref{eqgaugeflux}),
\bea && B_2 ^{cl} + F_{2} ^{cl} \propto {2\over 3 r^{6} } ( 
  k(r)  + {3 P \over  2 } ) ,\ [k(r) =(3/2)^4 g_s M\a ' \ln (r /r'_{ir} ) ] \eea
  (both of  same  dimension $ E^0$) can  be included  together
  by shifting   the  above radial profile
  $k (r) \equiv  3 r^6 b(r) \to k (r) + {3 P /2 } $. 

\end{appendix}
   
\end{document}